\documentclass[traditabstract]{aa} 
\usepackage{txfonts}
\usepackage{amsbsy}
\usepackage{psfig}
\usepackage[dvips]{graphicx,color}
\usepackage{natbib}
\usepackage{longtable,lscape} 
\bibpunct{(}{)}{;}{a}{}{,}
\newcommand{\lapp}{\mbox{\raisebox{-0.3em}{$\stackrel{\textstyle <}{\sim}$}}}
\newcommand{\gapp}{\mbox{\raisebox{-0.3em}{$\stackrel{\textstyle >}{\sim}$}}}
\newcommand{\be}{\begin{equation}}
\newcommand{\en}{\end{equation}}

\def\zabs{$z_{\rm abs}$}
\def\zem{$z_{\rm em}$~}
\def\lya{Ly$\alpha$}

\def\hi{H~{\sc i}}
\def\mgi{Mg~{\sc i}}
\def\mgii{Mg~{\sc ii}}
\def\feii{Fe~{\sc ii}}
\def\feiia{Fe~{\sc ii}$\lambda$2600}
\def\mgia{Mg~{\sc i}$\lambda$2852}
\def\mgiia{Mg~{\sc ii}$\lambda$2796}
\def\mgiib{Mg~{\sc ii}$\lambda$2803}

\def\kms{km~s$^{-1}$}
\def\tdv{$\int\tau$dv}

\begin{document}
\title{Search for cold gas in strong \mgii\ absorbers at 0.5$<$$z$$<$1.5: nature and evolution of 21-cm absorbers}
\author{N. Gupta \inst{1}
       \and R. Srianand \inst{2}
       \and P. Petitjean \inst{3} 
       \and J. Bergeron \inst{3}
       \and P. Noterdaeme \inst{3} 
       \and S. Muzahid \inst{2}
}

\institute{ASTRON, the Netherlands Institute for Radio Astronomy, Postbus 2, 7990 AA, Dwingeloo, The Netherlands
\\ \email{gupta@astron.nl} 
\and 
IUCAA, Post Bag 4, Ganeshkhind, Pune 411007, India 
\and 
UPMC-CNRS, UMR7095, Institut d'Astrophysique de Paris, F-75014 Paris, France
}

\date{Received 2 March 2012; accepted 17 May 2012}

\titlerunning{21-cm absorption in strong \mgii\ absorbers}

\abstract{

We report 4 new detections of 21-cm absorption from a systematic search of 21-cm
absorption in a sample of 17 strong (rest equivalent width,
$W_{\rm r}$(\mgiia)$\ge$1\,\AA) intervening \mgii\ absorbers at 0.5$<$\zabs$<$1.5.
We also present 20-cm milliarcsecond scale maps of 40 quasars having 42 intervening
strong \mgii\ absorbers for which we have searched for 21-cm
absorption. These maps are used to understand the dependence of 21-cm
detection rate on the radio morphology of the background quasar 
and address the issues related to the covering factor of absorbing gas.
Combining 21-cm absorption measurements for 50 strong \mgii\ systems
from our surveys with the measurements from literature, we obtain a sample
of 85 strong \mgii\ absorbers at 0.5$<$\zabs$<$1 and 1.1$<$\zabs$<$1.5.
We present detailed analysis of this 21-cm absorption sample, 
taking into account the effect of the varying 21-cm optical depth sensitivity and 
covering factor associated with the different quasar sight lines.  
%
We find that the 21-cm detection rate is higher towards the quasars
with flat or inverted spectral index at cm wavelengths.
About 70\% of 21-cm detections are towards the quasars with
linear size, LS$<$100\,pc.  The 21-cm absorption lines having
velocity widths, $\Delta$V$>$100\,\kms\ are mainly seen towards the quasars 
with extended radio morphology at arcsecond scales.
However, we do not find any correlation between the integrated 21-cm
optical depth, \tdv, or the width of 21-cm absorption line, $\Delta$V, 
with the LS measured from the milliarcsecond scale images.  All this can be understood
if the absorbing gas is patchy with a typical correlation length of 
$\sim$30-100\,pc.  
We confirm our previous finding that the 21-cm detection rate for a given optical depth threshold can be
increased by up to a factor 2 by imposing the following additional constraints:  
Mg~{\sc ii} doublet ratio $<$1.1, W(Mg~{\sc ii})/W(Fe~{\sc ii})$<$1.47 and W(Mg~{\sc i})/W(Mg~{\sc ii})$>$0.27. 
This suggests that the probability of detecting 21-cm 
absorption is higher in the systems with high $N$(\hi). 
We show that within the measurement uncertainty, the 21-cm detection rate 
in strong \mgii\ systems is constant over 0.5$<$\zabs$<$1.5, i.e., over  
$\sim$30\% of the total age of universe. 
We show that the detection rate can be underestimated by up to a factor 2 if 21-cm optical 
depths are not corrected for the partial coverage estimated using milliarcsecond 
scale maps. 
Since stellar feedback processes are expected to diminish the filling factor of 
cold neutral medium over 0.5$<z<$1, this lack of evolution in the 21-cm 
detection rate in strong \mgii\ absorbers is intriguing. 
Large blind surveys of 21-cm absorption lines with the upcoming Square
Kilometre Array pathfinders will provide a complete view of the evolution of 
cold gas in galaxies and shed light on the nature of \mgii\ systems and DLAs, 
and their relationship with stellar feedback processes.
}


\maketitle

\section{Introduction}
The diffuse interstellar medium (ISM) exhibits a wide range of  
physical conditions such as temperature, density and radiation field, 
and structures at tens of AU to kpc scales in the form  
of shells, filaments and spurs.   
The conditions in the ISM are intimately related to the local 
star-formation through various stellar feedback mechanisms.  
Therefore understanding the physical conditions 
and the structure of gas in the ISM, and the processes that maintain 
these is a key topic in the field of galaxy evolution.  
The absorption lines seen in the spectra of distant luminous sources such as 
quasi-stellar objects (QSOs) and Gamma-ray bursts (GRBs) provide a powerful tool to determine the physical and 
chemical state of the different phases of ISM in both the Galaxy and 
external galaxies.  
These absorbers provide a luminosity unbiased view of the galaxy evolution 
and measurements of the physical parameters of gas at the scales much smaller 
than that possible with the continuum and emission line studies.

Due to their observability with the ground based optical telescopes, 
the resonant absorption lines of \mgii\ are commonly used  
for detecting the interstellar media and/or gaseous halos surrounding the 
galaxies at $z\lapp2$.  The availability of large samples of low resolution quasar 
spectra from the Sloan Digital Sky Survey (SDSS) has led to the compilation of 
large homogeneous catalogues of \mgii\ absorbers at 
0.35$<$\zabs$<$2.3 that are complete down to 
$W_{\rm r}$(\mgiia)$=$1\AA\ \citep[see e.g.,][]{Prochter06, Quider11}.  
It has been shown using the Hubble Space Telescope (HST) spectroscopic observations that a significant 
fraction (36\%$\pm$6\%) of \mgii\ systems with the rest equivalent 
widths\footnote{Throughout this paper, $W_{\rm r} \equiv W_{\rm r}$(\mgiia) unless specified.},  
$W_{\rm r}>$0.5\AA\ and $W_{\rm r}$(\feiia)$>$0.5\AA\ 
have $N$(\hi)$\ge$2$\times$10$^{20}$\,cm$^{-2}$ 
and therefore are bonafide damped \lya\ (DLA) systems \citep[][]{Rao06}. 
The optical imaging and spectroscopic surveys have shown that the \mgii\ absorbers are 
associated with the galaxies of a wide range of morphological type and are routinely 
observed out to galactic radii of $\sim$50\,kpc.  
\citep[e.g.][]{Bergeron91, Steidel95, Churchill05, Chen10mg2sur, Rao11, Bowen11}.

While it is well established that the \mgii\ absorption lines arise from 
the gas associated with the galaxies, the exact nature and the underlying 
physical process driving the gas seen in absorption is still a matter 
of debate. 
Using the integral field spectroscopy, \citet{Bouche07mg2} have detected
H$\alpha$ emission associated with the 14 intervening \mgii\ absorbers 
($W_{\rm r}>2$\,\AA)
within the QSO impact parameters of 1-40\,kpc implying large star formation 
rates (SFRs) of 1$-$20 M$_\odot$ yr$^{-1}$. 
Similarly, the [O~{\sc ii}] emission lines corresponding to the absorbing galaxies 
have been detected in the individual as well as the stacked QSO spectra of $W_{\rm r}>$0.7\,\AA\ 
\mgii\ absorbers \citep{Noterdaeme10o3, Menard11}. 
These results point towards a strong physical connection between the starburst phenomena 
and the absorbing gas associated with the strong ($W_{\rm r}\gapp$1\,\AA) \mgii\ absorbers 
\citep[see also][]{Prochter06, Murphy07, Nestor11}.
On the other hand, the weaker \mgii\ absorbers are believed to be tracing the 
infalling/accreting gas \citep{Chen10, Kacprzak11}. 

The detection of 21-cm absorption in a well defined sample of 
\mgii\ absorbers towards the radio-loud quasars can be used 
to estimate the filling factor of cold gas in \mgii\ systems. 
Until blind surveys of radio absorption lines become a real 
possibility, this is also the only practical way to search for 
the 21-cm absorption at 0.2$<$\zabs$<$2 and trace the evolution of 
cold gas in galaxies. 
Since the cold neutral medium (CNM) filling factor of gas depends sensitively on the 
stellar feedback, such surveys can also provide insights  
into the physical processes driving the gas seen in \mgii\ absorption.  
There have been several searches for 21-cm absorption in \mgii\ systems 
\citep[e.g.][]{Briggs83, Lane_phd, Curran07, Kanekar09mg2}.    
A systematic survey of 21-cm absorption in a sample of 35 intervening \mgii\ 
absorbers at 1.1$<$\zabs$<$1.5 was presented in  
\citet{Gupta07} and \citet[][hereafter referred to as G09]{Gupta09}.  
The detection of 9 new 21-cm absorbers from this survey significantly increased 
the number of known 21-cm absorbers at the intermediate redshifts. 

G09 provided first estimate of the number density per unit redshift of 
21-cm absorbers ($n_{21}$) associated with the $W_{\rm r}$$\ge$1\,\AA\ \mgii\ absorbers. 
Hereafter, we will refer to $W_{\rm r}$$\ge$1\,\AA\ absorbers as the {\it strong} \mgii\ absorbers. 
Using the twenty-one 21-cm absorption measurements at $0.3<$\zabs$<1$ from the sample of 
\citet{Lane_phd}, they showed that the $n_{21}$ at  
$z$ = 1.3 may have decreased with respect to the $n_{21}$ at $z$ = 0.5. 
The evolution was found to be stronger for the stronger ($W_{\rm r}\ge$1.8\AA)  
\mgii\ systems. This  could be due to the evolution of the CNM filling factor   
of the gas in galaxies, and the fact that the stronger \mgii\ systems 
at higher-$z$ possibly arise in the outflows driven by 
starbursts \citep{Prochter06, Noterdaeme12}.  This interpretation of the results from the 
survey was limited by the availability of the milliarcsecond scale radio images of the  
background quasars and the lack of sensitive 21-cm absorption measurements 
at $z<1$.

As an extension of our work presented in G09, we present 
here the new 21-cm absorption measurements of the 17 \mgii\ systems at 
0.5$<$\zabs$<$1.5  and the 20-cm Very Long Baseline Array (VLBA) maps of the 40 quasars 
to study the evolution of CNM filling factor of \mgii\ systems and 
investigate the effect of radio source structure on the 
detectability of 21-cm absorption.  The 21-cm absorption measurements 
were made using 
the Green Bank Telescope (GBT), the Giant Metrewave Radio Telescope (GMRT) 
and the Westerbork Synthesis Radio Telescope (WSRT). 
The details of these 21-cm absorption and the VLBA observations, 
and the new 21-cm absorption detections are provided in Section~\ref{sec:data}.  
In Section~\ref{sec:fc}, we describe the assumptions and method used to determine 
the correction for partial coverage using the milliarcsecond scale VLBA maps.   
We define a homogeneous sample that is selected primarily on the basis 
of $W_{\rm r}\ge$1\AA\ criterion.  This sample is used to derive the 21-cm detection 
rate in strong \mgii\ absorbers as a function of redshift. 
In Section~\ref{sec:structure}, we discuss the relationship between the 21-cm absorption and 
the radio structure of the background quasar. 
The relationship of 21-cm absorption with the properties of 
metal absorption lines and the dust content of the absorbing gas derived from the 
optical spectra are discussed in Section~\ref{sec:opt}. 
We discuss the implications of our results in Section~\ref{sec:disc}. 
The results are summarised in Section~\ref{sec:summ}.

Throughout this paper we use the $\Lambda$CDM cosmology with 
$\Omega_m$=0.27, $\Omega_\Lambda$=0.73 and H$_{\rm o}$=71\,\kms\,Mpc$^{-1}$.


\section{Data}
\label{sec:data}

\begin{table*}[!t]
\caption{The sample of strong \mgii\ systems ($W_{\rm r}$(\mgiia)$\ge$1\,\AA) observed with the GBT, GMRT and WSRT. 
The systems for which no useful 21-cm absorption spectra could be obtained are also given.  
Listed from left to right are the names of the quasars, their emission redshifts, redshifts of the intervening \mgii\ system, 
rest frame equivalent widths of the \mgiia, \mgiib, \mgia\ and \feiia\ absorption lines, and 1.4\,GHz peak flux densities of the 
quasars as determined from the FIRST or NVSS surveys. 
The last column specifies whether the quasar is compact (C) or resolved (R) at the FIRST resolution.
The quasars with the deconvolved sizes $\le$2$^{\prime\prime}$ are classified as compact \citep{White97first}.   
}
\begin{center}
\begin{tabular}{ccccccccc}
\hline
\hline
{\large \strut} Quasar      &   \zem   & \zabs  &  $W_{\rm r}$(\mgiia)  & $W_{\rm r}$(\mgiib)   & $W_{\rm r}$(\mgia) & $W_{\rm r}$(\feiia) &  Flux       & Mor. \\
            &          &        &        (\AA)          &       (\AA)           &       (\AA)        &      (\AA)          & (mJy/beam)  &      \\
\hline
\multicolumn{9}{c}{\bf {\large \strut} Systems with 21-cm absorption spectra}\\
J045703$-$232452         & 1.003  & 0.8922   & 2.20$\pm$0.02  & 1.90$\pm$0.02  &  0.61$\pm$0.02  & 1.87$\pm$0.02    & 1640  & C\tablefootmark{\dag}     \\
J080036$+$501044         & 1.615  & 1.4146   & 1.94$\pm$0.04  & 1.70$\pm$0.04  &  0.45$\pm$0.05  & 1.26$\pm$0.05    & 113.9 & C            \\
J081710$+$235223         & 1.732  & 1.3060   & 1.70$\pm$0.13  & 1.53$\pm$0.13  &  0.80$\pm$0.15  & 1.38$\pm$0.17    & 210.4 & C            \\
J093035$+$464408         & 2.033  & 0.6216   & 3.18$\pm$0.14  & 2.87$\pm$0.14  &  1.21$\pm$0.14  & 2.44$\pm$0.13    & 313.9 & C            \\
J095631$+$404628         & 1.510  & 1.3237   & 5.02$\pm$0.37  & 4.19$\pm$0.35  &  2.43$\pm$0.36  & 1.97$\pm$0.57    & 30.6  & R            \\
J100718$+$225126         & 2.308  & 0.5602   & 1.19$\pm$0.32  & 1.89$\pm$0.24  &  0.64$\pm$0.27  & 1.03$\pm$0.19    & 341.3 & C            \\
J114856$+$525425         & 1.632  & 0.8306   & 2.77$\pm$0.04  & 2.39$\pm$0.05  &  0.76$\pm$0.06  & 1.66$\pm$0.05    &  95.6 & C            \\
J121604$+$584333         & 1.454  & 0.7487   & 1.55$\pm$0.12  & 1.44$\pm$0.12  &  0.47$\pm$0.12  & 0.97$\pm$0.11    & 376.8 & R            \\ 
J125227$+$442737         & 1.347  & 0.9112   & 1.39$\pm$0.15  & 1.33$\pm$0.14  &  0.36$\pm$0.13  & 1.24$\pm$0.14    & 248.7 & R            \\ 
J132901$+$105304         & 1.933  & 0.6715   & 1.47$\pm$0.18  & 1.45$\pm$0.18  &$<$0.24          & 1.00$\pm$0.25    & 111.2 & C?           \\
J133335$+$164903         & 2.089  & 0.7448   & 1.80$\pm$0.04  & 1.46$\pm$0.04  &  0.18$\pm$0.04  & 0.53$\pm$0.05    & 409.1 & C            \\
J140856$-$075227         & 1.500  & 1.2753   & 2.08$\pm$0.05  & 1.90$\pm$0.05  &  0.62$\pm$0.05  & 1.53$\pm$0.05    & 604.0 & C            \\ 
J141030$+$614136         & 2.247  & 0.7596   & 1.31$\pm$0.16  & 1.48$\pm$0.16  &  0.57$\pm$0.10  & 0.84$\pm$0.19    & 116.1 & C            \\
J150124$+$561949         & 1.467  & 1.2788   & 1.38$\pm$0.10  & 1.02$\pm$0.09  &  0.37$\pm$0.10  & 0.72$\pm$0.12    & 148.8 & C            \\
J163638$+$211255         & 1.801  & 0.8000   & 2.26$\pm$0.20  & 2.41$\pm$0.21  &  0.53$\pm$0.24  & 1.62$\pm$0.20    & 365.8 & C            \\
J203155$+$121941         & 1.215  & 1.1158   & 1.29$\pm$0.05  & 1.16$\pm$0.05  &  0.54$\pm$0.05  & 1.12$\pm$0.05    & 941.0 & R\tablefootmark{\dag}    \\
J212912$-$153841         & 3.268  & 0.6628   & 2.49$\pm$0.003 &     $-$        &      $-$        &     $-$          & 589.7 & R\tablefootmark{\dag}    \\
\multicolumn{9}{c}{\bf Systems with VLBA image but no 21-cm absorption spectra}\\
J081534$+$330529         & 2.426  & 0.8515   & 2.67$\pm$0.28  & 2.19$\pm$0.31  &  0.55$\pm$0.15  & 1.12$\pm$0.27    & 327.9 & C            \\
J110021$+$162914         & 3.380  & 0.8540   & 1.99$\pm$0.11  & 1.64$\pm$0.11  &  0.66$\pm$0.12  & 1.08$\pm$0.13    & 256.5 & C            \\
J115734$+$163859         & 1.061  & 0.7624   & 1.19$\pm$0.07  & 1.05$\pm$0.07  &  0.25$\pm$0.08  & 0.56$\pm$0.07    & 734.3 & C            \\
J121332$+$130720         & 1.139  & 0.7718   & 1.19$\pm$0.08  & 1.10$\pm$0.08  &  0.36$\pm$0.09  & 0.81$\pm$0.08    &1281.0 & C            \\
J130036$+$082802         & 1.090  & 0.8665   & 1.39$\pm$0.08  & 1.07$\pm$0.08  &  0.48$\pm$0.10  & 0.78$\pm$0.11    & 101.3 & C            \\
J143009$+$104326         & 1.710  & 1.2431   & 1.64$\pm$0.09  & 1.53$\pm$0.09  &$<$0.10          & 0.64$\pm$0.17    & 303.9 & C            \\
\hline
\end{tabular}
\end{center}
\tablefoot{
\tablefoottext{\dag} Not covered in FIRST.  
The flux densities for J0457$-$2324, J2031+1219 and J2129$-$1538 are taken from the NVSS. 
The morphology classification for J0457$-$2324 and J2129$-$1538 is 
based on the 1.4\,GHz VLA images with the resolution of $\sim$3$^{\prime\prime}$ 
and $\sim$1.4$^{\prime\prime}$ respectively \citep{Ulvestad81, Neff90}.  
More than 90\% of the flux density of J0457$-$2324 is contained in 
an unresolved component whereas J2129$-$1538 exhibits a core-jet 
morphology with linear size of 4.6$^{\prime\prime}$ and 95\% of the total flux 
density in the core component. The quasar J2031$+$1219 is extended 
in the GMRT image at 610\,MHz (resolution$\sim$5$^{\prime\prime}$) 
presented in Fig.~\ref{gmrtmaps}. 
}
\label{mg2sample}
\end{table*}

\begin{figure}
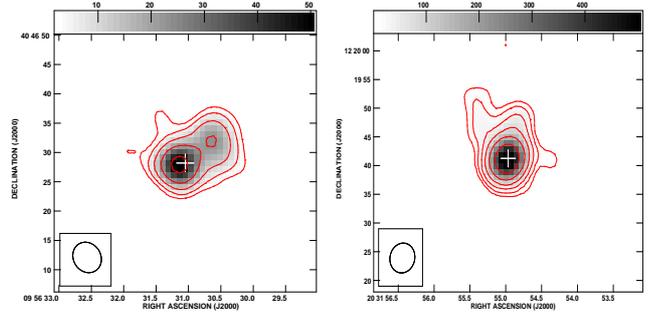

\centerline{
\vbox{
\hbox{
\psfig{figure=j0956_2_mapwithoutlabel.ps,height=4.2cm,width=4.2cm,angle=-90}
\psfig{figure=j2031map_forpaper_nolabels.ps,height=4.2cm,width=4.2cm,angle=-90}
}
}
}
\caption[]{GMRT maps of J0956+4046 ({\it left}) and J2031+1219 that are
resolved in our observations at $\sim$610\,MHz with contour
levels as n$\times$(-1,1,2,4, ...) mJy beam$^{-1}$, where n=2.8 and 10 
respectively.  The maps have rms of 0.36 and 1.4 mJy beam$^{-1}$, and 
the beams are 5.43$^{\prime\prime}$$\times$4.76$^{\prime\prime}$ 
(position angle PA=37$^{\circ}$), and  5.34$^{\prime\prime}$$\times$4.96$^{\prime\prime}$ (PA=$-$39$^{\circ}$) respectively.  
The position of the optical source is marked as `+' \citep[SDSS for J0956+4046;][for J2031+1219]{Condon77}.
}
\label{gmrtmaps}
\end{figure}

\begin{figure*}
\centerline{{
\psfig{figure=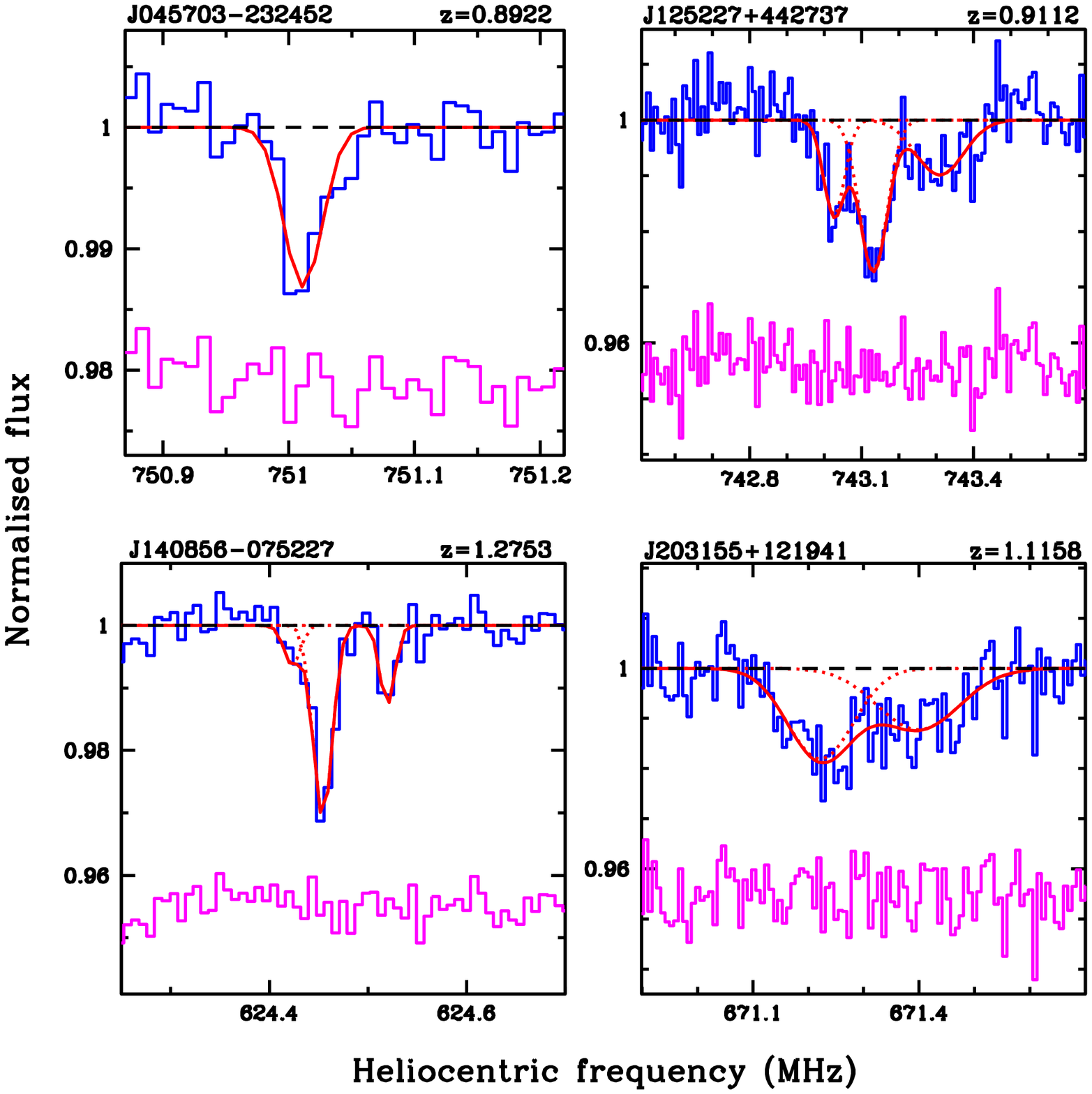,height=13.50cm,width=15.3cm,angle=0}
}}
\caption[]{Spectra of 21-cm absorption detections from the GBT ({\it top}) and GMRT ({\it bottom}). 
Individual Gaussian components and resultant fits to the absorption profiles are overplotted 
as dotted and continuous lines respectively.  
Residuals, with an arbitrary flux offset for clarity, are also shown.
}
\label{mg2det}
\end{figure*}

%

\subsection{Redshifted 21-cm line observations}

\subsubsection{Sample}
\label{sec:21samp}
Our sample of \mgii\ absorbers to search for 21-cm absorption  
is listed in Table~\ref{mg2sample}. The sample is mostly 
drawn from the SDSS Data Release\,7 using an automatic procedure 
that fits the QSO continuum redwards of the Ly$\alpha$ emission using 
Savitzky-Golay filtering and identifies \mgii\ doublets at 0.35$<$\zabs$<$2.3   
through a correlation analysis. 
Similar to G09, we confine our sample to the \mgii\ systems with 
$W_{\rm r}$$\ge$1\,\AA.
The sample was designed for observations with the GMRT 610-MHz and WSRT UHF-high 
receivers (see Section~\ref{sec:obs}). Therefore we selected \mgii\ systems at 
0.5$<$\zabs$<$1.0 and 1.1$<$\zabs$<$1.5. 
Although \mgii\ systems down to the redshifts of $\sim$0.35 can be detected in 
the SDSS spectra and the corresponding redshifted 21-cm absorption is observable 
with the WSRT UHF-high receiver, we chose the lower redshift cut-off of our sample to be at 
\zabs=0.5 to avoid the frequencies around 950\,MHz that are known to be affected 
by strong radio frequency interference (RFI).  
By cross correlating the position of the QSOs having strong intervening \mgii\ 
absorption with radio sources in the Faint Images of the Radio Sky at 
Twenty-Centimeters (FIRST) catalogue  we identified \mgii\ systems that 
are in front of the quasars brighter than 100\,mJy at 20-cm.  
After visual inspection of the optical spectra 
and radio images to ensure that there are no false identifications, 
we selected 19 \mgii\ systems that are not covered in G09 
for our observations. 

In addition, we selected 4 strong \mgii\ systems towards well known  
blazars with the metal absorption line properties suggesting a high probability of them 
being DLAs.  
The \zabs = 0.8922 system towards the BL Lac J0457$-$2324 (PKS\,0454$-$234),    
the \zabs = 1.2753 system towards the FSRQ J1408$-$0752 (PKS\,1406$-$076), and 
the \zabs = 1.1158 system towards the BL Lac J2031$+$1219 (PKS\,2029$+$121), 
were selected from the sample of \mgii\ systems towards blazars 
given in \citet{Bergeron11}.  
The \zabs = 0.6628 system towards the FSRQ J2129$-$1538 (PKS\,2126$-$158) 
was identified in the UVES/VLT large programme data \citep{Bergeron04}. 

In total, we have a sample of 23 strong \mgii\ systems. 
The rest equivalent widths of \mgia, \mgiia, \mgiib\ and \feiia\ for these  
are given in the Table~\ref{mg2sample}. The equivalent widths have been 
estimated by integrating over the absorption profile as done in G09. 
The \mgia, \mgiib, and \feiia\ absorption lines for the system towards J2129$-$1538 
fall in the \lya\ forest, and their profiles are blended with the other absorption features. 
For this reason, the equivalent widths for these absorption lines for this system 
are not given in Table~\ref{mg2sample}.  
A detailed analysis of this system will be presented in Boiss\'e et al. (in prep.).  

\subsubsection{Observations and data reduction}
\label{sec:obs}

The systems at 0.5$<$\zabs$<$1.0 were observed using the WSRT UHF-high receiver 
whereas the systems at 1.1$<$\zabs$<$1.5 were observed using  the GMRT 610-MHz receiver. 
The latter corresponds to the frequency coverage of the 610-MHz receiver.  
The systems at \zabs$<$1 that could not be observed with the WSRT or resulted in 
the spectra severely affected by the RFI were subsequently observed with the 
GBT prime-focus PF1-800\,MHz receiver.  
In total, we observed  17 systems from our sample. 
The Observing log is provided in Table~\ref{obslog}. 
For the GBT observations of 9 \mgii\ systems,  we used the Spectral Processor as 
the backend with a bandwidth of 2.5\,MHz split into 1024 spectral 
channels.  In addition, fast time sampling of 1\,sec was used.  
The observations were done in standard position-switching mode with 
typically 5\,min spent on-source and same amount of time spent at the 
off-source position. The data were acquired in two linear polarization 
channels i.e. XX and YY.  
The GMRT and WSRT observations were done using baseband bandwidth of 
1\,MHz split into 128 and 2.5\,MHz split into 1024 frequency channels 
respectively. 
For flux density and bandpass calibration, the standard flux density calibrators such 
as 3C48, 3C147 and 3C286 were observed.  In GMRT observations, a phase 
calibrator was also observed for 10\,min after every $\sim$45\,min to get 
reliable phase solutions. The data were acquired in the two polarization 
channels RR and LL. The total on-source time after excluding telescope 
set-up time and calibration overheads are also provided in Table~\ref{obslog}.

\begin{figure*}
\centerline{{
\psfig{figure=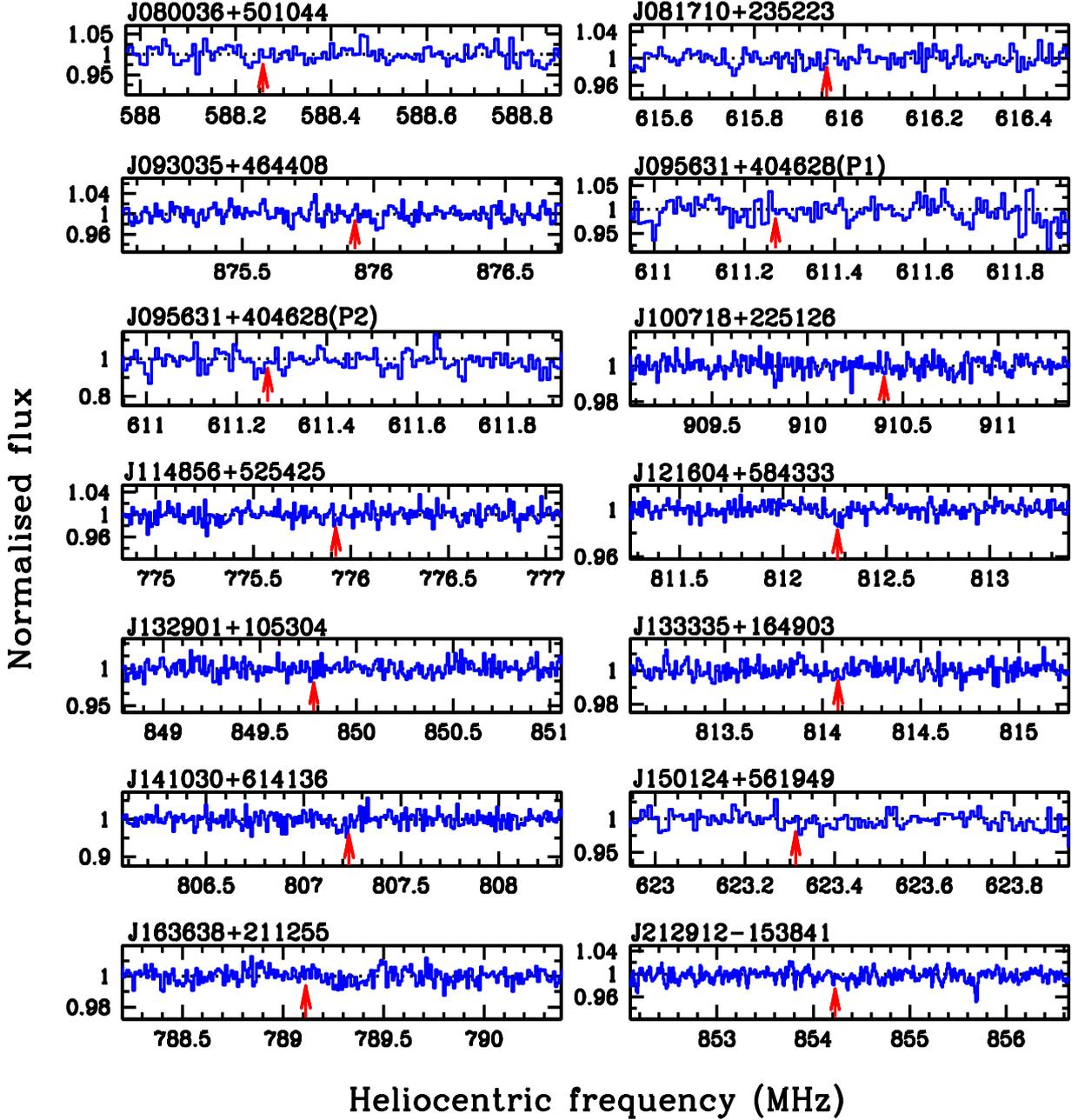,height=18.0cm,width=17.0cm,angle=0}
}}
\caption[]{GBT, GMRT and WSRT spectra of the \mgii\ systems with 21-cm 
non-detection.  Arrows mark the expected positions of 21-cm absorption 
lines based on metal absorption lines.  For J0956+4046, P1 and P2 
correspond to spectra towards the southern and the northern peaks respectively 
(see Fig.~\ref{gmrtmaps} and the text for details).  
}
\label{mg2ndet}
\end{figure*}

The GMRT and WSRT data were reduced using NRAO Astronomical 
Image Processing System (AIPS) following the standard procedures as 
described in \citet{Gupta06}. Unfortunately, the WSRT data for the systems 
towards  J1007+2251, J1148+5254, J1216+5843 and J1329+1053 were severely 
affected by RFI.  Although no useful spectrum could be obtained for 
these systems, the data were still useful for measuring the flux 
densities of J1007+2251, J1216+5843 and J1329+1053 at the redshifted 21-cm 
frequencies, and estimating the accurate optical depths from the GBT data.  
The GMRT and WSRT observations resulted in 2 new 21-cm absorption detections 
and 21-cm optical depth limits for 6 systems. The continuum images of  
two quasars that are resolved in the GMRT observations are presented in Fig.~\ref{gmrtmaps}. 
The quasar J0956+4046 is resolved into two peaks.  We refer to the southern component 
which is coincident with the optical QSO as P1 and the other component as P2 in the following text. 

We used NRAO's GBTIDL package to develop a pipeline to automatically 
analyse the GBT spectral-line datasets.
After excluding the time ranges for which no useful data were obtained, the 
data were processed through this pipeline.
The pipeline calibrates each data record individually and 
flags the spectral channels with deviations larger than 5$\sigma$ seen in 
either XX or YY as affected 
by RFI.  After subtracting a first order baseline these data are  
averaged to produce the baseline (i.e. continuum) subtracted spectra  
for XX and YY.    
The baseline fit and statistics for flagging are determined using the spectral 
region that excludes central 25\% and last 10\% channels at both ends 
of the spectrum.
If necessary a first-order cubic spline was fitted to the averaged XX and YY 
spectra obtained from the pipeline, which were then combined to produce the stokes 
I spectrum.  The spectrum was then shifted to the heliocentric frame.  
The multiple-epoch spectra for a system were then resampled onto the same frequency 
scale and combined to produce the final spectrum.  
These GBT observations resulted in 2 new 21-cm absorption detections 
and 21-cm optical depth limits for 7 systems.

The confusion due to other radio sources in the primary beam is not an 
obstacle in determining accurate flux densities of the background quasars and optical depths 
for the systems observed with the GMRT and WSRT. 
But for the systems observed with the GBT, the rms confusion due to other sources 
in the beam can be the dominant effect that limits the accuracy with which 
the 21-cm optical depth can be determined.  
We used interferometric images from our WSRT observations and 
literature to address this issue.  The details are provided in 
Appendix~\ref{sec:conf} and the flux densities and optical depth 
values for all the systems are listed in Table~\ref{mg2obsres}. 

In summary, our spectroscopic observations have resulted in 4 new 
21-cm absorption detections that  are shown in Fig.~\ref{mg2det}. 
The spectra of 13 21-cm non-detections are presented in Fig.~\ref{mg2ndet}.  
The GMRT spectra are at a spectral resolution of $\sim$4\,km\,s$^{-1}$ whereas 
GBT and WSRT spectra have been smoothed to $\sim$4\,km\,s$^{-1}$.  
For the exact resolution of the spectra plotted in Figs.~\ref{mg2det} and \ref{mg2ndet} 
see Table~\ref{mg2obsres}.  

\subsection{Details of new 21-cm absorption detections}
\label{sec:new21}
Now we present the details of 4 new 21-cm absorption detections (Fig.~\ref{mg2det}).  
In Table~\ref{21cmfit} we provide the details of Gaussian components 
i.e. the absorption redshift (\zabs), the full width at half maximum (FWHM) and 
the peak optical depth ($\tau_p$) fitted to these spectra.  These Gaussian components provide 
convenient parametrization of absorption lines even if they do not 
necessarily represent the actual physical components.   

\begin{table}
\caption{Details of multiple Gaussian fits to 21-cm absorption profiles. 
Listed from left to right are the names of quasars, and the absorption 
redshifts, FWHMs and peak optical depths of the Gaussian components.
}
\begin{center}
\begin{tabular}{cccc}
\hline
\hline
{\large \strut} Quasar             &    \zabs  & FWHM  & $\tau_p$ \\
                   &           &    (\kms)         &                  \\
\hline
{\large \strut} J0457$-$2324       &      0.89132    &  14$\pm$2   &  0.013$\pm$0.002 \\
~J1252$+$4427       &      0.91138    &  35$\pm$6   &  0.027$\pm$0.002 \\
                   &      0.91092    &  61$\pm$16  &  0.010$\pm$0.002 \\ 
                   &      0.91165    &  27$\pm$6   &  0.017$\pm$0.003 \\ 
~J1408$-$0752\tablefootmark{\dag}&      1.27464    &  11$\pm$1   &  0.031$\pm$0.002 \\ 
                   &      1.27440    &   9$\pm$2   &  0.013$\pm$0.002\\ 
                   &      1.27475    &  10$\pm$6   &  0.006$\pm$0.002 \\ 
~J2031$+$1219       &      1.11558    &  78$\pm$24  &  0.013$\pm$0.002 \\
                   &      1.11614    &  62$\pm$12  &  0.018$\pm$0.002 \\ 
\hline
\end{tabular}
\end{center}
\tablefoot{
\tablefoottext{\dag} Corresponds to the spectrum taken on 21/6/2009 (see Fig.~\ref{mg2det}). 
} 
\label{21cmfit}
\end{table}

\begin{enumerate}
\item{\bf \zabs=0.8922 system towards J0457$-$2324:}
The background quasar, also known as PKS0454$-$234, is a BL Lac object 
and highly polarized with optical polarization of 27\% \citep{Wills92}.  
The 21-cm absorption detected in our GBT spectrum has velocity 
width,\footnote{Velocity width of absorption profile corresponding to 
the 5\% and 95\% percentiles of the apparent optical depth distribution estimated 
using the method described by \citet{Ledoux06a}.}
$\Delta$V=23.4\,\kms\  
and integrated 21-cm optical depth, \tdv=0.20$\pm$0.02\,\kms.  
The \hi\ column density of an optically thin cloud covering the fraction 
$f_c$ of the background radio source is related to the integrated 21-cm 
optical depth and the spin temperature (T$_s$) through  
\begin{equation}
N{\rm (H~{\scriptstyle I})}=1.823\times10^{18}~{{\rm T}_{\rm s}\over f_{\rm c}}\int~\tau({\rm v})~{\rm dv}~{\rm cm^{-2}}.
\label{eq:t21}
\end{equation}
%
This corresponds to an \hi\ column density of 
$N$(\hi)=5.8$\times 10^{19}$(T$_s$/100)(0.63/$f_c$)\,cm$^{-2}$ for the 
absorber towards J0457$-$2324. 
Here, we have assumed $f_c = c_f$ where $c_f$ is defined as the ratio 
of the flux density of the VLBA `core' to the total arcsecond scale 
flux density (see  Section~\ref{sec:fc} for the details regarding $f_c$). 

\item{\bf \zabs=0.9112 system towards J1252$+$4427:}
The background quasar is resolved in FIRST survey with a deconvolved size 
of 3.68$^{\prime\prime}\times$1.01$^{\prime\prime}$.  
The 21-cm absorption detected in our GBT spectrum has $\Delta$V=134\,\kms  
and \tdv=2.06$\pm$0.13\,\kms.  This corresponds to $N$(\hi)=3.8$\times 10^{20}$(T$_s$/100)(1/$f_c$)\,cm$^{-2}$. 
No subarcsecond scale images are available for this quasar to constrain $f_c$.

\item{\bf \zabs=1.2753 system towards J1408$-$0755:}  
The background quasar is a blazar at $z$=\,1.494 and   
the associated radio jet at milliarcsecond scale exhibits the highest 
superluminal velocity ($>$25c) amongst the known blazars \citep{Piner06}.  
The 21-cm absorption detected in our GMRT spectrum has $\Delta$V=45.0\,\kms\  
and \tdv=0.51$\pm$0.05\,\kms.  This corresponds to 
$N$(\hi)=2.5$\times 10^{20}$(T$_s$/100)(0.38/$f_c$)\,cm$^{-2}$, 
where we have assumed $f_c = c_f$. 
We reobserved this system with the GMRT after 7 months (Table~\ref{obslog}) 
to obtain the higher resolution spectrum and investigate the variability in 
the 21-cm absorption profile due to jet proper motion.  This higher resolution 
spectrum is presented in the top panel of Fig.~\ref{j1408}.  
It shows three well detached components: A1, A2 and A3. The detection of
multiple components in this absorber could be related to the complex morphology 
exhibited by this radio source at milliarcsecond scales. While the components A1 and A2 are 
clearly detected at both the epochs, the same cannot be said about the component A3. 
The component A3 has \tdv=0.07$\pm$0.02\,\kms\ and is detected with a significance
of 3.7$\sigma$ at the second epoch. 
Unlike the features in the RFI affected spectral range shown in  
Fig.~\ref{j1408}, the component A3 is also consistently reproduced in the 
spectra from the two polarisations. 
However, due to the lower resolution and S/N we cannot be certain of the 
presence of this component at the first epoch. If this component is indeed 
real and was not present at the first epoch then it would be related to the small scale structure 
in the CNM gas being traced by the jet components \citep[e.g. C2, C3 or C4 as defined by][]{Piner06}. 
The proper motion of $\sim$0.3\,mas\,yr$^{-1}$ as measured by \citet{Piner06}
for these components would correspond to a scale of $\sim$2\,pc at the absorber redshift 
for a period of seven months.  
The follow-up GMRT observations to investigate this further are in progress.

\begin{figure}
\centerline{{
\psfig{figure=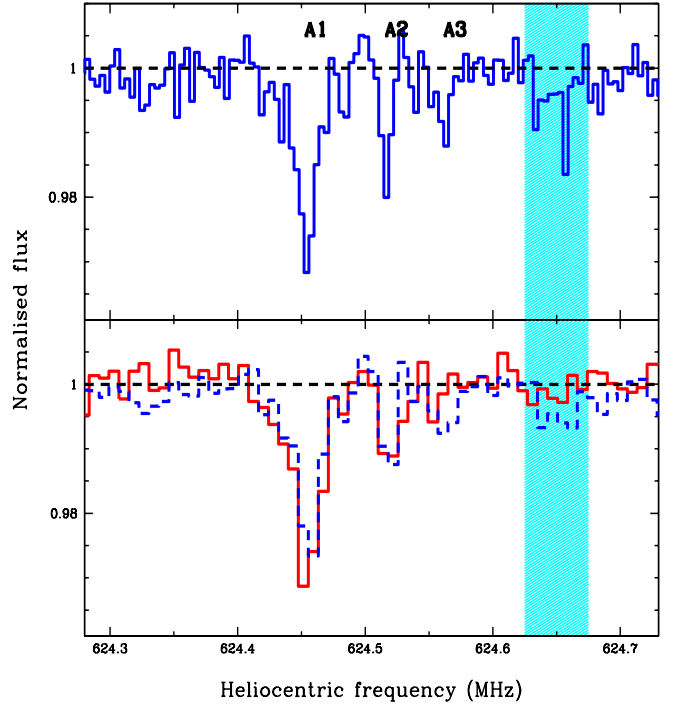,height=10.0cm,width=9.0cm,angle=0}
}}
\caption[]{21-cm absorption spectra of \zabs = 1.2753 \mgii\ absorber towards J1408$-$0752.  Top panel shows 
the higher resolution (1.9\,\kms) spectrum taken on 22/1/2010 
and bottom panel shows the lower resolution (3.8\,\kms) 
spectrum taken on 21/6/2009.   
The 22/1/2010 spectrum smoothed to the resolution of 21/6/2009 spectrum is 
also plotted (dashed line) in the bottom panel. 
The shaded region corresponds to the RFI affected spectral range in the higher 
resolution spectrum.    
}
\label{j1408}
\end{figure}

\item{\bf \zabs=1.1158 system towards J2031$+$1219:}
The background radio source is a BL Lac object at $z$=1.215. 
This radio source is extended in our GMRT image (Fig.~\ref{gmrtmaps}).
The VLA A-array map from \citet{Rector01} shows this quasar to be core dominated 
at the arcsecond scales.  
A broad and shallow 21-cm absorption is detected towards this quasar.  
The 21-cm absorption detected in our GMRT spectrum has $\Delta$V=143\,\kms  
and \tdv=2.11$\pm$0.15\,\kms.  This corresponds to 
$N$(\hi)=8.6$\times 10^{20}$(T$_s$/100)(0.45/$f_c$)\,cm$^{-2}$, 
where we have assumed $f_c = c_f$. 

\end{enumerate}

\subsection{Defining a homogeneous sample of 21-cm absorbers: measurements from literature}
\label{sec:unbiased}

\begin{table*}
\caption{Summary of all the $W_{\rm r}\ge$1\,\AA\ \mgii\ absorbers at 0.5$< z <$1 and 1.1$< z <$1.5 with 21-cm 
absorption measurements considered here.
The samples S1 and S2 discussed in the text are comprised of the subsamples B-E and A-E, respectively.}
\begin{center}
\begin{tabular}{ccccc}
\hline
\hline
{\large \strut} Sample     &     Number        & 21-cm absorption  &   Selection       &  Reference                 \\
           &   of systems  &   detections      &   criterion       &                            \\
\hline                                  
{\large \strut}    A       &    4             &                    3       &   $W_{\rm r}\ge$1\AA\ + Zn~{\sc ii}\,/Cr~{\sc ii}\,/Mn~{\sc ii}                 &  This paper                \\
                  ~B       &   13             &                    1       &   $W_{\rm r}\ge$1\AA          &     ''                     \\
                  ~C       &   32\tablefootmark{\dag}      &       8       &     ''            &  \citet{Gupta09}           \\
                  ~D       &   23             &                    1       &     ''            &  \citet{Kanekar09mg2}      \\
                  ~E       &   13             &                    2       &     ''            &  \citet{Lane_phd}          \\
\hline
\end{tabular}
\end{center}
\tablefoot{
\tablefoottext{\dag} Measurements of 33 $W_{\rm r}\ge$1\,\AA\ \mgii\ systems are available from G09. 
However, the optical depth value for the system towards J1145+0455 is taken 
from the more sensitive spectrum of \citet{Kanekar09mg2}.  
} 
\label{sum}
\end{table*}

\begin{figure}
\centerline{{
\psfig{figure=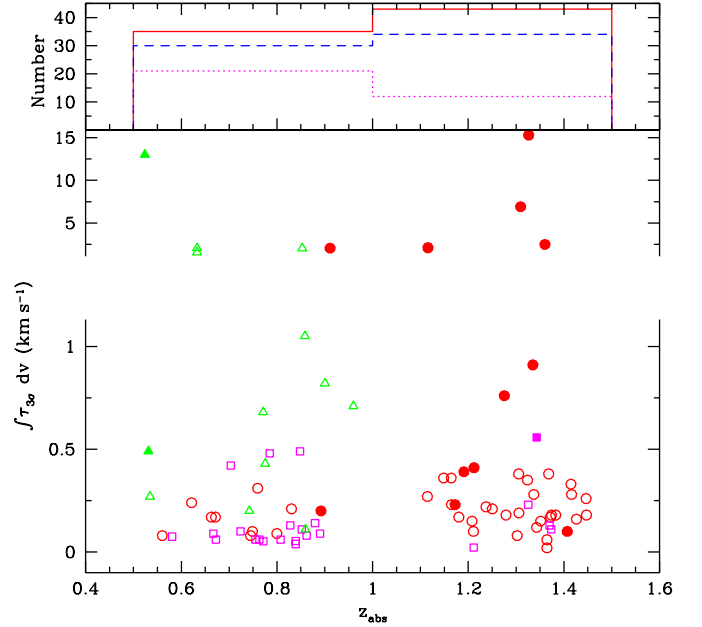,height=9.0cm,width=9.0cm,angle=0}
}}
\caption[]{Redshift distribution of 21-cm optical depths for the strong 
\mgii\ absorbers from this paper and G09 (circles), \citet{Kanekar09mg2} (squares) and 
\citet{Lane_phd} (triangles).  The filled symbols are for 21-cm absorption 
detections and the open symbols correspond to 3$\sigma$ optical depth limits for  
21-cm non-detections.  In the {\it top} panel, dotted, dashed and solid line histograms are for the 
sample S2 with $\int\tau_{3\sigma}$dv$\le\,$0.1, 0.3 and 0.5\,\kms\ respectively.  
}
\label{tdvzabs}
\end{figure}

The purpose of this Section is to define a sample that includes only the 
21-cm absorption measurements of the \mgii\ systems that have been primarily 
selected on the basis of $W_{\rm r}\ge1$\,\AA\ criterion, and is therefore 
suitable for determining the 21-cm detection rate in the strong \mgii\ systems 
and its redshift evolution in an unbiased way. 
Hereafter, we refer to this homogeneous sample as S1.
From our survey presented in this paper and G09, in total, we have 21-cm 
absorption measurements for 50 strong \mgii\ systems over the redshift range 
$0.5<z<1.5$ albeit with a small gap at $z\sim1$. 
The redshift distribution of these systems is plotted in the Fig.~\ref{tdvzabs}.  
Only 46 of these systems that are selected primarily on the basis of $W_{\rm r}\ge1$\,\AA\ \mgii\ 
absorption should be included in the sample S1.  
We do not include 4 systems towards the blazars mentioned in Section~\ref{sec:21samp} 
in S1 as these were specifically selected on the basis of 
metal absorption line properties suggesting the \hi\ column densities 
similar to the DLAs (see Section~\ref{sec:drr1}). 
It is probably not just a coincidence that 3 of these systems are detected in 
the 21-cm absorption (Section~\ref{sec:new21}).   

Next we enlarge the sample S1 by considering  
those 21-cm absorption measurements from the literature that match our 
sample definition i.e. 
(1) fall in the redshift range: $0.5<z<1.0$ and $1.1<z<1.5$, and 
(2) selected primarily on the basis of $W_{\rm r}\ge1$\,\AA\ criterion.
Since the inclusion of individually reported detections from the literature can bias 
our detection rate estimates, we take measurements only from the \mgii\ selected 
samples of \citet{Lane_phd} and \citet{Kanekar09mg2} for which 
both the 21-cm detections and non-detections have been systematically reported.   
From \citet{Kanekar09mg2}, the independent measurements for 4 systems 
at $1.1<z<1.5$ and 18 systems at $0.5<z<1.0$ are available.  
These are listed in Table~\ref{kansamp}.  
We have measured metal absorption line equivalent widths for these systems 
from the SDSS and the VLT archival spectra by integrating over the absorption profile 
in the same way as for our systems.   
Measurements for 13 $W_{\rm r}\ge$1\,\AA\ systems  at $0.5<z<1.0$ are available from 
\citet{Lane_phd}.  These are listed in Table~\ref{lanesamp}.

Thus, we have a homogeneous sample S1 of 81 strong \mgii\ systems 
with 21-cm absorption measurements. 
This sample having 12 21-cm detections will be used 
for investigating the redshift evolution of 21-cm detection rate amongst 
the \mgii\ absorbers. 
The full sample, hereafter referred as S2, of 85 systems that also includes 4 above  
mentioned systems towards the blazars will be used when examining the 
relationship of 21-cm absorption line properties with the radio structure and 
metal absorption line ratios.
A summary of these samples is provided in Table~\ref{sum}.  
In Fig.~\ref{tdvzabs}, we plot 21-cm optical depths as a 
function of redshift for the sample S2. 
For uniformity and ease of comparison with the measurements from literature, 
the 21-cm optical depth limits ($\int\tau_{3\sigma}$dv) for all the systems 
have been computed for a velocity resolution of 10\,\kms.
The median $\int\tau_{3\sigma}$dv achieved for the systems 
from \citet{Lane_phd} is $\sim$0.64\,\kms.  This is at 
least a factor three lower sensitivity as compared to that achieved for 
our measurements or for the systems from \citet{Kanekar09mg2}. 
The method we use to estimate the 21-cm detection rates by taking 
into account the variation in optical depth sensitivities towards different 
quasar sight lines is described in Section~\ref{sec:detrate}.

\subsection{VLBA continuum observations}

\subsubsection{Motivation and sample}

The detectability of 21-cm absorption, in addition to the optical depth, 
depends directly on the covering factor, $f_c$, of the absorbing gas 
(see Equation~\ref{eq:t21}).  
Since radio sources in general exhibit emission over a wide range of 
physical scales ranging from pc to kpc the $f_c$ will depend both on the 
overall extent of the background quasar and the small-scale 
structure (i.e. clumpiness) of absorbing gas.  
The motivation behind these VLBA observations is to understand the 
relationship between the detectability of 21-cm absorption and the structure 
in the radio source and absorbing gas at the scales of tens of pc. 
For this purpose we selected all the systems from our sample 
that are towards the quasars compact at arcsecond scales.  
This list of  42 \mgii\ systems towards 40 quasars is 
drawn from Table~1 of G09 and Table~\ref{mg2sample} of this paper, and   
it includes all the systems towards the compact quasars from our sample with 21-cm absorption 
spectra.  
Exceptions are J0457$-$2324 and J2031+1219.
The former, although compact at arcsecond scales, could not be observed 
whereas observations of BL Lac J2031+1219, extended at arcsecond 
scales, could be included within our observing schedule. 

\subsubsection{Observations and data reduction}
The VLBA 20-cm observations were done as part of a larger survey to obtain milliarcsecond  
scale images of the quasars with DLAs and \mgii\ systems along the sight line.  
Observations were carried out on  
28/10/2008 ($\sim$5\,hrs), 7/11/2008 ($\sim$14\,hrs), 21/2/2010 ($\sim$11\,hrs) and 
10/6/2010 ($\sim$17\,hrs) \citep{Srianand12dla}. 
We used eight 8\,MHz baseband channels i.e. the total bandwidth of 64\,MHz.  
Each baseband channel was split into 32 spectral points. 
Both the right-hand and left-hand circular polarization 
channels were recorded.  
Two bit sampling and the time resolution of 2\,seconds were used. 
  
The observations were done using nodding-style phase-referencing 
with a cycle time of $\sim$5\,min i.e.$\sim$3\,min on 
the source and $\sim$1.5\,min on the phase-referencing calibrator. 
The phase-referencing calibrators were selected from  
the VLBA Calibrator Survey (VCS)\footnote{http://www.vlba.nrao.edu/astro/calib/} 
at 13-cm and 3.6-cm.  
In order to improve the {\sl uv}-coverage, the total time 
on each source was split into {\it snapshots} over a 
number of different hour angles. Each source was 
typically observed for a total of $\sim$30\,min.
Strong fringe finders/ bandpass calibrators 
such as J0555+3948, J0927+3902, J1800+3848 and J2253+1608  
were also observed every $\sim$3\,hrs for 4-5\,min. 

The data were calibrated and imaged using AIPS and DIFMAP in a standard way 
\citep[e.g.][]{Beasley02}. 
The global fringe fitting was performed on the phase-referencing calibrators.  
The delays, rates and phases estimated from these were transferred to the 
target source which was then self-calibrated until the final image was obtained. 
The VLBA maps of these 40 quasars are presented in Fig.~\ref{vlbamap}.

The milliarcsecond scale structure was characterised by fitting Gaussian models to the 
self-calibrated visibilities.  The results of the model fitting are provided in columns\#\,5-11 
of Table~\ref{vlbares}.  
We define $f_{VLBA}$ as the ratio of total flux density detected in the VLBA and the arcsecond scale 
image. Similarly, $c_f$ is defined as the ratio of the flux density of the VLBA  `core' component 
to the total arcsecond scale flux (see Section~\ref{sec:fc} for details).  
The largest linear size (LS) represents separation between the farthest components of the 
radio source at the absorber redshift.  For the compact sources represented by single component 
we take the major axis of the deconvolved component as the upper limit on source size.  
The $f_{VLBA}$ and $c_f$ are discussed in detail in the 
following sections where we use them to estimate $f_c$ and correct observed 21-cm optical depths 
for partial coverage.       

\addtocounter{table}{1}

\subsection{Milliarcsecond scale images from literature and covering factor}
\label{sec:masmap}
The radio morphology of all the quasars in the 21-cm absorption sample S2   
are important to our later discussion.  Therefore, for the quasars that were not 
covered in our VLBA 20\,cm survey we have compiled this information from literature.  
For J0457$-$2324, J1145+0455 and J2129$-$1538 from our sample, 
the $c_f$, $f_{VLBA}$ and largest linear size (LS) 
obtained using maps at 13-cm from the VCS are provided at the end of 
Table~\ref{vlbares}. 

In addition, the milliarcsecond images for the 14 quasars with 15 intervening systems taken 
from \citet{Kanekar09mg2}  are available at 13-cm from the VCS, and for another quasar B0812+332 
from our VLBA survey.  The $f_{VLBA}$, $c_f$ and LS for these 
sources are given in the last three columns of Table~\ref{kansamp}. 

From the sample of 13 systems taken from \citet{Lane_phd}, we will 
consider here only six systems that have been observed with      
the optical depth sensitivity, $\int\tau_{3\sigma}$dv$<$0.6\,\kms.  
From this, the milliarcsecond images at 13-cm are available for B0235+164, 
B0454+039 and B1218+339 from VCS, and for B1629+139 from \citet{Dallacasa98} (Table~\ref{lanesamp}).    
As we shall see later in the Section~\ref{sec:arcsec}, the remaining 7 systems 
that are all 21-cm non-detections, have been observed with the optical depth sensitivity 
that is too poor to address the covering factor related issues.

In total, we have information on milliarcsecond scale radio structure for 
54 quasars with 57 intervening systems from the sample S2.  
The VLBA information for the 34 quasars (with 36 systems along their line
of sight) are based on 20-cm VLBA images and for the remaining quasars it is
based on 13-cm VLBA images. 
Unlike using the source sizes and flux densities obtained from the Gaussian 
parametrization of 20-cm maps from our survey, the $f_{VLBA}$ and $c_f$ at 13-cm have 
been estimated using the total and peak flux densities from the VCS.  
However, these values of the $c_f$ and $f_{VLBA}$ for the cases with the `core' 
identifications, especially in the 13/18 sightlines where the background quasar can be 
represented by a single component, should not differ\footnote{Based on the data from our 20-cm VLBA survey.}  
by more than $\sim$10\% relative to the estimates from the Gaussian parametrization technique. 
The total arcsecond scale flux densities at 13-cm in all these cases have been 
estimated by interpolating flux densities available from the NASA extragalactic database.    
The upper limits on the size of quasars that are represented by single 
component have been estimated using the major axis of point spread function. 
These are conservative compared to those obtained using the Gaussian deconvolution.    

\section{Estimating the gas covering factor and 21-cm absorption detection rate}
\label{sec:fc}

Radio source structure often consists of a combination of compact 
components and extended diffuse emission.  The latter due to its 
low surface brightness sensitivity will only contribute to the 21-cm 
optical depth sensitivity if the large scale structure  of the
absorbing gas is well aligned to the radio emission of the quasar.
Such fortuitous combinations are rare.  Therefore by using the flux densities 
from our spectroscopic observations that at the best have a spatial 
resolution of $\sim$30-40\,kpc at the absorber redshift, in general, 
we are overestimating optical depth sensitivities.  

\begin{figure}
\centerline{
\vbox{
\hbox{
\psfig{figure=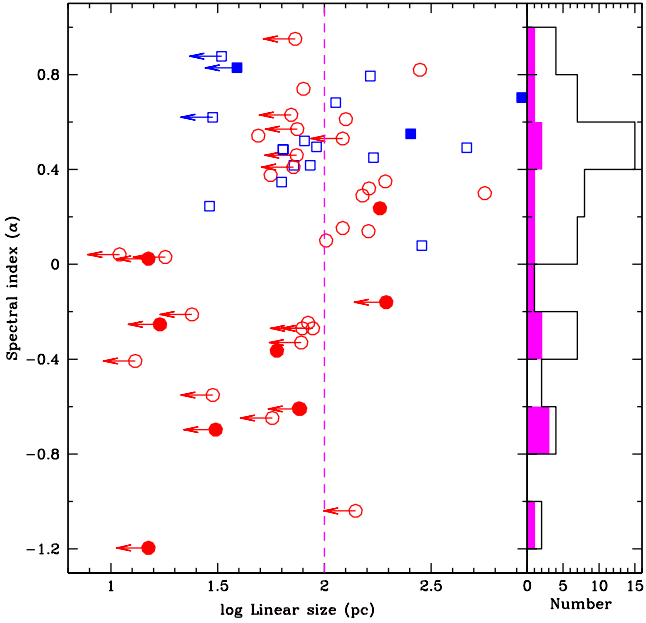,height=9.0cm,width=9.0cm,angle=00}
}
}
}
\caption[]{Spectral index ($\alpha$) vs linear size (in pc) at the absorber redshift 
for the quasars with the milliarcsecond scale images.  The $\alpha$ are from the arcsecond scale 
images.  Systems towards the quasars with no `core' 
identifications are plotted as squares. 
The point for the system towards the quasar B1629+120 that has $\alpha$=+0.54 and LLS=7.2\,kpc 
is not plotted to keep the x-axis short. 
The histogram for the $\alpha$ is also shown. The filled symbols and shaded histogram are 
for 21-cm detections.   
}
\label{alphalls}
\end{figure}

To show that this is indeed the case with the measurements considered in 
Table~\ref{sum}, in Fig.~\ref{alphalls}, we plot the spectral 
index\footnote{We adopt the definition that the flux density at 
frequency $\nu$, S$_\nu\propto$$\nu^{-\alpha}$.} ($\alpha$) 
vs linear size (LS) for all the quasars from the sample S2 
with milliarcsecond scale images.  
These spectral indices have been estimated between 20-cm and 
50-cm/90-cm using the flux densities from the maps with arcsecond scale resolution 
and range from $-$1.2 to 0.7.  
It is clear from Fig.~\ref{alphalls} that the 21-cm detection rate is higher towards 
the quasars with the flat/inverted spectral indices. 
The 44$^{+24}_{-16}$\% (7/16) of systems towards the quasars with $\alpha\le$0 
from the sample S2 are detected in 21-cm absorption as 
compared to the 15$^{+9}_{-6}$\% (6/41) towards the quasars with $\alpha >$0. 
It is also clear that 73\% of the 21-cm detections are towards the quasars with LS$\le$100\,pc 
and amongst the quasars with LS $\le$ 100\,pc 
the detection rate is less towards the quasars with steeper spectrum 
(i.e $\alpha > 0$).  This may imply that the characteristic size of the
absorbing gas is less than 100\,pc. The low detection rate
in the quasars with the steeper spectrum will then be related to the  
poor optical depth sensitivity to detect 21-cm absorption towards the `cores'. 
The term `core' is used here to represent the radio component coincident with the 
optical QSO.  

Thus, identifying the `core' component in the sample S2  
is important to confirm these suggestions and estimate $f_c$ of the absorbing gas.
For this, we need milliarcsecond scale maps at multiple frequencies 
to estimate the spectral indices of the various components detected in the 20-cm or 13-cm 
VLBA maps discussed in Section~\ref{sec:masmap}.  For the 42 systems, images of the background quasar at 
more than one frequency are available through our observations, the VCS and 
the VLBA Imaging and Polarization Survey (VIPS).  
Using these images at  20-cm, 13-cm, 6-cm and 3.6-cm we find that in 
36 cases the dominant VLBA component has a flat spectrum at milliarcsecond 
scales and can be identified as the `core'.  
Although for the quasars J0154$-$0007, J0845+4257 and J0852+3435 only 
images at 20-cm are available but all three have more than $\sim$70\% 
of the arcsecond scale flux contained in the dominant VLBA component. 
We identify this component as the `core' on the basis of $\alpha\lapp$0 estimated 
at the arcsecond scale.  Therefore, in total, for the 39 systems we identify 
the dominant VLBA component of the background quasar as `core'.  

In Fig.~\ref{alphalls}, we plot 17 quasars with 18 intervening systems that have no 
`core' identification as squares.   
In 6 cases, the multifrequency VLBA images are available, but the dominant component at the milliarcsecond 
scales has a steep spectrum and cannot be identified as `core'. 
No multifrequency VLBA images are available for the remaining 11 quasars with 12 intervening systems.  
These 17 quasars with no `core' identifications also have on the average steeper spectral index. 
The $c_f$ values for these are marked as `*' in Tables~\ref{vlbares} and \ref{kansamp}.  
This subset of 18 systems includes three 21-cm detections which are towards 
J1058+4939, J1145+0455 and J1623+0718.  
The former two exhibit complex morphology at milliarcsecond scales.  The third is compact in 20-cm 
VLBA map but unlike most of the other quasars that are compact at milliarcsecond scales has a 
low $c_f$ value of 0.42.  We will discuss the nature of these systems in detail 
in the subsequent sections.

Thus, we have `core' identification for 39 quasars from the sample S2. 
We will consider the following two scenarios and use $c_f$ and $f_{VLBA}$ to estimate
$f_c$ for the 21-cm absorption systems towards these.
In the first scenario we define $f_c=c_f$. 
The assumption in this case is that the 21-cm absorbing gas only covers the `core'. 
If the filling factor of the absorbing gas at these scales is unity, this provides a 
conservative lower limit on the actual value of $f_c$.
In the second scenario we use $f_c=f_{VLBA}$.  The assumptions in this 
case are that the 21-cm absorbing gas extends well beyond $\sim$300\,pc 
to cover all the milliarcsecond scale components and the diffuse extended ($>$kpc scale) 
emission that got resolved out in the VLBA maps does not contribute to 
the optical depth sensitivity (see also Section~\ref{sec:cav}).

\subsection{Detection rate}
\label{sec:detrate}

In this Section, we determine 21-cm detection rates in the strong \mgii\ 
absorbers.  Due to the reasons mentioned in Section~\ref{sec:unbiased}, we 
will use the homogeneous sample S1 (Table~\ref{sum}) for this purpose. 
In the sample S1,  
there are 40 systems at $0.5<z<1$ and 41 at $1.1<z<1.5$.  
There are 3 and 9 detections in the low- and high-$z$ bins respectively.  
The {\it top} panel of Fig.~\ref{tdvzabs} shows the distribution 
of optical depth sensitivities in these two redshift bins. 
We will determine the 21-cm detection rates following the method 
described in Section~7 of G09 that takes into account the variation 
of optical depth sensitivities towards different quasar sight lines. 
The  detection rate for 21-cm absorption, $C$, (defined as the
fraction of \mgii\ systems that show detectable 21-cm absorption
with the integrated optical depths  ${\cal{T}}_{21}$ greater than
the 3$\sigma$ limiting value ${\cal{T}}_{\rm 0}$) for the 
$0.5<z<1$ and $1.1<z<1.5$ bins are given in Table~\ref{n21}.  
Since the lower limiting values of ${\cal{T}}_{\rm 0}$ select 
the weaker 21-cm absorbers and higher values select the stronger 
absorbers, in Table~\ref{n21}, we give detection rates for a 
range of ${\cal{T}}_{\rm 0}$ values suitable for the sample considered 
here (see the histograms in Fig.~\ref{tdvzabs}).

\begin{table*}
\caption{The fraction of 21-cm absorption detections (C) and the number density of 21-cm absorbers 
($n_{21}$) for the different 21-cm optical depth limits 
(${\cal T}_0$) using the homogeneous sample S1. 
N is the total number of systems in the sample with $\int\tau_{3\sigma,10}dv\le{\cal T}_0$, 
and N$_{21}$ is the number of 21-cm detections with $\int\tau_{3\sigma,10}dv\le{\cal T}_0$ and $\int\tau dv\ge{\cal T}_0$. 
}
\begin{center}
\begin{tabular}{cccccccc}
\hline
\hline
{\large \strut} $z$ range& Sample &${\cal T}_0$   & Systems (N)  &  Detections (N$_{21}$)   &  C      & $n_{21}$   \\
         &      & (\kms) &   &    &        &    \\
\hline
{\large \strut} 1.1-1.5 & Ours (i.e. This paper + G09)         & 0.10&     8  &     4       &  0.50$^{+0.40}_{-0.24}$     & 0.13$^{+0.11}_{-0.06}$   \\
        &                                      & 0.20&    22  &     4       &  0.18$^{+0.14}_{-0.09}$     & 0.05$^{+0.04}_{-0.02}$   \\
        &                                      & 0.30&    27  &     3       &  0.11$^{+0.11}_{-0.06}$     & 0.03$^{+0.03}_{-0.02}$   \\
        &                                      & 0.50&    36  &     4       &  0.11$^{+0.09}_{-0.05}$     & 0.03$^{+0.02}_{-0.01}$   \\
~1.1-1.5 & Ours + literature                    & 0.10&    10  &     5       &  0.50$^{+0.34}_{-0.22}$     & 0.13$^{+0.09}_{-0.06}$   \\
        &                                      & 0.20&    26  &     5       &  0.19$^{+0.13}_{-0.08}$     & 0.05$^{+0.04}_{-0.02}$   \\
        &                                      & 0.30&    32  &     4       &  0.13$^{+0.10}_{-0.06}$     & 0.03$^{+0.03}_{-0.02}$   \\
        &                                      & 0.50&    41  &     5       &  0.12$^{+0.08}_{-0.05}$     & 0.03$^{+0.02}_{-0.01}$   \\
~1.1-1.5 & Ours + literature {\it with} VLBA    & 0.10&      4  &     2      &  0.50$^{+0.50}_{-0.32}$     & 0.13$^{+0.17}_{-0.09}$   \\
        &(assuming $f_c=c_f$)                  & 0.20&      8  &     1      &  0.13$^{+0.29}_{-0.10}$     & 0.03$^{+0.07}_{-0.03}$   \\
        &                                      & 0.30&      10 &     1      &  0.10$^{+0.23}_{-0.08}$     & 0.03$^{+0.06}_{-0.02}$   \\
        &                                      & 0.50&      16 &     2      &  0.13$^{+0.17}_{-0.08}$     & 0.03$^{+0.04}_{-0.02}$   \\
~1.1-1.5 & Ours + literature {\it with} VLBA    & 0.10&     4  &     2      &  0.50$^{+0.50}_{-0.32}$     & 0.13$^{+0.17}_{-0.09}$   \\
        &(assuming $f_c=f_{VLBA}$)             & 0.20&     8  &     1      &  0.13$^{+0.29}_{-0.10}$     & 0.03$^{+0.07}_{-0.03}$   \\
        &                                      & 0.30&     13 &     1      &  0.08$^{+0.18}_{-0.06}$     & 0.02$^{+0.05}_{-0.02}$   \\
        &                                      & 0.50&     17 &     3      &  0.18$^{+0.18}_{-0.10}$     & 0.05$^{+0.05}_{-0.03}$   \\
~0.5-1.0 & Ours                                 & 0.10&     4  &     1      &  0.25$^{+0.58}_{-0.21}$     & 0.04$^{+0.10}_{-0.04}$   \\
        &                                      & 0.20&     6  &     1      &  0.17$^{+0.38}_{-0.14}$     & 0.03$^{+0.07}_{-0.02}$   \\
        &                                      & 0.30&     8  &     1      &  0.13$^{+0.29}_{-0.10}$     & 0.02$^{+0.05}_{-0.02}$   \\
        &                                      & 0.50&     9  &     1      &  0.11$^{+0.26}_{-0.09}$     & 0.02$^{+0.04}_{-0.02}$   \\
~0.5-1.0 & Ours + literature                    & 0.10&     20 &     3      &  0.15$^{+0.15}_{-0.08}$     & 0.03$^{+0.03}_{-0.02}$   \\
        &                                      & 0.20&     25 &     3      &  0.12$^{+0.12}_{-0.07}$     & 0.02$^{+0.02}_{-0.01}$   \\
        &                                      & 0.30&     28 &     3      &  0.11$^{+0.10}_{-0.06}$     & 0.02$^{+0.02}_{-0.01}$   \\
        &                                      & 0.50&     33 &     2      &  0.06$^{+0.08}_{-0.05}$     & 0.01$^{+0.01}_{-0.01}$   \\
~0.5-1.0 & Ours + literature {\it with} VLBA    & 0.10&     5  &     2      &  0.40$^{+0.58}_{-0.32}$     & 0.07$^{+0.10}_{-0.06}$   \\
        &(assuming $f_c=c_f$)                  & 0.20&     10 &     2      &  0.20$^{+0.29}_{-0.16}$     & 0.04$^{+0.05}_{-0.03}$   \\
        &                                      & 0.30&     12 &     2      &  0.17$^{+0.24}_{-0.14}$     & 0.03$^{+0.04}_{-0.02}$   \\
        &                                      & 0.50&     15 &     2      &  0.13$^{+0.20}_{-0.11}$     & 0.02$^{+0.03}_{-0.02}$   \\
~0.5-1.0 & Ours + literature {\it with} VLBA    & 0.10&     5  &     2      &  0.40$^{+0.58}_{-0.33}$     & 0.07$^{+0.10}_{-0.06}$   \\
        &(assuming $f_c=f_{VLBA}$)             & 0.20&     11 &     2      &  0.18$^{+0.27}_{-0.15}$     & 0.03$^{+0.05}_{-0.03}$   \\
        &                                      & 0.30&     14 &     2      &  0.14$^{+0.21}_{-0.12}$     & 0.03$^{+0.04}_{-0.02}$   \\
        &                                      & 0.50&     15 &     2      &  0.13$^{+0.19}_{-0.11}$     & 0.02$^{+0.03}_{-0.02}$   \\
\hline
\end{tabular}
\end{center}
\label{n21}
\end{table*}

It is clear from Table~\ref{n21} that the 21-cm detection rate $C$  
for any value of ${\cal T}_o$ is only slightly lower for the low-$z$ bin. 
Since the number of detections involved are small in both the 
redshift bins, the errors are large and the noted difference is not statistically 
significant (i.e say at more than 3$\sigma$ level).  
The maximum difference is seen for ${\cal T}_o$=0.1\,\kms\ and is only 
significant at the level of $\sim$0.6$\sigma$.   
To test this further, we perform the generalized rank correlation test
that takes into account the upper limits \citep{Isobe86}. 
We find a correlation coefficient of 1.37 and the probability that
this correlation occurs by chance of 17\%. Therefore the apparent 
trend is also hinted by the correlation test but once again
without high statistical significance.

Thus we conclude that the 21-cm absorption detection rate `C' for the strong 
\mgii\ absorbers is constant over 0.5$< z <$1.5 that is $\sim$30\% 
of the age of Universe. 
As the radio emission from quasars originate from a much larger spatial extent than the optical emission,
one must pay attention to the issues related to radio source structure.
Therefore, before discussing the evolution of `C' any further, we investigate the impact of 
the radio structure of background quasar on the 21-cm detection rate.

\section{21-cm absorption and radio source structure}
\label{sec:structure}

\citet[][]{Curran10mg2} has used angular diameter distance ratios (DA$_{abs}$/DA$_{qso}$) 
at the absorber and background quasar redshift to investigate the effect of 
line-of-sight geometry on the coverage of background radio source by the 21-cm absorbing 
gas.  He suggests that since the background radio source can be more effectively 
covered at the smaller DA$_{abs}$/DA$_{qso}$, the larger values ($\ge$0.8) of DA$_{abs}$/DA$_{qso}$ 
at \zabs$\ge$1 can explain the lower 21-cm detection rates at the higher redshifts.  

In the homogeneous sample S1, only 4 out of 81 systems have 
$0.6<$DA$_{abs}$/DA$_{qso}$$<0.8$ and two of these are detected in 21-cm absorption. 
The remaining 95\% of the \mgii\ systems have DA$_{abs}$/DA$_{qso}$$>$0.8 and dominate the 
statistics.  
As expected, all the 4 systems with DA$_{abs}$/DA$_{qso}$$<0.8$ are at \zabs$<$1. 
For ${\cal T}_{\rm 0}$=0.1\,\kms, they correspond to $C$=0.50$^{+50}_{-32}$. 
Interestingly, this is same as the $C$ for ${\cal T}_{\rm 0}$=0.1\,\kms\ 
we obtained in the previous Section for the $z>1$ bin with all the systems 
having DA$_{abs}$/DA$_{qso}$$>0.8$.    
Therefore, we do not see any significant influence of the geometric effects as quantified 
by \citet{Curran10mg2} on the detectability of 21-cm absorption in our sample.          
This also implies that the redshift evolution of 21-cm detection rates over the 
low- and high-$z$ bins in Table~\ref{n21} is not affected by any evolution in DA$_{abs}$/DA$_{qso}$.

However, this does not mean that the 21-cm detection rate estimates are not affected 
by the partial coverage of the background quasar. 
As discussed in Section~\ref{sec:fc}, the 21-cm optical depth sensitivities from 
the GBT, GMRT and WSRT observations are overestimated.   
Our purpose here is to estimate the 21-cm optical depth sensitivities and detection rates 
corrected for partial coverage.  For this we will use $f_c$ as estimated from the 
arcsecond and milliarcsecond scale images under reasonable assumptions of the gas filling 
factor and the extent of absorbing gas (cf. Section\,\ref{sec:fc}). 

\subsection{Arcsecond scale radio structure}
\label{sec:arcsec}

\begin{figure}
\centerline{
\vbox{
\hbox{
\psfig{figure=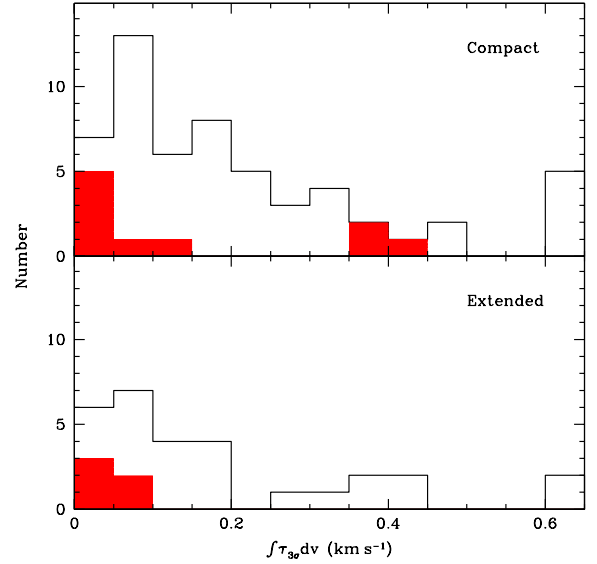,height=8.0cm,width=8.0cm,angle=0}
}
}
}
\caption[]{
Distribution of 21-cm optical depth sensitivity for the quasars that are
compact and extended at arcsecond scales in the sample S2. 
Shaded histogram shows the sensitivity achieved in case of 21-cm absorption detections. 
The 7 systems from \citet{Lane_phd} with $\int\tau_{3\sigma}$dv$>$0.6\,\kms\ are placed 
in the last bin.  
}
\label{arcsec}
\end{figure}

The sample S2 has 56 and 31 systems 
towards the quasars that are compact (deconvolved sizes 
$<$2$^{\prime\prime}$) and extended at arcsecond scales, respectively.  
The distribution of optical depth sensitivities for these are plotted 
in Fig.~\ref{arcsec}. 
The 21-cm absorption detection rates for ${\cal T}_o$=0.3\,\kms\  
towards the compact and the extended quasars are 10$^{+8}_{-5}$\% 
and 14$^{+13}_{-7}$\% respectively. 
The detection rates are also similar for ${\cal T}_o$=0.1\,\kms\ and 0.5\,\kms, and 
do not show any significant difference when the sample is split into 
the two redshift bins as given in Table~\ref{n21}. 
However, it is interesting to note from Fig.~\ref{arcsec} that while 
in the case of systems towards the extended quasars the 21-cm absorption 
is detected only when $\int\tau_{3\sigma}$dv$<0.2$\,\kms, in case of 
compact quasars the 21-cm absorption is also detected for 
$\int\tau_{3\sigma}$dv$>$0.4\,\kms\ (without any correction for the partial 
coverage). To understand this, we now discuss these detections in detail.    

Amongst the systems towards the compact quasars, the three 21-cm detections in the tail of 
$\int\tau_{3\sigma}$dv distribution are towards J0850+5159, J0852+3435 and 
J2340$-$0053. The optical depth sensitivities achieved are lower as these 
systems are towards relatively weaker radio sources in our sample.  
Still, the 21-cm absorption is detected due to the large 21-cm optical depth.  
The rare 2175\AA\ UV-bump is detected in the first two cases suggesting the 
presence of cold gas with a large column density of dust \citep[][see also Section~\ref{sec:dust}]{Srianand08bump, Kulkarni11}. 
In the third case, the 21-cm line width is consistent with the absorbing gas having 
T$_{\rm s}<$200\,K resulting in large 21-cm optical depth (G09; Rahmani et al. in prep.).  

%

Now we discuss the 21-cm detections towards the quasars that are extended at 
arcsecond scales.
The 5 detections in this case are towards B0235+164, J0804+3012, J1145+0455, 
J1252+4427 and J2031+1219.
The subarcsecond scale images are available for 4 of these quasars  
(i.e. except J1252+4427) that are also amongst the brightest radio sources
in our sample.
As expected the estimated correction to account for the partial coverage 
is also large ($c_f<$0.5) for these. 
Despite this the 21-cm optical depth sensitivity achieved towards the 
`core' component is exceptionally good (i.e. $\int\tau_{3\sigma}$dv$\le$0.2 \kms 
even after correcting for the $f_c$ assuming $f_c=c_f$) for these 4 systems. 
The good optical depth sensitivity towards the milliarcsecond scale `core' 
coupled with the large $f_c$ correction ($c_f<$0.5) can be taken to imply that 
these quasars also have good optical depth sensitivity at the physical scales beyond 
($>$30\,pc)\footnote{Based on the typical resolution at the absorber redshift achieved in the VLBA images 
used for `core' identification.} 
the `core' component. 
We notice that the 21-cm absorption in all these 5 cases exhibits large velocity 
widths i.e. $\Delta$V=100-150\,\kms. 
Compared to this the widths of absorption lines towards the 
compact quasars plotted in the {\it top} panel of Fig.~\ref{arcsec} are less than 100\,\kms. 
This favours the idea that larger $\Delta$V observed towards the extended sources arise from 
the 21-cm absorbing gas that has transverse extent of $>$30\,pc. 
We examine this further in the next Section.

\subsection{Correlations with VLBA morphological parameters}
\label{sec:vlbacor}
\begin{figure}
\centerline{
\vbox{
\hbox{
\psfig{figure=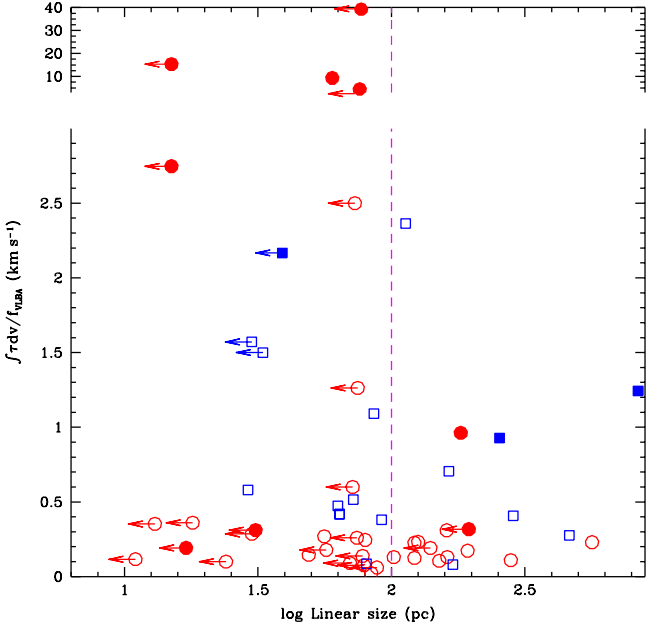,height=9.0cm,width=9.0cm,angle=00}
}
}
}
\caption[]{Integrated 21-cm optical depth corrected for partial coverage assuming $f_c=f_{VLBA}$ vs 
linear size (in pc) at the absorber redshift for the sample S2. Systems towards the quasars 
with no `core' identifications are plotted as squares.  
21-cm detections are shown as filled symbols where 21-cm non-detections are represented by open symbols.  
The point for the system towards B1629+120 with LS=7.2\,kpc is not plotted to keep LS-axis short. 
}
\label{tdvlls}
\end{figure}
Now we explore the correlations between the VLBA morphological and 21-cm 
absorption parameters using the sample S2.  
In Fig.~\ref{tdvlls} we plot $\int\tau$dv/$f_{VLBA}$ as the function of 
linear size. If we consider only those systems that are towards the quasars with 
the clear `core' identification 
(plotted as circles in Fig.~\ref{tdvlls}), we see that almost all the
detections (except one) are towards the quasars with the linear size, LS$<$100\,pc.
However, the generalised rank correlation test suggests no correlation 
between the linear size and the $\int\tau$dv/$c_f$, or  $\int\tau$dv/$f_{VLBA}$.  
The correlation between the linear size and 
$\int\tau$dv/$f_{VLBA}$ is only significant at the level of 0.4$\sigma$ with 
a probability that it can arise due to a chance of 71\%.  The significance 
is even lower (0.1$\sigma$) for the correlation between the linear size and 
$\int\tau$dv/$c_f$.

\begin{figure}
\centerline{
\vbox{
\hbox{
\psfig{figure=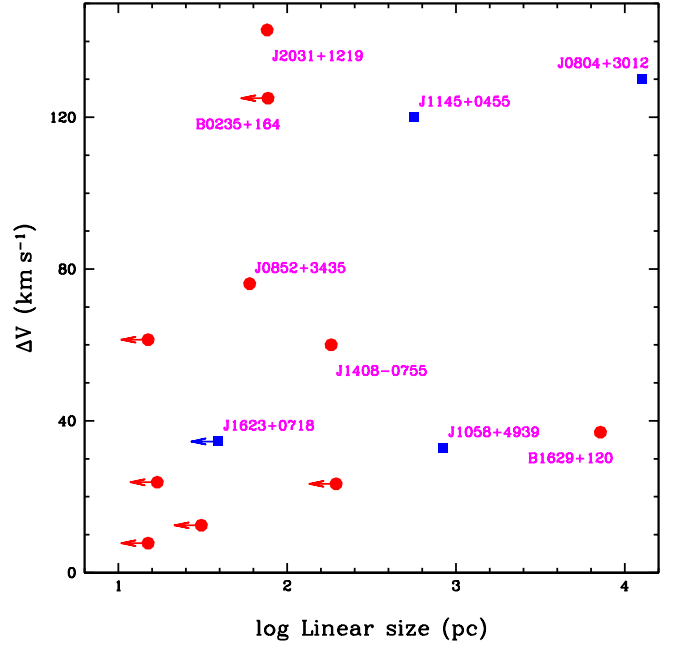,height=9cm,width=9.0cm,angle=00}
}
}
}
\caption[]{
Width ($\Delta$V) of 21-cm absorption line in \kms\ is plotted against the linear size in pc measured from the VLBA 
maps. The systems towards the quasars with the `core' identification are plotted as circles and the remaining are shown as 
squares.  
}
\label{21dvvlba}
\end{figure}

Now we return to the issue of relationship between the radio structure at 
subarcsecond scales and the width of 21-cm absorption lines.  
In Fig.~\ref{21dvvlba} we plot the width of 21-cm absorption lines as the function 
of linear size measured from the VLBA maps for all the 21-cm detections from the 
sample S2.  The only exception is the system towards J0804+3012 for which the 
linear size is measured from the subarcsecond scale images of \citet{Kunert02}.   
There are four systems in this figure showing $\Delta$V$>$100\,\kms.
These four are identified in Section~\ref{sec:arcsec} as the ones towards the radio sources
having extended morphology at arcsecond scales. It is also interesting to note
that there are two systems with  $\Delta$V$\le$40\,\kms\ towards the quasars that
show extended morphology in the VLBA images (i.e J1058+4939 and B1629+120).
The correlation analysis suggests only a weak correlation at the level 
of 1.54$\sigma$ with the probability that it can be by chance of 0.12.  

The lack of strong correlation between $\Delta V$ and projected linear size is not
surprising. This is because, even if we only look at compact radio sources
we expect a spread in the line of sight $\Delta V$ as measured by the distribution
of Mg~{\sc ii} equivalent widths.  Such line of sight velocity spread will dilute 
any possible correlation between $\Delta V$ and LS that is expected to originate
from the velocity gradients in the transverse direction as traced by the 
extended radio sources. In our sample, we do not find any correlation between
$W_{\rm r}$ and $\Delta V$.  The correlation analysis suggests a significance of 
0.27$\sigma$ with a chance probability of 0.74. Note that Curran (2010) also finds 
only 1.73$\sigma$ correlation between $W_{\rm r}$ and $\Delta V$ for a sample 
comprising both DLAs and $W_{\rm r}\ge$1\,\AA\ i.e. strong \mgii\ absorbers.

In summary,  we find a large fraction of 21-cm detections to be towards
the quasars having linear size $\le$ 100\,pc in the VLBA images. The largest
velocity widths are seen towards the quasars that show extended structure 
at arcsecond scales. However, we do not find strong correlation between
either \tdv\ or $\Delta$V with the linear size measured from the
milliarcsecond scale images. All this can be understood if the absorbing gas
is patchy with a typical correlation length of $\sim$30-100\,pc.  
It is quite likely that the narrow absorption components with the larger  
optical depths are associated with even smaller clouds 
(Srianand et al. in prep.).

\subsection{Milliarcsecond scale structure and 21-cm detection rate}
\label{sec:fccor}

In this Section, we determine the 21-cm detection rates for the strong \mgii\ 
absorbers after correcting the 21-cm optical depths for partial coverage.  
For this purpose, we will use the sample S1 that only consists 
of the systems primarily selected on the basis of $W_{\rm r}\ge1$\,\AA\ criterion 
as described in the Section~\ref{sec:unbiased}.
As noted above, if the 21-cm absorbing gas indeed extends beyond 
30\,pc then under the assumption of unit filling factor we can 
use $c_f$ and $f_{VLBA}$ to estimate $f_c$ and obtain a 
realistic estimate of the optical depth sensitivity. 
As discussed in Section.~\ref{sec:fc}, we have two possible values
of $f_c$ for 35 systems with `core' identifications in the 
sample S1. Just to recollect, in the first case we assume 
$f_c = c_f$.  The $f_c$ in this case provides a conservative lower limit on 
the actual value of $f_c$.  
In the second case of $f_c=f_{VLBA}$,
we assume that the absorbing gas extends over 300\,pc to cover all the VLBA 
components. We determine the 21-cm absorption detection rates for 
these two scenarios using the corrected optical depth sensitivity
limits. These detection rates for the different values of ${\cal T}_0$ 
are summarised in Table~\ref{n21}. 

It is clear from Table~\ref{n21} that the detection rates in the 
low-$z$ bin are slightly higher in this case compared to the results obtained with 
no correction for the partial coverage. 
The slight gap noted between the low- and high-$z$ detection rates, when no 
correction for partial coverage is applied, is also reduced.  
The maximum increase in $C$ after $f_c$ correction is 
by a factor 2.7 for the low-$z$ bin.  
Thus covering factor is an important issue in measuring the
redshift distribution of 21-cm absorption systems accurately. 
One may underestimate detection rate by up to factor $\sim$2 due to this 
effect alone. 

\section{21-cm absorption and properties from their optical spectra}
\label{sec:opt}

\subsection{Metal absorption lines}
\label{sec:drr1}

\begin{figure*}
\centerline{
\vbox{
\hbox{
\psfig{figure=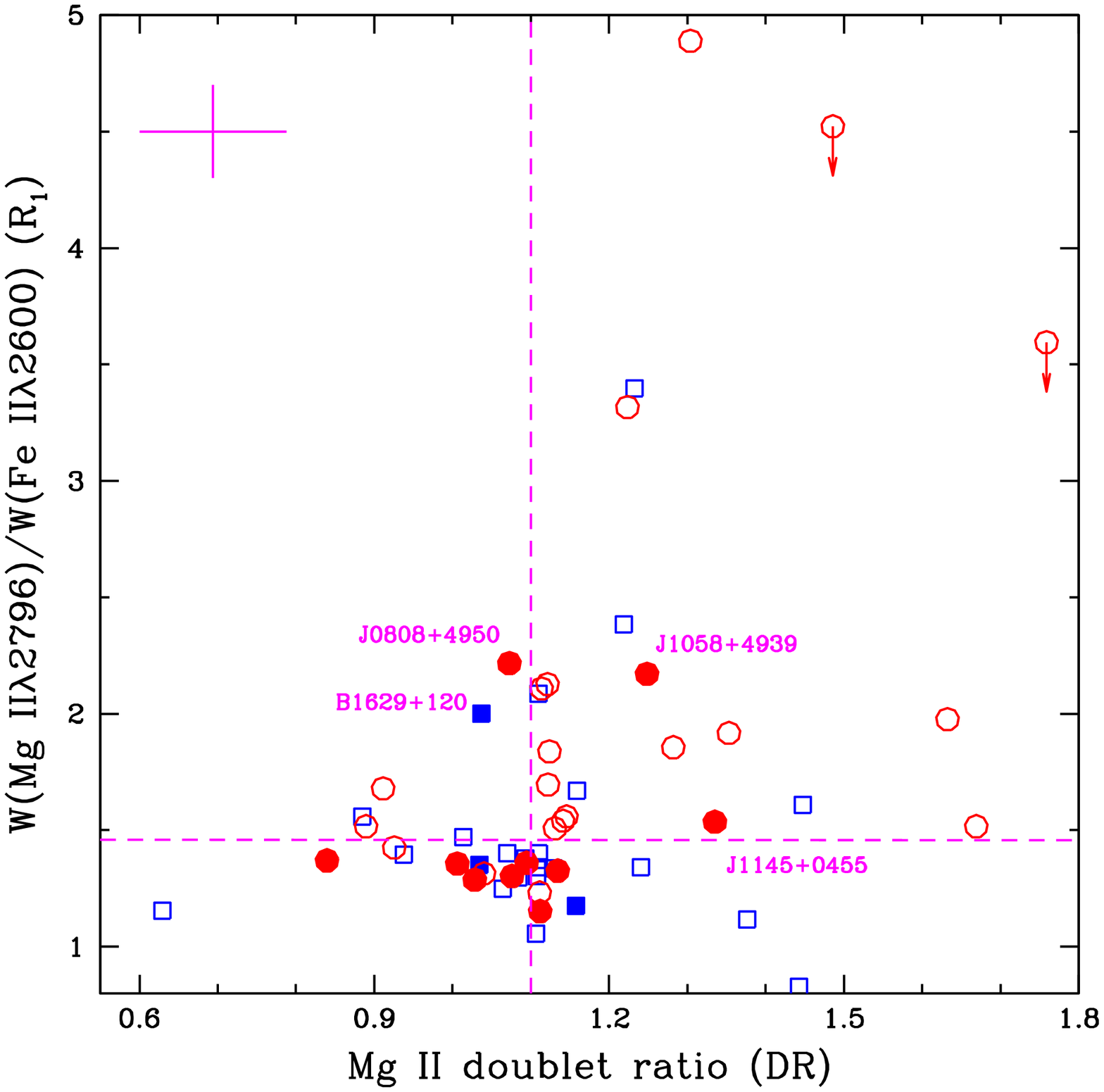,height=9.0cm,width=9.0cm,angle=00}
\psfig{figure=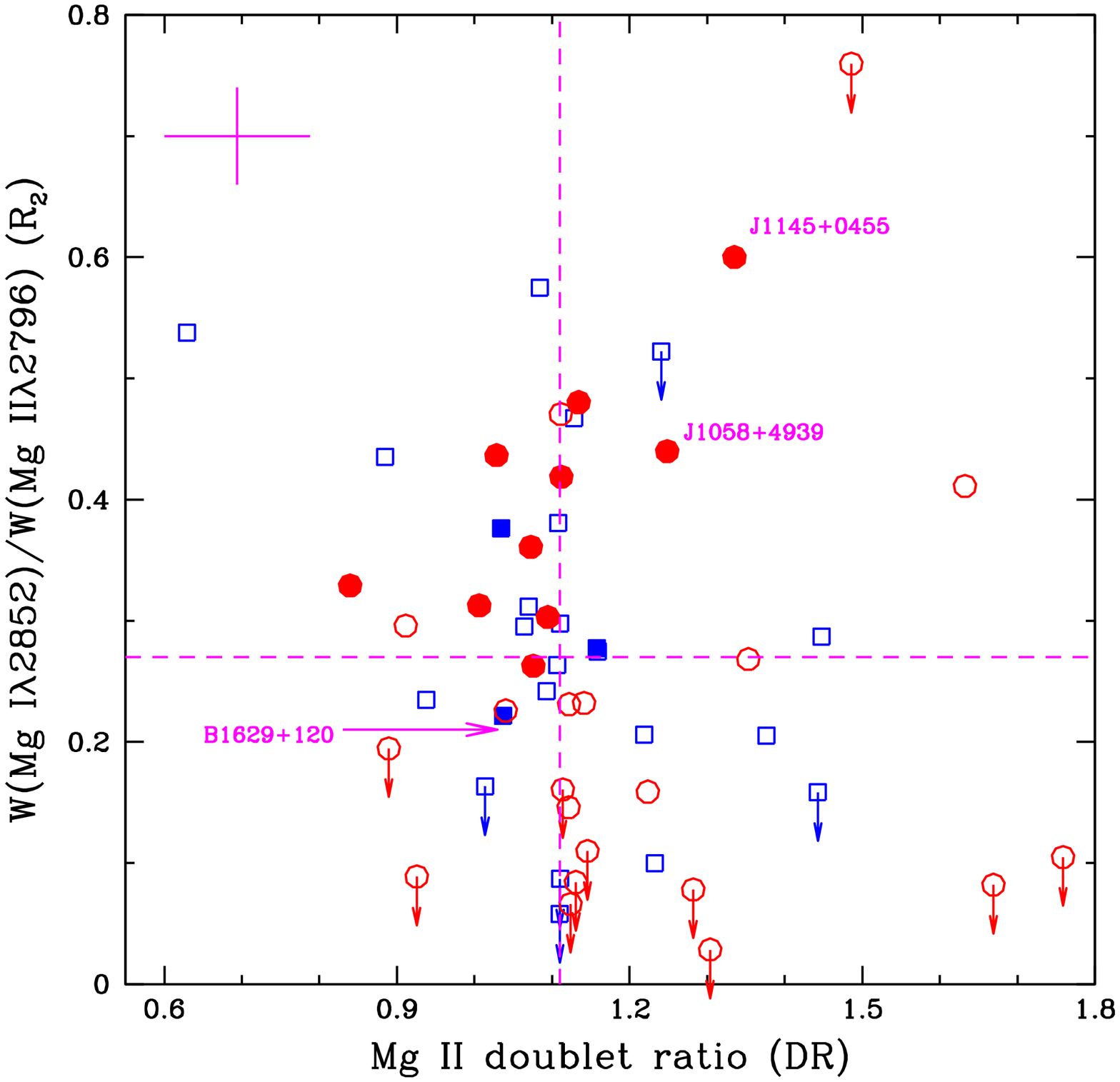,height=9.0cm,width=9.0cm,angle=00}
}
}
}
\caption[]{
R$_1$ and R$_2$ vs DR are plotted for the systems towards quasars with the VLBA images.  
Dashed lines in the panels are for median values of DR=1.1, R$_1$=1.47 and R$_2$=0.27.  
Systems at $0.5<z<1$ are plotted as squares and $1.1<z<1.5$ as circles.  
Filled symbols are for 21-cm detections. 
Values for system towards J0808+4950 (G09), J1058+4939 (G09), J1145+0455 (Table~\ref{kansamp}) and 
B1629+120 \citep{Lane_phd} are labelled. 
Median error bars are shown at the top-left corner. 
System towards J2129$-$1538 is omitted due to large uncertainties on equivalent widths.  
}
\label{drr1r2}
\end{figure*}

\begin{figure*}
\centerline{{
\psfig{figure=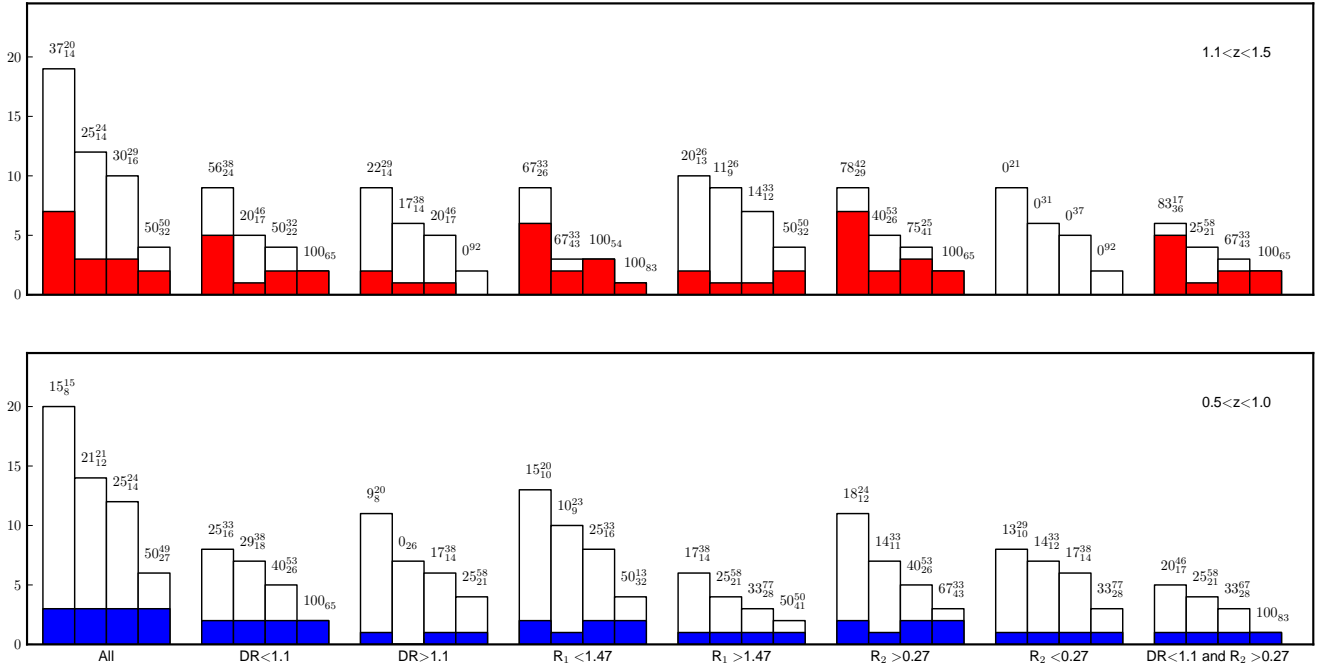,height=10.0cm,width=18.0cm,angle=0}
}}
\caption[]{Distribution of the number of \mgii\ systems for different subsamples 
from the sample S2 with the `core' identification.  
Systems at $1.1<z<1.5$ are plotted in the {\it top} panel and systems at $0.5<z<1$ in the {\it bottom} panel. 
Shaded histogram corresponds to detections and optical depths have been corrected for partial coverage assuming $f_c=c_f$.  
For each subsample four bars correspond to no-cutoff, ${\cal T}_o$ = 0.3, 0.2 
and 0.1\,\kms\ respectively going from left to right. Detection rate, $C$, in each case is given at the top of bar.  
}
\label{bar}
\end{figure*}
%

It has been well established that the fraction of \mgii\ systems that are 
DLAs increases with the $W_{\rm r}$ \citep{Rao06}.  However, there 
exists a large scatter in the $W_{\rm r}$ and $N$(\hi) relationship, 
and $W_{\rm r}$ cannot be used as an indicator of the $N$(\hi) \citep{Rao11}.  
On the other hand, the \mgii\ doublet ratio (DR), $W$(\mgiia)/$W$(\feiia) (defined as R$_1$) and 
$W$(\mgia)/$W$(\mgiia) (defined as R$_2$) can be used as a robust indicator of the presence 
of high $N$(\hi).  
From the HST sample of \citet{Rao06}, we notice that 27$^{+9}_{-7}$\% and 38$^{+15}_{-11}$\% 
of the strong \mgii\ systems are DLAs at 0.5$< z <$1.0 and 1.0$< z <$1.5 respectively. 
The DLA fraction can be as high as 50\% if further restrictions on the DR, R$_1$ and R$_2$ 
are imposed. 
G09 also noticed that a higher 21-cm detection rate can be obtained with the systems 
limited to a restricted range of these parameters.

In order to confirm if the 21-cm absorption detections occupy any preferred location 
in the parameter space defined by the metal absorption line ratios, we plot in Fig.~\ref{drr1r2}  
the ratios R$_1$ and R$_2$ as a function of DR.  
The \mgi, \mgii\ and \feii\ equivalent widths for the systems taken from 
\citet{Kanekar09mg2} (see Table~\ref{kansamp}) have been estimated using the SDSS 
and VLT archival spectra following the same procedure as done in G09 and 
Table~\ref{mg2sample}.  The equivalent widths for the systems that are from \citet{Lane_phd} 
are given in Table~\ref{lanesamp}.  These have been taken from \citet{Lane_phd}.
When comparing the properties of the radio and optical absorption lines it is important to 
be able to address the issues that could arise due to the different gas volumes being 
probed by the radio and optical sight lines.  
For this reason, from the sample S2, we will consider here only the \mgii\ systems 
towards the quasars with the VLBA images.

It is clear from Fig.~\ref{drr1r2} that if we split the sample along the 
median values of DR, R$_1$ and R$_2$, plotted as dashed lines, 
then $\sim$75\% of the 21-cm detections have DR$<$1.1, R$_1<$1.47 and R$_2>$0.27.  
The detection rate amongst the \mgii\ systems in the $1.1<z<1.5$ sample, without any optical 
depth cut-off, is as high as 80\% for these values of DR, R$_1$ and R$_2$.  This trend 
for $1.1<z<1.5$ was previously noted by G09. Since 75\% of the \mgii\ systems in   
the sample S2 at $1.1<z<1.5$ considered here come from G09, this finding is not surprising.  
This result implies that the probability of detecting cold gas via 
21-cm absorption is higher in systems with the high values of $N$(\hi), 
and is consistent with the 3$\sigma$ level correlation between $N$(\hi) and 
T$_s$/$fc$ noted by \citet{Curran10}.
A similar trend has also been noted for 21-cm detections in a sample of 28 bonafide DLAs at 
$z>2$ by \citet{Srianand12dla}.

In this context, it is interesting to recall that in the Section~\ref{sec:data}, 
we also presented 21-cm absorption measurements of the 4 \mgii\ systems i.e. towards 
the blazars J0457$-$2324, J1408$-$0752, J2031+1219 and J2129$-$1538, that were specifically 
selected on the basis of the metal absorption line properties 
that suggest that these systems may have \hi\ column densities similar to the DLAs.
The first three not only have $W$(\mgiia)/$W$(\feiia)$<$3/2, 
a criterion satisfied by the DLAs, they also have strong \mgi\ and 
Mn~{\sc ii} absorption. The first two systems also show absorption 
lines of Zn~{\sc ii} and Cr~{\sc ii}. In the case of system 
towards J2129$-$1538, the absorption lines of Na~{\sc i}$\lambda\lambda$3303.3,3303.9 
are also detected.  
Detection of these species together with the \mgi, \mgii\ and \feii\ 
suggest that the \hi\ column densities should be significantly higher than the 
threshold used to define DLAs. 
Therefore, high 21-cm detection rates towards these 4 blazars could be due to 
these systems having high $N$(\hi).

Now we discuss the two deviations from the above mentioned trends.  
First, there are 3 21-cm detections in Fig.~\ref{drr1r2}, labelled by the quasar name, 
that do not satisfy both the DR-R$_1$ and DR-R$_2$ values satisfied by the other 21-cm detections.  
The doublet ratio for system towards J1058+4939 is only 1.5$\sigma$ from 
the median dashed line of DR=1.1.  
It is interesting to note that the quasars J1058+4939, 
J1145+0455 and B1629+120 exhibit multiple radio components at sub-arcsecond scales. 
The dominant VLBA component for these three quasars is not `core'.  Therefore 
it is possible that the radio and optical sight lines in these cases are probing 
different gas volumes resulting in the deviations seen in Fig.~\ref{drr1r2}.   

Second deviation is that at $0.5<z<1$ only 20-30\% of the systems with 
the DR$<1.1$, R$_1<$1.47 and R$_2>$0.27 are detected in the 21-cm absorption 
whereas at $1.1<z<1.5$ this fraction can be as high as 80\%. 
This cannot be attributed to the lower optical depth sensitivities achieved 
in the low-$z$ bin or the evolution of \hi\ column density amongst the 
strong \mgii\ absorbers as the fraction of strong \mgii\ systems that are 
DLAs in the HST sample of \citet{Rao06} are same at $0.5<z<1$ and $1.1<z<1.5$.
To examine this in detail we plot in Fig.~\ref{bar}, the distribution of systems 
in the sample S2 for the different subsets of 
\mgii\ systems defined on the basis of DR, R$_1$ and R$_2$.
For each subsample, the four histograms provide the number of systems and 
21-cm detections for no optical depth cut-off,  and ${\cal T}_o$ = 0.3, 0.2 
and 0.1\,\kms\ respectively.  
We have considered here only the subset of 39 systems with clear `core'  
identifications, and the optical depths have been corrected for partial coverage assuming
$f_c = c_f$. The trends observed in Fig.~\ref{bar} are similar if we consider all the 56 systems 
with the VLBA maps and assume $f_c=f_{VLBA}$ to correct for the partial coverage.  

We stress here that the metal absorption line ratios measured from low resolution 
spectra such as SDSS considered here provide the properties averaged over 
several absorption components for a system.  21-cm absorption often arise 
from certain specific components \citep[see e.g. Fig.~8 of][]{Gupta09}.  
Also the error bars on individual DR, R$_1$ and R$_2$ measurements are 
large (see Fig.~\ref{drr1r2}), 
and the observed deviations could simply be due to statistical fluctuations.  
Higher resolution optical spectra and most importantly $N$(\hi) measurements 
are required to address these issues. 

\subsection{Dust content}
\label{sec:dust}

Most of the \mgii\ systems and DLAs in SDSS trace diffuse ISM and 
rarely show reddening due to dust in the spectra of individual 
quasars \citep{Budzynski11, Wild06, York06}.
However in the 21-cm absorption sample discussed in this paper, 
there are 4 \mgii\ systems selected from the SDSS that produce strong reddening signatures in the
optical spectrum of the background quasar. \citet{Srianand08bump} have
discussed two of these systems (\zabs = 1.3265 system towards J0850+5159 and
\zabs = 1.3095 towards J0852+3435) that show both the 2175\,\AA\ UV bump and 21-cm
absorption. 
The other two systems that show strong signatures of the reddening due to dust 
but interestingly are not detected in the 21-cm absorption are discussed below.

\begin{figure}
\centerline{{
\vbox{
\psfig{figure=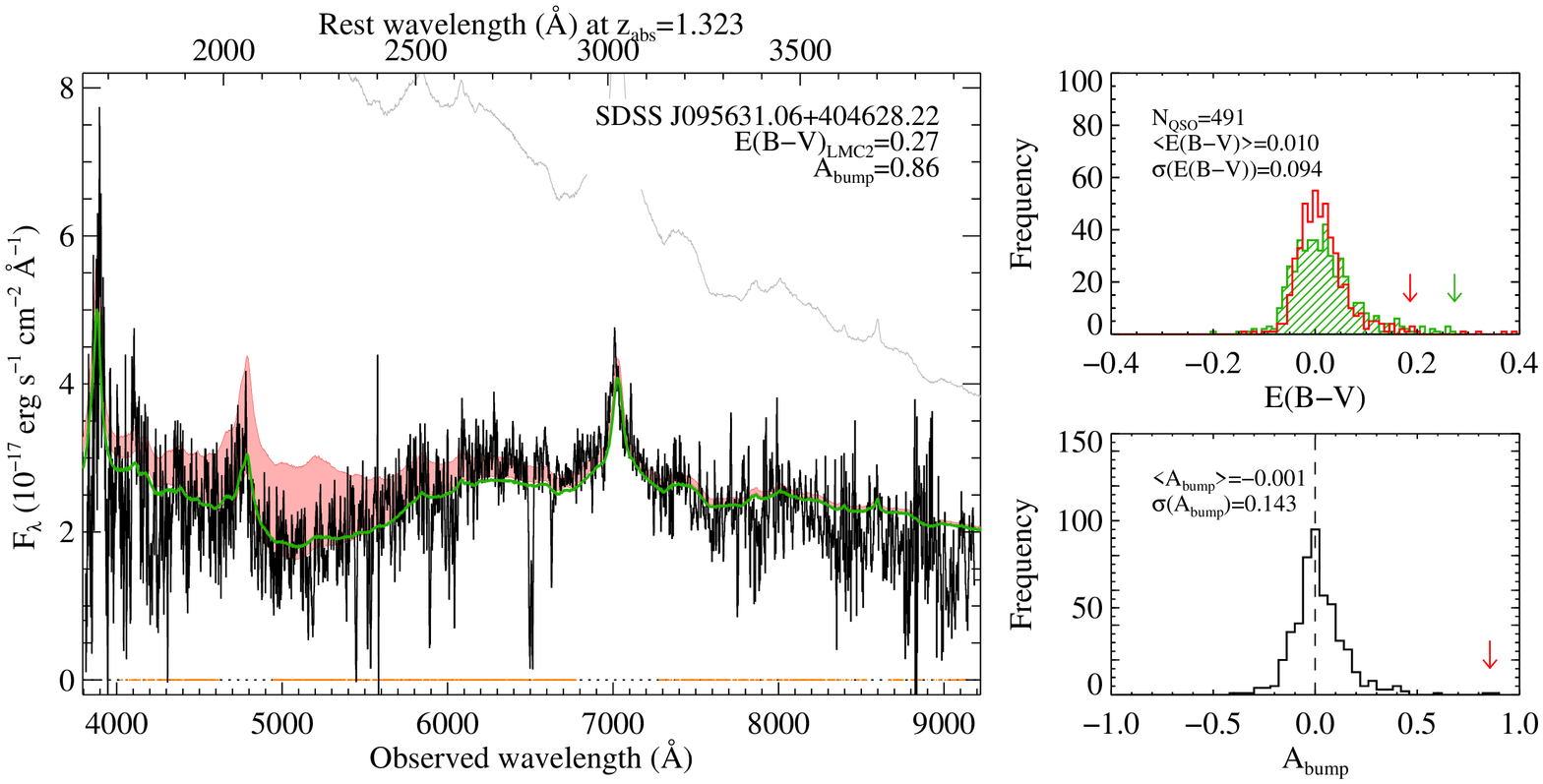,width=8.0cm,angle=90,bbllx=100pt,bblly=20pt,bburx=500pt,bbury=700pt} 
\psfig{figure=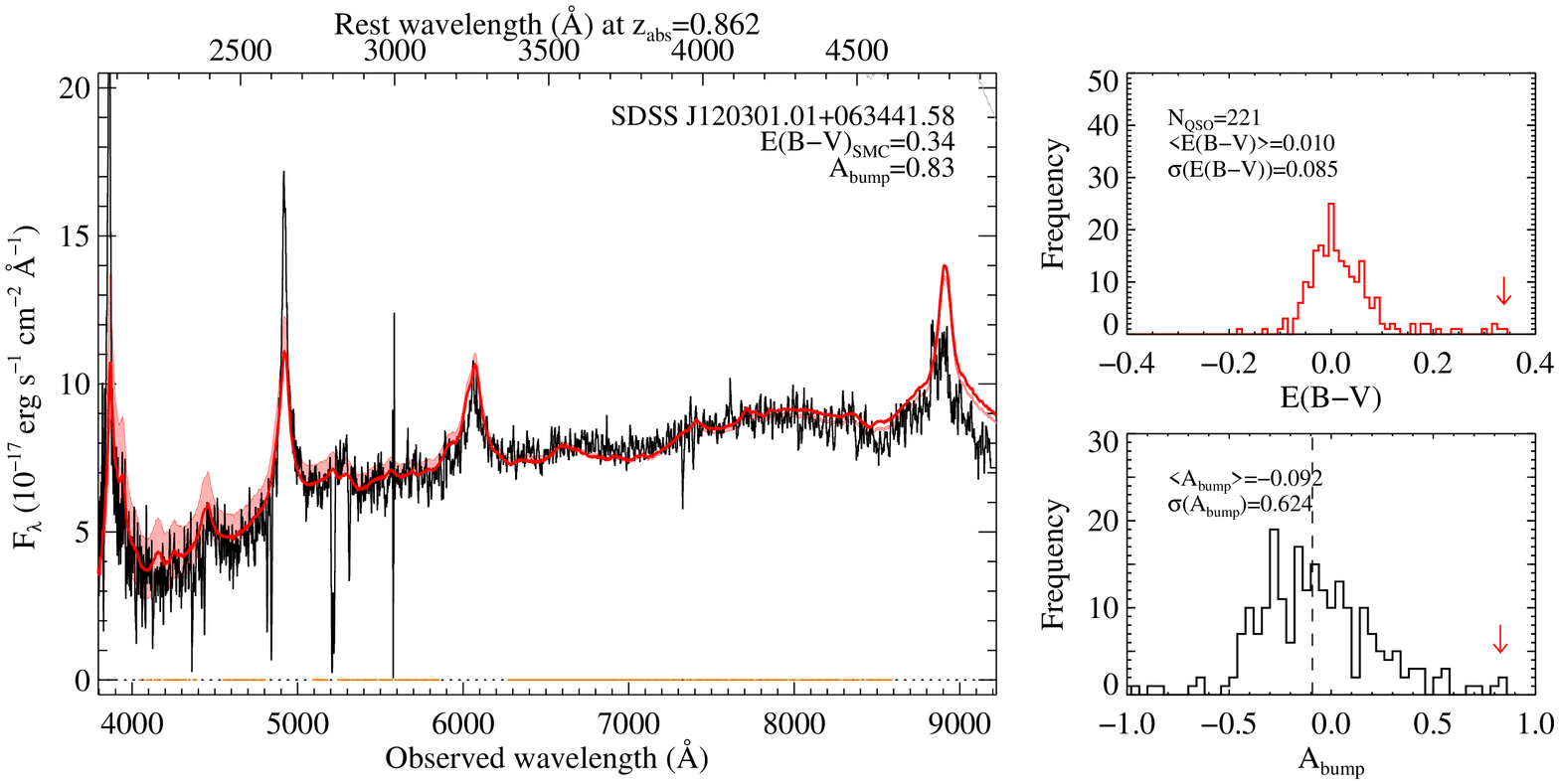,width=8.0cm,angle=90,bbllx=100pt,bblly=20pt,bburx=500pt,bbury=700pt}
}
}}
\caption[]{Fitting the spectral energy distribution of J0956+4046 (top) 
and J1203+0634 (bottom). The SDSS spectrum (black)
and the unreddened SDSS composite (gray)  from \citet{VandenBerk01}.
The best extinction modelling is obtained for E(B-V)=0.27 and a LMC2 
extinction law for J0956+4046 (resp. E(B-V)=0.34 and SMC law for 
J1203+0634) at the redshift of the absorber. The shaded
area represents $A_{\rm bump}$, the difference between the best model
using \citet{Jiang11} parametrisation with and without bump. The orange
regions at y=0 indicate the spectral range considered in the fitting 
process. The right panels show the distributions of E(B-V) 
(using SMC-law: red, LMC2: green hashed) and $A_{\rm bump}$ obtained 
for a control sample of SDSS QSOs \citep{Schneider10} within $\Delta z = \pm 0.004$ from
J0956+4046 (resp. J1203+0634). The arrows indicate the positions of the
considered QSO.}
\label{uvbump}
\end{figure}

\begin{enumerate}

\item{\bf \zabs = 1.3237 system towards J0956+4046:}
This system is amongst one of the 12  high-confidence 2175\AA\ absorber 
candidates of \citet{Jiang11}.
In the top panel of Fig~\ref{uvbump}, we show extinction curve fits to the SDSS 
spectrum of this object using the method described in \citet{Noterdaeme09co}. 
The observed spectrum is well fitted using LMC2 extinction curve and 
E(B-V) = 0.27 and the bump strength \citep[as defined by][]{Jiang11} of 
${\rm A_{bump}=0.86}$. The main difference in our procedure is that
we have fixed the wavelength and width of the 2175\,\AA\ feature as given
by \citet{Fitzpatrick07}. Using all the 491 QSOs with \zem\ within 0.004 
to J0956+4046 we confirm the presence of bump at the high significance level 
(see left panels in Fig~\ref{uvbump}). If extinction per hydrogen
atom in this system is similar to LMC then we expect $N$(\hi) 
$\sim 5\times10^{21}$\,cm$^{-2}$. The estimated E(B-V) value and rest 
equivalent widths of various species measured in this system are identical to
the 21-cm absorber at \zabs = 1.3265  towards J0850+5159 \citep{Srianand08bump}
where the measured integrated 21-cm optical depth is 15.3\,\kms\ (see Table 3 of G09).
As shown in Fig.~\ref{gmrtmaps}, J0956+4046 is resolved into two
components in our GMRT image. The brightest component having a flux density 
of 53\,mJy coincides with the optical position. A 21-cm absorber like 
the one we have seen towards J0850+5159 would have been easily detectable towards it.  
From Table~\ref{mg2obsres} we see that the 3$\sigma$ limit on \tdv\ is 
0.35\,\kms\ for this component. This together with the above estimated $N$(\hi) 
value suggest that $T_s/f_c >$ 7800\,K.  In the 
case of high-$z$ CO absorbers E(B-V)/$N$(\hi) is found to be
much higher (upto an order of magnitude) compared to what is seen in 
LMC \citep[][Ledoux et al. in prep.]{Noterdaeme10co}.
If we use the mean E(B-V)/$N$(\hi) found for the CO-DLAs whose SED is also well
fitted with LMC2 extinction curve we get $N$(\hi)$\sim$4$\times10^{20}$\,cm$^{-2}$ 
and $T_s/f_c > 600$\,K. Therefore, the lack of 21-cm absorption
could either be due to the covering factor of the gas being much less
than 1 or the extinction per hydrogen atom in this absorber being
much higher than that has been seen in LMC. 
As dusty regions are usually cold  and the fact that our spectra
is sensitive enough to detect the cold gas even when $f_c$ is as low
as 0.1 favours the second possibility.
The multifrequency VLBA imaging of this source to constrain $f_c$ and 
the $N$(\hi) measurement will provide important insights into this dusty Mg~{\sc ii} system.

\item{\bf \zabs = 0.8620 systems towards J1203+0634:}  
Based on the presence of strong Ca~{\sc ii} absorption and reddening, 
\citet{Wild06} have suggested this system as a possible DLA candidate.
We find that the SED can be well fitted with
the SMC like extinction curve without any need for the presence of
2175\,\AA\ bump (See {\it lower} panel in Fig.~\ref{uvbump}). The equivalent
width of \mgii\ and other species are like the other three systems
mentioned above. The observed E(B-V) for SMC like dust-to-gas ratio
corresponds to $N$(\hi)$>10^{22}$\,cm$^{-2}$. \citet{Kulkarni11}
have reported  a possible detection (3.6$\sigma$ level) of 10\,$\mu$m 
Silicate absorption in this system.
Interestingly \citet{Kanekar09mg2} have not detected
21-cm absorption despite having a very good $\int \tau dv$  limit of
0.07 \kms. Lack of 21-cm detection in this system that has a high probability
of having very large $N$(\hi) is interesting. 
Unlike  J0956+4046, J1203+0634 is a strong radio source and all
the flux density at arcsecond scales is recovered in the milliarcsecond scale 
VCS images at 2.3\,GHz. The strongest VLBA component contains 66\% of the total flux density.
Rest of the flux is extended within 170\,pc. Interestingly in
the 8.4\,GHz VLBA image the extended component is undetected and
the unresolved component seen in the 2.3\,GHz VLBA image resolves in to a core-jet
structure over a LS of $\le 60$\,pc. 
Even if only 10\% of the radio emission passes through the absorbing cloud
the \tdv\ limit achieved is sensitive enough to detect a T$\sim$100\,K gas 
having $N$(\hi)$\sim$$10^{20}$\,cm$^{-2}$.
All this suggests that the lack of 21-cm absorption in this system 
may also be related to the extinction per \hi\ atom being much higher 
than the Galaxy or SMC.  
\end{enumerate}

\section{Discussion}
\label{sec:disc}

\subsection{Evolution of 21-cm detection rate and $n_{21}$}
\label{sec:n21}

The nature of \mgii\ absorbers and their relationship with the galaxies  
has been a matter of debate.  
Recently [O~{\sc ii}] emission, a measure of the star formation rate, 
has been detected in the individual as well as stacked SDSS spectra of the  
$W_{\rm r}>$0.7\,\AA\  \mgii\ absorbers \citep{Noterdaeme10o3,Menard11}.  
The average emission line luminosities are found to correlate with $W_{\rm r}$. 
\citet{Menard11} show that \mgii\ absorbers not only trace a substantial 
fraction of the global [O~{\sc ii}] luminosity density but also recover the 
overall star formation history of the Universe up to $z\sim2$.  
They argue that this is suggestive of the scenario where most of the strong \mgii\ 
absorbers arise from the starburst driven galactic outflows 
\citep[however see,][ { who suggest that the 
$W_{\rm r}$ vs. [O~{\sc ii}] luminosity correlation can also 
be explained by the $W_{\rm r}$ vs. galaxy impact parameter 
anti-correlation and does not necessarily support the `outflow' scenario} ]{Lopez12}. 

\begin{figure}
\centerline{{
\psfig{figure=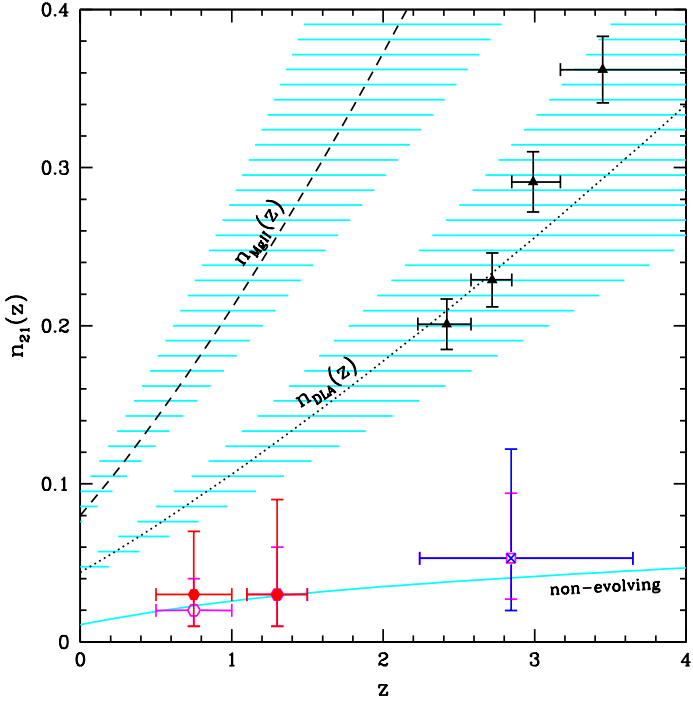,height=10.0cm,width=9.5cm,angle=0}
}}
\caption[]{Number of 21-cm absorbers per unit redshift range, $n_{21}(z)$, 
for ${\cal T}_o=$0.3\,\kms. 
Open circles are for no correction for partial coverage.
Filled circles correspond to $n_{21}$ after the optical depths  
have been corrected for partial coverage assuming $f_c=c_f$. 
The $n_{21}$ based on DLAs at $z>2$ are plotted as squares and crosses with the 
latter corresponding to values after correcting for partial coverage.  
The curve for non-evolving population of 21-cm absorbers normalized at 
$n_{21}$($z=1.3$) is also plotted. 
Lines and dashed areas show the number of 
absorbers per unit redshift for DLAs \citep[][; dotted line]{Rao06} and 
strong \mgii\ absorbers \citep[][; dashed line]{Prochter06}.  Triangles are
the number per unit redshift of DLAs measured from the sample of \citet{Noterdaeme09dla}.
}
\label{dndz}
\end{figure}
%

Since the volume filling factor of different phases of ISM depends sensitively on 
the stellar feedback, in the `outflow' scenario we expect CNM filling 
factor and the 21-cm detection rate of strong \mgii\ absorbers to evolve 
strongly with the starburst intensity \citep{deAvillez04}. 
But in our analysis of the 21-cm absorption sample presented in the previous sections, 
we find that the 21-cm detection rate, $C$, is constant over the 
redshift range of $0.5<z<1.5$ i.e. 30\% of the age of Universe. 

As described in G09, we can estimate the number density per unit redshift, 
$n_{21}({\cal{T}}_{\rm 0},$W$_{\rm o},z)$, of 21-cm systems with integrated
optical depth ${\cal{T}}_{21}\ge {\cal{T}}_{\rm 0}$ and Mg~{\sc ii} equivalent width
$W_{\rm r}\ge W_{\rm o}$ 
from the number density per unit redshift of Mg~{\sc ii} absorbers, 
$n_{\rm Mg{\sc II}}(W_{\rm o},z)$, using the equation,
\begin{equation}
n_{21}({\cal{T}}_{21}\ge {\cal{T}}_0, W_{\rm r}\ge W_{\rm o}, z)= C(z) \times n_{\rm Mg{\sc II}}(W_r\ge W_o, z).
\label{eq:n21}
\end{equation}
The $n_{21}$ for different values of ${\cal T}_0$ are given\footnote{
$n_{21}$ for strong \mgii\ absorbers at $0.5<z<1.5$ have been estimated by 
approximating $n_{\rm MgII}$ as, $n_{\rm Mg{\sc II}}(W_{\rm o},z)$ = $n_0$ $\times$ $(1+z)^\gamma$
with $n_0 = 0.080^{+0.015}_{-0.005}$ and $\gamma = 1.40\pm0.16$ for $W_{\rm o}=1$\,\AA\
\citep{Prochter06}.
} 
in the last column of Table~\ref{n21}.  
In Fig.~\ref{dndz} we also plot the $n_{21}$ for ${\cal T}_0$ =0.3\,\kms.  
The constancy of $C$ as a function of redshift implies that the $n_{21}$ increases with 
redshift with a same slope as $n_{\rm MgII}$.
Due to the small number of positive 21-cm detections in \mgii\ systems the 
errors on $n_{21}$ are large and on its own the estimates would also be 
consistent with the non-evolving population of 21-cm absorbers (Fig.~\ref{dndz}).  
We conclude that the current data does not show 
any sign of the evolution of CNM filling factor of strong \mgii\ absorbers 
and $n_{21}$ simply follows $n_{\rm MgII}$ over $0.5<z<1.5$.  
This is puzzling if the strong \mgii\ absorbers are indeed 
driven by the stellar feedback processes that tend to diminish the CNM filling 
factor.  

\subsection{Implications for 21-cm detection rates based on samples of DLAs at $z>2$}

Unlike strong \mgii\ absorbers that are often favoured to be associated with 
the galactic winds, DLAs are believed to probe the gas confined to the 
disks of galaxies.  
It is also interesting to note that, in contrast to the several \mgii\ host galaxy 
detections, the searches to detect DLA host galaxies have mostly resulted in 
null detections \citep[e.g.][]{Moller93, Fynbo10, Rahmani10, Fynbo11, Bouche12, Noterdaeme12}.  
Consequently DLAs are believed to correspond to low star-forming galaxies or trace 
the outer regions of extended \hi\ disks far away from the star-forming regions 
\citep{Rahmani10}.   
Given that a large fraction ($\sim$40\%) of strong \mgii\ absorbers are likely to be DLAs, 
it is interesting to view the 21-cm detection rates derived for the \mgii\ systems 
in the context of 21-cm absorption searches based on bonafide DLAs.  

Since at $z\gapp2$, \lya\ can be observed from the ground based facilities 
the systematic searches of 21-cm absorption in DLAs have only been possible at 
$z>2$.  
A summary of all the 28 DLAs searched for 21-cm absorption 
at 2.0$\le$\zabs$\le$3.5 is provided in Table~6 of \citep{Srianand12dla}. 
Using the subsample of DLAs towards the quasars with the VLBA `core' 
identifications from this Table, we estimate the 21-cm detection 
rate, C = $0.22^{+0.29}_{-0.14}$ for ${\cal T}_0$ = 0.3\,\kms.
In Fig~\ref{n21} we plot $n_{21}$ corresponding to this. 
The method used is exactly similar to the one used for estimating 
$n_{21}$ from \mgii\ systems in Section~\ref{sec:n21} except that 
the number per unit redshift range of DLAs ($n_{\rm {DLA}}$) is 
used in the place of $n_{\rm {MgII}}$ in Equation~\ref{eq:n21}.   
The $n_{\rm {DLA}}$ required for this purpose has been measured from 
the DLA sample of \citet{Noterdaeme09dla}. 

\citet{Kanekar09mg2} used the detection rate of 21-cm absorption
in \mgii\ systems and the fraction of \mgii\ systems that are DLAs
to argue that the 21-cm detection rate in DLAs is strongly decreasing
with redshift. 
In the $W_{\rm r}\ge$1\,\AA\ \mgii\ subsample with the VLBA `core' 
identifications presented in this paper,  the 21-cm detection rates at 
$0.5<z<1$ and $1.1<z<1.5$ for  ${\cal T}_0$ = 0.3\,\kms\ 
and $f_c = c_f$ correction are   
0.17$^{+0.24}_{-0.14}$ and 0.10$^{+0.23}_{-0.08}$ respectively (Table~\ref{n21}).   
From the HST sample of \citet{Rao06}, the fraction of strong \mgii\ 
systems that are DLAs in these redshift bins are 
27$^{+9}_{-7}$\% and 38$^{+15}_{-11}$\% respectively. 
If we assume that all the 21-cm detections are DLAs then the
fraction of DLAs showing 21-cm absorption will be 0.63 and 0.26 
respectively for the low- and  high-$z$ bins. 
Since all the 21-cm detections need not be associated with the DLAs,  
these fractions should only be considered as upper limits.
The 21-cm absorption detection rate in the DLAs at $z>2$ is lower by 
at most a factor 3 compared to the upper limits we obtained using the 
$W_{\rm r}\ge$1\,\AA\ \mgii\ absorbers at $0.5<z<1.0$. 
However, at 1.1$< z <$1.5
the detection rate or the upper limit from the \mgii\ systems 
are consistent with the 21-cm detection rates from the DLAs at $z>2$.    
This would imply that either the cross section of the 21-cm 
absorbing gas amongst DLAs has increased from $z=3.5$ to $z=0.5$ or 
that the significant fraction of 21-cm absorbers at $0.5<z<1.0$ 
arise from sub-DLAs. 
The $N$(\hi) measurements of \mgii\ systems with good 21-cm optical depth 
limits will be useful to distinguish between these possibilities 
\citep[see also][]{Curran06, Curran12}.

\subsection{Caveats and future work}
\label{sec:cav}

In previous sections, we presented results from 21-cm absorption line 
searches based on the sample of 85 strong \mgii\ absorbers at $0.5<z<1.5$.  
When no correction for the effect of partial coverage is incorporated
we find that the 21-cm detection rate can be underestimated by as much as 
factor 2.  Using milliarcsecond scale VLBA maps for 53 quasars with 57 
intervening absorption systems from this sample, we determined the realistic 
21-cm optical depths and detection rates. 
This is the largest sample of 21-cm absorbers assembled at any redshift range, 
and we systematically take into account the effect of varying 21-cm 
optical depth sensitivity and partial coverage towards different quasar sight lines.  
Despite this the small number statistics still play a significant role and 
results are not without caveats.

First caveat is related to our assumption that $f_c$ can be estimated from 
the VLBA continuum images at 20-cm and 13-cm, whereas the redshifted 21-cm 
observations are at the wavelengths longer by a factor 2-4 compared to these. 
The actual values of $f_c$ are likely to be different at the observing 
frequency.  It is not obvious if the effect of this assumption is to 
simply increase the scatter or produce systematic wavelength dependent 
offsets as one also needs to take into account the effect of finite resolution 
of the VLBA images when identifying `core' and determining $f_c$.  
Given the sensitivity of current receivers and the difficulties involved with calibration, 
the low-frequency milliarcsecond scale imaging and spectroscopy is possible only 
for a very few strong targets.  The VLBI imaging at 20-cm or shorter wavelengths 
will be the optimal way to estimate $f_c$ for the large samples as presented here.  
Therefore it will be valuable to determine the appropriateness of this method by 
measuring the filling factor and extent of the 21-cm absorbing gas via VLBI spectroscopy 
of $z<0.2$ 21-cm absorbers that is possible with the currently available VLBI receivers.

Second caveat is related to the fact that the 21-cm absorption searches in the different 
redshift ranges are based on the samples of different atomic or ionic species 
which complicates the overall interpretation.  
For example, the samples at $z>2$ are based on the DLAs whereas at $z<2$ on the 
\mgii\ systems which selects both the DLAs and sub-DLAs.
Measuring $N$(\hi) in the \mgii\ systems is important to address this issue.    
More importantly it is well known that the optical spectroscopic surveys such
as SDSS are performed after the color selection of QSO candidates and are biased 
against the dusty sight lines that are relevant for the cold atomic and molecular 
absorption lines that can be searched at radio wavelengths 
\citep[e.g.,][]{Carilli98}.  
Therefore, radio absorption line surveys based on the samples of optical absorption 
lines provide a very incomplete view of the evolution of the cold atomic and molecular 
gas in the galaxies.  
The blind searches of radio absorption lines are required to overcome this barrier 
and eliminate the biases due to the pre-selection based on different atomic/ionic 
species.  
Such large spectroscopic surveys will be possible with 
the upcoming Square Kilometre Array (SKA) Pathfinders such as 
APERTIF, ASKAP, EVLA and MeerKAT. 

We finish with the following question: what is the relationship of the 21-cm 
absorbers with the galaxies? 
This link is required to establish the direct connection with the nearby galaxies that are 
detectable in \hi\ 21-cm emission. Using 21-cm emission line maps of nearby galaxies, 
\citet{Zwaan05} have concluded that sight-lines with log\,$N$(\hi)(cm$^{-2}$)$\ge$20.3 
occur with a median impact parameter, $\rho$$<$7.8\,kpc. It is also clear from their 
Fig.~14 that when 10\,kpc$< \rho <$15\,kpc the probability of having high $N$(\hi) is 
roughly between 50\% to 70\%. Galaxies have been found within 20\,kpc to the DLAs and 
21-cm absorbers at z$\le$1.0 \citep{Rao03, Rao11}.
The redshift distribution of DLAs typically require $\rho$$<$30\,kpc to be consistent with 
the observed luminosity function of galaxies \citep{Steidel95}. 
While these studies constrain typical impact parameter ranges where 21-cm
absorption is likely to be detected, they do not tell much about the nature of 21-cm absorbers itself due to the
lack of information on spin temperature. Therefore, observations of 21-cm absorption towards the bright 
radio sources at small impact parameters to the gas disks/halos of nearby galaxies are required to 
determine the 21-cm absorption cross section of the galaxies and understand the nature of 
21-cm absorbers \citep{Carilli92, Gupta10, Borthakur11}.

\section{Summary}
\label{sec:summ}


We have presented results from a systematic search of 
21-cm absorption in a sample of 17 strong \mgii\ 
absorbers at 0.5$<$ \zabs $<$1.5 using the GBT, GMRT and WSRT.  
This resulted in 4 new 21-cm detections.  
The VLBA 20-cm maps of 40 quasars with 42 intervening \mgii\ systems 
are also presented.  

Combining 21-cm absorption measurements for 50 strong \mgii\ systems 
presented in this paper and \citet{Gupta09} with the measurements from 
literature \citep{Lane_phd, Kanekar09mg2}, we assemble a sample of 85 strong 
\mgii\ absorbers at $0.5<z<1$ and $1.1<z<1.5$. 
Using the milliarcsecond scale VLBA maps for the 54 quasars with 57 
intervening absorption systems from this sample, we estimated the actual  
21-cm optical depths and detection rates after correcting for the partial coverage. 
In our detailed analysis of this 21-cm absorption sample, we have 
taken into account the effect of varying 21-cm optical depth sensitivity and the partial 
coverage towards different quasar sight lines.  
Our main results are summarized as follow: 
\begin{enumerate}

\item The 21-cm detection rate is found to be higher towards the quasars 
with flat/inverted spectral index, $\alpha$$<$0.  More than 73\% of the 
21-cm detections are towards the sources with linear size, LS$<$100\,pc.  
This may imply that the characteristic size of the absorbing gas is  
less than 100\,pc.  The low detection rate towards the quasars with  
the steeper $\alpha$ are due to poor optical depth sensitivity towards 
the `core' i.e. low covering factor, $f_c$ of the absorbing gas.

\item  We find that the largest velocity widths 
($\Delta$V$>$100\,\kms) are mainly seen towards the quasars that 
show extended radio structure at arcsecond scales. 
However, we do not find any correlation between the integrated 21-cm 
optical depth, \tdv\ or the width of 21-cm absorption line, $\Delta$V 
with the LS measured from the VLBA images.  This suggests that 
the absorbing gas is patchy with a typical correlation length 
$\sim$30-100\,pc.   

\item We find that when no correction for the effect of partial 
coverage i.e. as estimated from the milliarcsecond maps is incorporated the 
21-cm detection rate can be underestimated by as much as a factor of 2.  

\item The 21-cm detection rate is constant over $0.5<z<1.5$ i.e. 30\% 
of the age of Universe.  
The constancy of $C$ as a function of redshift implies that  
$n_{21}$ increases with the redshift with a same slope as $n_{\rm MgII}$.
Due to the small number of positive 21-cm detections in \mgii\ systems the 
errors on $n_{21}$ are large and on its own the estimates would also be 
consistent with a non-evolving population of 21-cm absorbers (Fig.~\ref{dndz}).  
We conclude that the current data do not show 
any sign of the evolution of the CNM filling factor of strong \mgii\ absorbers 
and $n_{21}$ simply follows $n_{\rm MgII}$ over $0.5<z<1.5$.  
This is intriguing given that the strong \mgii\ absorbers are believed 
to be driven by the stellar feedback processes that tend to diminish the CNM filling factor.  

\item We find that the 21-cm absorption detection rate in the DLAs at $z>2$ is at most 
a factor of 3 less than the upper limits we obtained using the \mgii\ absorbers at 
$0.5<z<1.0$.  This would imply that either the cross section of 21-cm 
absorbing gas amongst DLAs has increased from $z=3.5$ to $z=0.5$ or 
that a significant fraction of 21-cm absorbers at $0.5<z<1.0$ 
arise from sub-DLAs. 

\item We find that 75\% of 21-cm detections have \mgii\ doublet ratio $\le$1.1, 
$W$(\mgia)/$W$(\mgiia)$\ge$0.27 and 
$W$(\mgiia)/$W$(\feiia)$\le$1.47.  
This confirms our previous finding that the probability of detecting 21-cm 
absorption is higher in the systems with high $N$(\hi) \citep{Gupta09, Srianand12dla}, 
and is consistent with the 3$\sigma$ level correlation between $N$(\hi) and 
T$_s$/$fc$ noted by \citet{Curran10}.

\item We present a detailed discussion on the non-detection of 21-cm absorption 
in the two dusty absorbers that produce substantial reddening to the background QSOs. 
This could be either related to the gas covering factor being
low or to the extinction per hydrogen atom in these systems being much higher 
than what is seen in the SMC and LMC. We prefer the second possibility on the 
basis of larger extinction per hydrogen atom seen in some of the high-$z$ DLAs 
with CO detections.

\end{enumerate}

We discuss caveats of our survey results. The analysis is mainly limited 
by the small number of 21-cm detections.  
21-cm detections of the order of SDSS DLA sample \citep{Noterdaeme09dla}  
will be possible with the upcoming SKA pathfinders. 
Blind searches of 21-cm absorption lines with these instruments  
will provide a complete view of the evolution of cold gas in galaxies and 
shed light on the nature of \mgii\ systems and DLAs, and their relationship 
with stellar feedback processes.

\begin{acknowledgements}
We thank C. Carilli, E. de Blok, D.J. Saikia and an anonymous referee for useful suggestions and comments.
We thank GBT, GMRT, VLBA and WSRT staff for their support during the observations.  
GBT and VLBA are run by National Radio Astronomy Observatory.
The VLBA data from 2010 were correlated using NRAO's implementation of 
the DiFX software correlator that was developed as part of the 
Australian Major National Research facilities Programme and 
operated under license. 
The National Radio Astronomy Observatory is a facility of the National Science Foundation 
operated under cooperative agreement by Associated Universities, Inc. 
GMRT is run by the National Centre for Radio Astrophysics of the Tata Institute of Fundamental Research.
The WSRT is operated by the ASTRON (Netherlands Institute for Radio Astronomy) with 
support from the Netherlands Foundation for Scientific Research (NWO).
We acknowledge the use of SDSS spectra from the archive (http://www.sdss.org/).
Funding for the SDSS and SDSS-II has been provided by the Alfred P. Sloan Foundation, 
the Participating Institutions, the National Science Foundation, the U.S. Department of Energy, 
the National Aeronautics and Space Administration, the Japanese Monbukagakusho, the Max Planck Society, 
and the Higher Education Funding Council for England. 
RS and PPJ gratefully acknowledge support from the Indo-French
Centre for the Promotion of Advanced Research (Centre Franco-Indien pour
la promotion de la recherche avanc\'ee) under Project N.4304-2.
\end{acknowledgements}

\bibliographystyle{aa}
\bibliography{mybib}

\begin{thebibliography}{80}
\expandafter\ifx\csname natexlab\endcsname\relax\def\natexlab#1{#1}\fi

\bibitem[{{Aldcroft} {et~al.}(1994){Aldcroft}, {Bechtold}, \&
  {Elvis}}]{Aldcroft94}
{Aldcroft}, T.~L., {Bechtold}, J., \& {Elvis}, M. 1994, \apjs, 93, 1

\bibitem[{{Barthel} {et~al.}(1988){Barthel}, {Miley}, {Schilizzi}, \&
  {Lonsdale}}]{Barthel88}
{Barthel}, P.~D., {Miley}, G.~K., {Schilizzi}, R.~T., \& {Lonsdale}, C.~J.
  1988, \aaps, 73, 515

\bibitem[{{Barthel} {et~al.}(1990){Barthel}, {Tytler}, \&
  {Thomson}}]{Barthel90}
{Barthel}, P.~D., {Tytler}, D.~R., \& {Thomson}, B. 1990, \aaps, 82, 339

\bibitem[{{Barthel} {et~al.}(2000){Barthel}, {Vestergaard}, \&
  {Lonsdale}}]{Barthel00}
{Barthel}, P.~D., {Vestergaard}, M., \& {Lonsdale}, C.~J. 2000, \aap, 354, 7

\bibitem[{{Beasley} {et~al.}(2002){Beasley}, {Gordon}, {Peck}, {Petrov},
  {MacMillan}, {Fomalont}, \& {Ma}}]{Beasley02}
{Beasley}, A.~J., {Gordon}, D., {Peck}, A.~B., {et~al.} 2002, \apjs, 141, 13

\bibitem[{{Bergeron} \& {Boiss{\'e}}(1991)}]{Bergeron91}
{Bergeron}, J. \& {Boiss{\'e}}, P. 1991, \aap, 243, 344

\bibitem[{{Bergeron} {et~al.}(2011){Bergeron}, {Boiss{\'e}}, \&
  {M{\'e}nard}}]{Bergeron11}
{Bergeron}, J., {Boiss{\'e}}, P., \& {M{\'e}nard}, B. 2011, \aap, 525, A51+

\bibitem[{{Bergeron} {et~al.}(2004){Bergeron}, {Petitjean}, {Aracil}, {Pichon},
  {Scannapieco}, {Srianand}, {Boisse}, {Carswell}, {Chand}, {Cristiani},
  {Ferrara}, {Haehnelt}, {Hughes}, {Kim}, {Ledoux}, {Richter}, \&
  {Viel}}]{Bergeron04}
{Bergeron}, J., {Petitjean}, P., {Aracil}, B., {et~al.} 2004, The Messenger,
  118, 40

\bibitem[{{Borthakur} {et~al.}(2011){Borthakur}, {Tripp}, {Yun}, {Bowen},
  {Meiring}, {York}, \& {Momjian}}]{Borthakur11}
{Borthakur}, S., {Tripp}, T.~M., {Yun}, M.~S., {et~al.} 2011, \apj, 727, 52

\bibitem[{{Bouch{\'e}} {et~al.}(2012){Bouch{\'e}}, {Murphy}, {P{\'e}roux},
  {Contini}, {Martin}, {Forster Schreiber}, {Genzel}, {Lutz}, {Gillessen},
  {Tacconi}, {Davies}, \& {Eisenhauer}}]{Bouche12}
{Bouch{\'e}}, N., {Murphy}, M.~T., {P{\'e}roux}, C., {et~al.} 2012, \mnras,
  419, 2

\bibitem[{{Bouch{\'e}} {et~al.}(2007){Bouch{\'e}}, {Murphy}, {P{\'e}roux},
  {Davies}, {Eisenhauer}, {F{\"o}rster Schreiber}, \& {Tacconi}}]{Bouche07mg2}
{Bouch{\'e}}, N., {Murphy}, M.~T., {P{\'e}roux}, C., {et~al.} 2007, \apjl, 669,
  L5

\bibitem[{{Bowen} \& {Chelouche}(2011)}]{Bowen11}
{Bowen}, D.~V. \& {Chelouche}, D. 2011, \apj, 727, 47

\bibitem[{{Briggs} \& {Wolfe}(1983)}]{Briggs83}
{Briggs}, F.~H. \& {Wolfe}, A.~M. 1983, \apj, 268, 76

\bibitem[{{Budzynski} \& {Hewett}(2011)}]{Budzynski11}
{Budzynski}, J.~M. \& {Hewett}, P.~C. 2011, \mnras, 416, 1871

\bibitem[{{Carilli} {et~al.}(1998){Carilli}, {Menten}, {Reid}, {Rupen}, \&
  {Yun}}]{Carilli98}
{Carilli}, C.~L., {Menten}, K.~M., {Reid}, M.~J., {Rupen}, M.~P., \& {Yun},
  M.~S. 1998, \apj, 494, 175

\bibitem[{{Carilli} \& {van Gorkom}(1992)}]{Carilli92}
{Carilli}, C.~L. \& {van Gorkom}, J.~H. 1992, \apj, 399, 373

\bibitem[{{Chen} {et~al.}(2010{\natexlab{a}}){Chen}, {Helsby}, {Gauthier},
  {Shectman}, {Thompson}, \& {Tinker}}]{Chen10mg2sur}
{Chen}, H.-W., {Helsby}, J.~E., {Gauthier}, J.-R., {et~al.} 2010{\natexlab{a}},
  \apj, 714, 1521

\bibitem[{{Chen} {et~al.}(2010{\natexlab{b}}){Chen}, {Wild}, {Tinker},
  {Gauthier}, {Helsby}, {Shectman}, \& {Thompson}}]{Chen10}
{Chen}, H.-W., {Wild}, V., {Tinker}, J.~L., {et~al.} 2010{\natexlab{b}}, \apjl,
  724, L176

\bibitem[{{Churchill} {et~al.}(2005){Churchill}, {Kacprzak}, \&
  {Steidel}}]{Churchill05}
{Churchill}, C.~W., {Kacprzak}, G.~G., \& {Steidel}, C.~C. 2005, in IAU Colloq.
  199: Probing Galaxies through Quasar Absorption Lines, ed. P.~{Williams},
  C.-G. {Shu}, \& B.~{Menard}, 24--41

\bibitem[{{Condon} {et~al.}(1977){Condon}, {Hicks}, \& {Jauncey}}]{Condon77}
{Condon}, J.~J., {Hicks}, P.~D., \& {Jauncey}, D.~L. 1977, \aj, 82, 692

\bibitem[{{Curran}(2010)}]{Curran10mg2}
{Curran}, S.~J. 2010, \mnras, 402, 2657

\bibitem[{{Curran}(2012)}]{Curran12}
{Curran}, S.~J. 2012, \apjl, 748, L18

\bibitem[{{Curran} {et~al.}(2010){Curran}, {Tzanavaris}, {Darling}, {Whiting},
  {Webb}, {Bignell}, {Athreya}, \& {Murphy}}]{Curran10}
{Curran}, S.~J., {Tzanavaris}, P., {Darling}, J.~K., {et~al.} 2010, \mnras,
  402, 35

\bibitem[{{Curran} {et~al.}(2007){Curran}, {Tzanavaris}, {Murphy}, {Webb}, \&
  {Pihlstr{\"o}m}}]{Curran07}
{Curran}, S.~J., {Tzanavaris}, P., {Murphy}, M.~T., {Webb}, J.~K., \&
  {Pihlstr{\"o}m}, Y.~M. 2007, \mnras, 381, L6

\bibitem[{{Curran} \& {Webb}(2006)}]{Curran06}
{Curran}, S.~J. \& {Webb}, J.~K. 2006, \mnras, 371, 356

\bibitem[{{Dallacasa} {et~al.}(1998){Dallacasa}, {Bondi}, {Alef}, \&
  {Mantovani}}]{Dallacasa98}
{Dallacasa}, D., {Bondi}, M., {Alef}, W., \& {Mantovani}, F. 1998, \aaps, 129,
  219

\bibitem[{{de Avillez} \& {Breitschwerdt}(2004)}]{deAvillez04}
{de Avillez}, M.~A. \& {Breitschwerdt}, D. 2004, \aap, 425, 899

\bibitem[{{Ellison} {et~al.}(2004){Ellison}, {Churchill}, {Rix}, \&
  {Pettini}}]{Ellison04}
{Ellison}, S.~L., {Churchill}, C.~W., {Rix}, S.~A., \& {Pettini}, M. 2004,
  \apj, 615, 118

\bibitem[{{Fitzpatrick} \& {Massa}(2007)}]{Fitzpatrick07}
{Fitzpatrick}, E.~L. \& {Massa}, D. 2007, \apj, 663, 320

\bibitem[{{Fynbo} {et~al.}(2010){Fynbo}, {Laursen}, {Ledoux}, {M{\o}ller},
  {Durgapal}, {Goldoni}, {Gullberg}, {Kaper}, {Maund}, {Noterdaeme},
  {{\"O}stlin}, {Strandet}, {Toft}, {Vreeswijk}, \& {Zafar}}]{Fynbo10}
{Fynbo}, J.~P.~U., {Laursen}, P., {Ledoux}, C., {et~al.} 2010, \mnras, 408,
  2128

\bibitem[{{Fynbo} {et~al.}(2011){Fynbo}, {Ledoux}, {Noterdaeme}, {Christensen},
  {M{\o}ller}, {Durgapal}, {Goldoni}, {Kaper}, {Krogager}, {Laursen}, {Maund},
  {Milvang-Jensen}, {Okoshi}, {Rasmussen}, {Thorsen}, {Toft}, \&
  {Zafar}}]{Fynbo11}
{Fynbo}, J.~P.~U., {Ledoux}, C., {Noterdaeme}, P., {et~al.} 2011, \mnras, 413,
  2481

\bibitem[{{Gupta} {et~al.}(2006){Gupta}, {Salter}, {Saikia}, {Ghosh}, \&
  {Jeyakumar}}]{Gupta06}
{Gupta}, N., {Salter}, C.~J., {Saikia}, D.~J., {Ghosh}, T., \& {Jeyakumar}, S.
  2006, \mnras, 373, 972

\bibitem[{{Gupta} {et~al.}(2010){Gupta}, {Srianand}, {Bowen}, {York}, \&
  {Wadadekar}}]{Gupta10}
{Gupta}, N., {Srianand}, R., {Bowen}, D.~V., {York}, D.~G., \& {Wadadekar}, Y.
  2010, \mnras, 408, 849

\bibitem[{{Gupta} {et~al.}(2007){Gupta}, {Srianand}, {Petitjean}, {Khare},
  {Saikia}, \& {York}}]{Gupta07}
{Gupta}, N., {Srianand}, R., {Petitjean}, P., {et~al.} 2007, \apjl, 654, L111

\bibitem[{{Gupta} {et~al.}(2009){Gupta}, {Srianand}, {Petitjean}, {Noterdaeme},
  \& {Saikia}}]{Gupta09}
{Gupta}, N., {Srianand}, R., {Petitjean}, P., {Noterdaeme}, P., \& {Saikia},
  D.~J. 2009, \mnras, 398, 201

\bibitem[{{Isobe} {et~al.}(1986){Isobe}, {Feigelson}, \& {Nelson}}]{Isobe86}
{Isobe}, T., {Feigelson}, E.~D., \& {Nelson}, P.~I. 1986, \apj, 306, 490

\bibitem[{{Jiang} {et~al.}(2011){Jiang}, {Ge}, {Zhou}, {Wang}, \&
  {Wang}}]{Jiang11}
{Jiang}, P., {Ge}, J., {Zhou}, H., {Wang}, J., \& {Wang}, T. 2011, \apj, 732,
  110

\bibitem[{{Kacprzak} \& {Churchill}(2011)}]{Kacprzak11}
{Kacprzak}, G.~G. \& {Churchill}, C.~W. 2011, \apjl, 743, L34

\bibitem[{{Kanekar} \& {Chengalur}(2003)}]{Kanekar03}
{Kanekar}, N. \& {Chengalur}, J.~N. 2003, \aap, 399, 857

\bibitem[{{Kanekar} {et~al.}(2009){Kanekar}, {Prochaska}, {Ellison}, \&
  {Chengalur}}]{Kanekar09mg2}
{Kanekar}, N., {Prochaska}, J.~X., {Ellison}, S.~L., \& {Chengalur}, J.~N.
  2009, \mnras, 396, 385

\bibitem[{{Kulkarni} {et~al.}(2011){Kulkarni}, {Torres-Garcia}, {Som}, {York},
  {Welty}, \& {Vladilo}}]{Kulkarni11}
{Kulkarni}, V.~P., {Torres-Garcia}, L.~M., {Som}, D., {et~al.} 2011, \apj, 726,
  14

\bibitem[{{Kunert} {et~al.}(2002){Kunert}, {Marecki}, {Spencer}, {Kus}, \&
  {Niezgoda}}]{Kunert02}
{Kunert}, M., {Marecki}, A., {Spencer}, R.~E., {Kus}, A.~J., \& {Niezgoda}, J.
  2002, \aap, 391, 47

\bibitem[{{Lane}(2000)}]{Lane_phd}
{Lane}, W. 2000, PhD thesis, University of Groningen

\bibitem[{{Ledoux} {et~al.}(2006){Ledoux}, {Petitjean}, {Fynbo}, {M{\o}ller},
  \& {Srianand}}]{Ledoux06a}
{Ledoux}, C., {Petitjean}, P., {Fynbo}, J.~P.~U., {M{\o}ller}, P., \&
  {Srianand}, R. 2006, \aap, 457, 71

\bibitem[{{L{\'o}pez} \& {Chen}(2012)}]{Lopez12}
{L{\'o}pez}, G. \& {Chen}, H.-W. 2012, \mnras, 419, 3553

\bibitem[{{M{\'e}nard} {et~al.}(2011){M{\'e}nard}, {Wild}, {Nestor}, {Quider},
  {Zibetti}, {Rao}, \& {Turnshek}}]{Menard11}
{M{\'e}nard}, B., {Wild}, V., {Nestor}, D., {et~al.} 2011, \mnras, 417, 801

\bibitem[{{Moller} \& {Warren}(1993)}]{Moller93}
{Moller}, P. \& {Warren}, S.~J. 1993, \aap, 270, 43

\bibitem[{{Murphy} {et~al.}(1993){Murphy}, {Browne}, \& {Perley}}]{Murphy93}
{Murphy}, D.~W., {Browne}, I.~W.~A., \& {Perley}, R.~A. 1993, \mnras, 264, 298

\bibitem[{{Murphy} {et~al.}(2007){Murphy}, {Curran}, {Webb}, {M{\'e}nager}, \&
  {Zych}}]{Murphy07}
{Murphy}, M.~T., {Curran}, S.~J., {Webb}, J.~K., {M{\'e}nager}, H., \& {Zych},
  B.~J. 2007, \mnras, 376, 673

\bibitem[{{Neff} \& {Hutchings}(1990)}]{Neff90}
{Neff}, S.~G. \& {Hutchings}, J.~B. 1990, \aj, 100, 1441

\bibitem[{{Nestor} {et~al.}(2011){Nestor}, {Johnson}, {Wild}, {M{\'e}nard},
  {Turnshek}, {Rao}, \& {Pettini}}]{Nestor11}
{Nestor}, D.~B., {Johnson}, B.~D., {Wild}, V., {et~al.} 2011, \mnras, 412, 1559

\bibitem[{{Noterdaeme} {et~al.}(2012){Noterdaeme}, {Laursen}, {Petitjean},
  {Vergani}, {Maureira}, {Ledoux}, {Fynbo}, {L{\'o}pez}, \&
  {Srianand}}]{Noterdaeme12}
{Noterdaeme}, P., {Laursen}, P., {Petitjean}, P., {et~al.} 2012, \aap, 540, A63

\bibitem[{{Noterdaeme} {et~al.}(2009{\natexlab{a}}){Noterdaeme}, {Ledoux},
  {Srianand}, {Petitjean}, \& {Lopez}}]{Noterdaeme09co}
{Noterdaeme}, P., {Ledoux}, C., {Srianand}, R., {Petitjean}, P., \& {Lopez}, S.
  2009{\natexlab{a}}, \aap, 503, 765

\bibitem[{{Noterdaeme} {et~al.}(2010{\natexlab{a}}){Noterdaeme}, {Petitjean},
  {Ledoux}, {L{\'o}pez}, {Srianand}, \& {Vergani}}]{Noterdaeme10co}
{Noterdaeme}, P., {Petitjean}, P., {Ledoux}, C., {et~al.} 2010{\natexlab{a}},
  \aap, 523, A80

\bibitem[{{Noterdaeme} {et~al.}(2009{\natexlab{b}}){Noterdaeme}, {Petitjean},
  {Ledoux}, \& {Srianand}}]{Noterdaeme09dla}
{Noterdaeme}, P., {Petitjean}, P., {Ledoux}, C., \& {Srianand}, R.
  2009{\natexlab{b}}, \aap, 505, 1087

\bibitem[{{Noterdaeme} {et~al.}(2010{\natexlab{b}}){Noterdaeme}, {Srianand}, \&
  {Mohan}}]{Noterdaeme10o3}
{Noterdaeme}, P., {Srianand}, R., \& {Mohan}, V. 2010{\natexlab{b}}, \mnras,
  403, 906

\bibitem[{{Piner} {et~al.}(2006){Piner}, {Bhattarai}, {Edwards}, \&
  {Jones}}]{Piner06}
{Piner}, B.~G., {Bhattarai}, D., {Edwards}, P.~G., \& {Jones}, D.~L. 2006,
  \apj, 640, 196

\bibitem[{{Prochter} {et~al.}(2006){Prochter}, {Prochaska}, \&
  {Burles}}]{Prochter06}
{Prochter}, G.~E., {Prochaska}, J.~X., \& {Burles}, S.~M. 2006, \apj, 639, 766

\bibitem[{{Quider} {et~al.}(2011){Quider}, {Nestor}, {Turnshek}, {Rao},
  {Monier}, {Weyant}, \& {Busche}}]{Quider11}
{Quider}, A.~M., {Nestor}, D.~B., {Turnshek}, D.~A., {et~al.} 2011, \aj, 141,
  137

\bibitem[{{Rahmani} {et~al.}(2010){Rahmani}, {Srianand}, {Noterdaeme}, \&
  {Petitjean}}]{Rahmani10}
{Rahmani}, H., {Srianand}, R., {Noterdaeme}, P., \& {Petitjean}, P. 2010,
  \mnras, 409, L59

\bibitem[{{Rao} {et~al.}(2011){Rao}, {Belfort-Mihalyi}, {Turnshek}, {Monier},
  {Nestor}, \& {Quider}}]{Rao11}
{Rao}, S.~M., {Belfort-Mihalyi}, M., {Turnshek}, D.~A., {et~al.} 2011, \mnras,
  1155

\bibitem[{{Rao} {et~al.}(2003){Rao}, {Nestor}, {Turnshek}, {Lane}, {Monier}, \&
  {Bergeron}}]{Rao03}
{Rao}, S.~M., {Nestor}, D.~B., {Turnshek}, D.~A., {et~al.} 2003, \apj, 595, 94

\bibitem[{{Rao} {et~al.}(2006){Rao}, {Turnshek}, \& {Nestor}}]{Rao06}
{Rao}, S.~M., {Turnshek}, D.~A., \& {Nestor}, D.~B. 2006, \apj, 636, 610

\bibitem[{{Rector} \& {Stocke}(2001)}]{Rector01}
{Rector}, T.~A. \& {Stocke}, J.~T. 2001, \aj, 122, 565

\bibitem[{{Reid} {et~al.}(1999){Reid}, {Kronberg}, \& {Perley}}]{Reid99}
{Reid}, R.~I., {Kronberg}, P.~P., \& {Perley}, R.~A. 1999, \apjs, 124, 285

\bibitem[{{Saikia} {et~al.}(1990){Saikia}, {Junor}, {Cornwell}, {Muxlow}, \&
  {Shastri}}]{Saikia90}
{Saikia}, D.~J., {Junor}, W., {Cornwell}, T.~J., {Muxlow}, T.~W.~B., \&
  {Shastri}, P. 1990, \mnras, 245, 408

\bibitem[{{Schneider} {et~al.}(2010){Schneider}, {Richards}, {Hall}, {Strauss},
  {Anderson}, {Boroson}, {Ross}, {Shen}, {Brandt}, {Fan}, {Inada}, {Jester},
  {Knapp}, {Krawczyk}, {Thakar}, {Vanden Berk}, {Voges}, {Yanny}, {York},
  {Bahcall}, {Bizyaev}, {Blanton}, {Brewington}, {Brinkmann}, {Eisenstein},
  {Frieman}, {Fukugita}, {Gray}, {Gunn}, {Hibon}, {Ivezi{\'c}}, {Kent}, {Kron},
  {Lee}, {Lupton}, {Malanushenko}, {Malanushenko}, {Oravetz}, {Pan}, {Pier},
  {Price}, {Saxe}, {Schlegel}, {Simmons}, {Snedden}, {SubbaRao}, {Szalay}, \&
  {Weinberg}}]{Schneider10}
{Schneider}, D.~P., {Richards}, G.~T., {Hall}, P.~B., {et~al.} 2010, \aj, 139,
  2360

\bibitem[{{Srianand} {et~al.}(2012){Srianand}, {Gupta}, {Petitjean},
  {Noterdaeme}, {Ledoux}, {Salter}, \& {Saikia}}]{Srianand12dla}
{Srianand}, R., {Gupta}, N., {Petitjean}, P., {et~al.} 2012, \mnras, 421, 651

\bibitem[{{Srianand} {et~al.}(2008){Srianand}, {Gupta}, {Petitjean},
  {Noterdaeme}, \& {Saikia}}]{Srianand08bump}
{Srianand}, R., {Gupta}, N., {Petitjean}, P., {Noterdaeme}, P., \& {Saikia},
  D.~J. 2008, \mnras, 391, L69

\bibitem[{{Steidel}(1995)}]{Steidel95}
{Steidel}, C.~C. 1995, in QSO Absorption Lines, ed. G.~{Meylan}, 139

\bibitem[{{Swarup} {et~al.}(1984){Swarup}, {Sinha}, \& {Hilldrup}}]{Swarup84}
{Swarup}, G., {Sinha}, R.~P., \& {Hilldrup}, K. 1984, \mnras, 208, 813

\bibitem[{{Ulvestad} {et~al.}(1981){Ulvestad}, {Johnston}, {Perley}, \&
  {Fomalont}}]{Ulvestad81}
{Ulvestad}, J., {Johnston}, K., {Perley}, R., \& {Fomalont}, E. 1981, \aj, 86,
  1010

\bibitem[{{Vanden Berk} {et~al.}(2001){Vanden Berk}, {Richards}, {Bauer},
  {Strauss}, {Schneider}, {Heckman}, {York}, {Hall}, {Fan}, {Knapp},
  {Anderson}, {Annis}, {Bahcall}, {Bernardi}, {Briggs}, {Brinkmann}, {Brunner},
  {Burles}, {Carey}, {Castander}, {Connolly}, {Crocker}, {Csabai}, {Doi},
  {Finkbeiner}, {Friedman}, {Frieman}, {Fukugita}, {Gunn}, {Hennessy},
  {Ivezi{\'c}}, {Kent}, {Kunszt}, {Lamb}, {Leger}, {Long}, {Loveday}, {Lupton},
  {Meiksin}, {Merelli}, {Munn}, {Newberg}, {Newcomb}, {Nichol}, {Owen}, {Pier},
  {Pope}, {Rockosi}, {Schlegel}, {Siegmund}, {Smee}, {Snir}, {Stoughton},
  {Stubbs}, {SubbaRao}, {Szalay}, {Szokoly}, {Tremonti}, {Uomoto}, {Waddell},
  {Yanny}, \& {Zheng}}]{VandenBerk01}
{Vanden Berk}, D.~E., {Richards}, G.~T., {Bauer}, A., {et~al.} 2001, \aj, 122,
  549

\bibitem[{{White} {et~al.}(1997){White}, {Becker}, {Helfand}, \&
  {Gregg}}]{White97first}
{White}, R.~L., {Becker}, R.~H., {Helfand}, D.~J., \& {Gregg}, M.~D. 1997,
  \apj, 475, 479

\bibitem[{{Wild} {et~al.}(2006){Wild}, {Hewett}, \& {Pettini}}]{Wild06}
{Wild}, V., {Hewett}, P.~C., \& {Pettini}, M. 2006, \mnras, 367, 211

\bibitem[{{Wills} {et~al.}(1992){Wills}, {Wills}, {Evans}, {Natta}, {Thompson},
  {Breger}, \& {Sitko}}]{Wills92}
{Wills}, B.~J., {Wills}, D., {Evans}, II, N.~J., {et~al.} 1992, \apj, 400, 96

\bibitem[{{Wolfe} {et~al.}(1978){Wolfe}, {Broderick}, {Johnston}, \&
  {Condon}}]{Wolfe78}
{Wolfe}, A.~M., {Broderick}, J.~J., {Johnston}, K.~J., \& {Condon}, J.~J. 1978,
  \apj, 222, 752

\bibitem[{{York} {et~al.}(2007){York}, {Kanekar}, {Ellison}, \&
  {Pettini}}]{York07}
{York}, B.~A., {Kanekar}, N., {Ellison}, S.~L., \& {Pettini}, M. 2007, \mnras,
  382, L53

\bibitem[{{York} {et~al.}(2006){York}, {Khare}, {Vanden Berk}, {Kulkarni},
  {Crotts}, {Lauroesch}, {Richards}, {Schneider}, {Welty}, {Alsayyad}, {Kumar},
  {Lundgren}, {Shanidze}, {Smith}, {Vanlandingham}, {Baugher}, {Hall},
  {Jenkins}, {Menard}, {Rao}, {Tumlinson}, {Turnshek}, {Yip}, \&
  {Brinkmann}}]{York06}
{York}, D.~G., {Khare}, P., {Vanden Berk}, D., {et~al.} 2006, \mnras, 367, 945

\bibitem[{{Zwaan} {et~al.}(2005){Zwaan}, {van der Hulst}, {Briggs},
  {Verheijen}, \& {Ryan-Weber}}]{Zwaan05}
{Zwaan}, M.~A., {van der Hulst}, J.~M., {Briggs}, F.~H., {Verheijen}, M.~A.~W.,
  \& {Ryan-Weber}, E.~V. 2005, \mnras, 364, 1467

\end{thebibliography}

\longtabL{4}{
\begin{landscape}
\begin{longtable}{cccccccccccccccc}
\caption{\label{vlbares} Results from the VLBA data}\\
\hline\hline
{\large \strut}   R.A.        &   Dec.      &  \zabs        & rms &  Id &  S  &  r  &$\theta$&  a & b/a &$\phi$& S$_T$ & $f_{\rm VLBA}/$ &LS & $\alpha$\\
             &              &                     &     &        &     &     &        &    &     &       &  & $c_f$      &         \\
     (J2000)       &      (J2000)       &               & (mJy\,beam$^{-1}$)  &  & (mJy)  & (mas)    & ($\deg$)    & (mas)     &  & ($\deg$)     & (mJy)  &  & (pc)      &         \\
   (1)       &   (2)        &   (3)               & (4) &  (5)   & (6) & (7) & (8)    &(9) &(10) & (11)  & (12)  & (13) &  (14)    &  (15)       \\
\hline
\endfirsthead
\caption{continued.}\\
\hline\hline
{\large \strut}   R.A.        &   Dec.      &  \zabs        & rms &  Id &  S  &  r  &$\theta$&  a & b/a &$\phi$& S$_T$ & $f_{\rm VLBA}/$ &LS & $\alpha$\\
             &              &                     &     &        &     &     &        &    &     &       &  & $c_f$      &         \\
     (J2000)   &       (J2000)      &               & (mJy\,beam$^{-1}$)  &  & (mJy)  & (mas)    & ($\deg$)    & (mas)     &  & ($\deg$)     & (mJy)  &  & (pc)      &         \\
   (1)       &   (2)        &   (3)               & (4) &  (5)   & (6) & (7) & (8)    &(9) &(10) & (11)  & (12)  & (13) &  (14)    &  (15)       \\
\hline
\endhead
\hline
\endfoot
\multicolumn{15}{c}{\bf {\large \strut} VLBA parameters at 20-cm}\\
~01 08 26.8432    & -00 37 24.060     & 1.3710  &1.8 &  1     & 489 &  0  &  $-$   & 6.36&0.54&  78  & 930   &  0.97/ &554 & 0.38\\
                  &                   &         &    &  2     & 162 &16.8 &  80    &29.90&0.12& -78  &       &  0.53  &   \\
                  &                   &         &    &  3     &  58 &33.1 &  88    &21.35&0.29& -33  &       &        &   \\
                  &                   &         &    &  4     &  90 &51.4 &  87    &13.61&0.39& -84  &       &        &   \\
                  &                   &         &    &  5     & 106 &65.3 &  79    &21.71&0.64& -64  &       &        &   \\
~01 54 54.3653    & -00 07 23.231     & 1.1803  &0.3 &  1     & 250 &  0  & $-$    & 6.83&0.16& -38  & 264   &  0.95  &$<$57  &-0.65\\
~02 14 52.2909    & +14 05 27.455     & 1.4463  &0.2 &  1     &  11 &  0  & $-$    & 4.24&0.00 & 68  & 125   &  0.11/ &113 & 0.68\\
                  &                   &         &    &  2     &   3 &13.3 & 57     &10.99&0.54 &-23  &       &  0.09* &   \\
~02 40 08.1758    & -23 09 15.727     & 1.3647  &2.3 &  1     &4080 &  0  & $-$    & 4.75&0.85&  39  & 6256  &  0.85/ &84  &-0.25\\
                  &                   &         &    &  2     &1224 & 9.9 & 22     & 4.85&0.29&  24  &       &  0.65  &   \\
~02 59 28.5155    & -00 19 59.983     & 1.3370  &0.3 &  1     & 231 &  0  & $-$    & 3.51&0.36 &-10  & 235   &  0.98  &$<$30  &-0.55\\
~07 42 37.3873    & +39 44 35.629     & 1.1485  &0.3 &  1     &  74 &  0  & $-$    & 5.98&0.17 &-27  & 114   &  0.76/ &63  & 0.35\\
                  &                   &         &    &  2     &  13 & 7.6 & 155    & 6.65&0.83 & 0   &       &  0.65* &   \\
~07 48 09.4683    & +30 06 30.533     & 1.4470  &0.3 &  1     & 105 &  0  & $-$    &28.33&0.50 &-85  & 231   &  0.65/ &464 & 0.49\\
                  &                   &         &    &  2     &  20 &30.1 &  0     &14.31&0.00 &-85  &       &  0.46* &   \\
                  &                   &         &    &  3     &  24 &28.8 & 135    &71.92&0.23 &-79  &       &        &   \\
~08 00 36.0269    & +50 10 44.290     & 1.4146  &0.3 &  1     &  81 &  0  & $-$    & 4.22&0.04 & 33   &118    & 0.81/ &286 & 0.08\\
                  &                   &         &    &  2     &  14 &33.6 & 24     & 16.1&0.19 & 19   &       & 0.69* &   \\
~08 02 48.4323    & +29 17 34.211     & 1.3648  &0.3 &  1     &   7 &  0  & $-$    & 3.94&0.42 &-72  & 195   &  0.04* &$<$33  & 0.88\\
~08 08 39.6666    & +49 50 36.529     & 1.4071  &2.1 &  1     & 482 &  0  & $-$    & 2.03&0.51 &-53  & 930   &  0.52  &$<$17  &-0.25 \\
~08 15 34.1624    & +33 05 29.010     & 0.8515  &0.2 &  1     &  25 &  0  & $-$    & 3.96&0.16 & 26   &343    & 0.07  &$<$30  &0.59 \\
~08 17 10.5482    & +23 52 23.970     & 1.3060  &0.3 &  1     & 105 &  0  & $-$    & 3.81&0.12 &-65   &215    & 0.84/ &122 &0.15 \\
                  &                   &         &    &  2     &  71 &5.92 &-44     & 4.73&0.00 & 66   &       & 0.49  &   \\ 
                  &                   &         &    &  3     &   5 &14.5 &-89     & 6.67&0.00 & 53   &       &       &   \\ 
~08 45 06.2503    & +42 57 18.393     & 1.1147  &0.3 &  1     & 164 &  0  & $-$    & 2.20&0.55 & 46  & 218   &  0.75  &$<$18  &0.03 \\
~08 50 42.2438    & +51 59 11.654     & 1.3265  &0.2 &  1     &  65 &  0  & $-$    & 1.80&0.10 &-89  &  63   &  1.03  &$<$15  &0.02\\
~08 52 44.7391    & +34 35 40.521     & 1.3095  &0.3 &  1     &  47 &  0  & $-$    & 2.39&0.67 & 64   & 69   &  0.74/ &60  &-0.36 \\
                  &                   &         &    &  2     &   4 & 7.1 &-124    & 8.69&0.00 &-57   &      &  0.68  &   \\ 
~09 30 35.0851    & +46 44 08.579     & 0.6216  &0.4 &  1     & 171 &  0  & $-$    & 11.4&0.11 & 23   &336    & 0.63/ &92  &0.49\\
                  &                   &         &    &  2     &  40 &13.5 &-154    & 13.6&0.39 & 15   &       & 0.51* &   \\ 
~09 53 27.9565    & +32 25 51.525     & 1.2372  &0.3 &  1     &  78 &  0  & $-$    & 4.13&0.28 &-1    &132   &  0.71/ &161 &0.14\\
                  &                   &         &    &  2     &  16 &19.3 &  0     & 11.0&0.34 &-1    &      &  0.59  &   \\
~10 07 18.0731    & +22 51 26.955     & 0.5602  &0.2 &  1     & 150 &  0  & $-$    & 8.16&0.52 &-64   &350    & 0.55/ &49  &0.54\\
                  &                   &         &    &  2     &  42 & 7.6 &-112    & 17.8&0.45 &-7    &       & 0.43  &   \\
~10 58 13.0427    & +49 39 36.046     & 1.2120  &0.2 &  1     &   5 &  0  & $-$    & 15.3&0.00 & 64   &244    & 0.33/ &840 &0.70\\
                  &                   &         &    &  2     &  27 &20.9 & 92     & 7.98&0.00 & 90   &       & 0.11* &   \\
                  &                   &         &    &  3     &  23 &35.8 & 97     & 5.26&0.00 &-81   &       &       &   \\
                  &                   &         &    &  4     &   6 &56.9 &-99     & 22.7&0.00 & 64   &       &       &   \\
                  &                   &         &    &  5     &  20 &66.7 &-107    & 12.7&0.98 &-62   &       &       &   \\
~11 00 21.0333    & +16 29 14.661     & 0.8540  &0.4 &  1     & 201 &  0  & $-$    & 3.20&0.41 & 9    &265    & 0.76  &$<$25  &0.22\\
{\large \strut} 11 26 57.6554    & +45 16 06.286     & 1.3022  &1.4 &  1     & 340 &  0  & $-$    & 3.42&0.66 & 21   &424    & 0.80  &$<$24  &-0.21\\
~11 48 56.5677    & +52 54 25.337     & 0.8306  &0.2 &  1     &  53 &  0  & $-$    & 1.97&0.65 &-69   &100    & 0.78/ &56 & 0.38\\
                  &                   &         &    &  2     &  25 &7.36 & 116    & 3.70&0.00 &-75   &       & 0.53  &   \\
~11 57 34.8393    & +16 38 59.769     & 0.7624  &0.6 &  1     & 845 &  0  & $-$    & 1.65&0.00 &-80   &747    & 1.00  &$<$12 &-1.13 \\
~12 08 54.2565    & +54 41 58.163     & 1.2110  &1.1 &  1     & 228 &  0  & $-$    & 2.24&0.34 &-51   &265    & 0.86  &$<$11 &0.04\\
~12 13 32.1710    & +13 07 20.892     & 0.7718  &1.4 &  1     & 145?&  0  & $-$    & 4.96&0.44 & 12   &1356   & 0.11  &$<$37 &0.45\\
~12 32 56.6101    & +57 22 14.208     & 1.3429  &0.2 &  1     &   4 &  0  & $-$    & 7.33&0.39 &-45   &115    & 0.17/ &164 &0.79 \\
                  &                   &         &    &  2     &  15 &19.3 &-150    &37.57&0.19 & 26   &       & 0.13* &   \\
~12 34 31.7244    & +64 55 56.530     & 1.3739/ &0.3 &  1     &  22 &  0  & $-$    & 3.95&0.00 & 56   & 93    & 0.43/ &64  &0.49\\
                  &                   & 1.3829  &    &  2     &  18 & 7.5 &  0     &10.14&0.74 & 23   &       & 0.24* &   \\
~13 00 36.4392    & +08 28 02.887     & 0.8665  &0.3 &  1     &  50 &  0  & $-$    & 4.86&0.21 &  3   &106    & 0.68/ &173 &0.34 \\
                  &                   &         &    &  2     &  18 &12.2 &-178    & 6.32&0.26 & 24   &       & 0.47  &   \\
                  &                   &         &    &  3     &   4 &22.4 &-170    & 7.60&0.00 &-20   &       &       &   \\
~13 29 01.4156    & +10 53 04.807     & 0.6715/ &0.2 &  1     &  34 &  0  & $-$    & 3.60&0.00 &-79   &121    & 0.33/ &72,86 &0.41 \\
                  &                   & 1.1645  &    &  2     &   6 &10.3 & 81     & 9.80&0.69 &  9   &       & 0.28* &   \\
~13 33 35.7804    & +16 49 04.109     & 0.7448  &0.4 &  1     & 217 &  0  & $-$    & 2.49&0.00 & 25   &421    & 0.76/ &151 & 0.29\\
                  &                   &         &    &  2     & 101 &20.6 & 21     & 12.8&0.14 & 22   &       & 0.52  &   \\
~14 08 56.4815    & -07 52 26.635     & 1.2753  &0.4 &  1     & 233 &  0  & $-$    & 2.24&0.53 & 43   &612    & 0.79/ &182 & 0.24\\
                  &                   &         &    &  2     & 150 &7.14 &-106    & 4.35&0.00 & 48   &       & 0.38  &   \\ 
                  &                   &         &    &  3     &  77 &13.5 &-118    & 11.6&0.00 &-85   &       &       &   \\ 
                  &                   &         &    &  4     &  25 &8.35 &  47    & 11.0&0.37 &-18   &       &       &   \\ 
~14 10 30.9982    & +61 41 36.908     & 0.7596  &0.2 &  1     & 104 &  0  & $-$    & 1.78&0.92 & 40   &118    & 0.88  &$<$13 &-0.41 \\
~14 30 09.7389    & +10 43 26.865     & 1.2431  &0.8 &  1     & 301 &  0  & $-$    & 1.69&0.46 & 65   &311    & 0.97  &$<$14 &-0.91\\
~15 01 24.6285    & +56 19 49.678     & 1.2788  &0.3 &  1     &  36 &  0  & $-$    & 1.69&0.70 & 86   &182    & 0.31/ &29 &0.25\\
                  &                   &         &    &  2     &  21 & 3.4 &-117    & 9.97&0.00 & 64   &       & 0.20* &   \\ 
~15 08 23.7169    & +33 47 00.762     & 1.1650  &0.3 &  1     & 123 &  0  & $-$    &12.48&0.15 &-45   &132    & 0.99/ &126 &0.61 \\
                  &                   &         &    &  2     &   7 &15.2 &-37     &10.44&0.00 &  0   &       & 0.93  &   \\%
~16 23 46.2300    & +07 18 54.891     & 1.3350        &0.3 &  1     &  30 &  0  & $-$    & 4.64&0.89  & 24    & 72    &$<$0.42* &39 &0.82 \\
~16 36 38.1833    & +21 12 55.558     & 0.8000  &0.4 &  1     & 147 &  0  & $-$    & 2.06&0.69 &-83   &392    & 0.52/ & 193 &0.34 \\
                  &                   &         &    &  2     &  38 &8.75 & 108    & 3.64&0.50 &-58   &       & 0.38  &   \\ 
                  &                   &         &    &  3     &   7 &9.11 &-121    & 8.63&0.00 &-47   &       &       &   \\ 
                  &                   &         &    &  4     &  13 &19.3 & 113    & 14.4&0.47 & 10   &       &       &   \\ 
~20 31 54.9948    & +12 19 41.363     & 1.1157  &0.2 &  1     & 447 &  0  & $-$    & 2.58&0.40 &  4   &985    & 0.48/ &76 & -0.60 \\
                  &                   &         &    &  2     &  23 &9.26 &-173    & 10.4&0.10 &-28   &       & 0.45  &   \\
~23 40 23.6695    & -00 53 27.009     & 1.3603  &0.3 &  1     & 106 &  0  & $-$    & 1.71&0.46 & 16   &117    & 0.91  &$<$15 & -1.20\\
~23 58 10.8819    & -10 20 08.624     & 1.1726  &0.4 &  1     & 571 &  0  & $-$    & 3.71&0.29 &-12   &770    & 0.74  &$<$31 & -0.70\\
\multicolumn{15}{c}{\bf VLBA parameters at 13-cm based on images from VCS}\\
~04 57 03.1792    & -23 24 52.020     & 0.8922  &4.4 & $-$    & $-$ & $-$ & $-$    & $-$ & $-$ & $-$  &1760   & 0.63  &$<$195& -0.16\\      
~11 45 21.3153    & +04 55 26.690     & 1.3433  &0.5 & $-$    & $-$ & $-$ & $-$    & $-$ & $-$ & $-$  &554    & 0.60/ &254& +0.55\\      
                  &                   &         &    &        &     &     &        &     &     &      &       & 0.32  &   &      \\      
~21 29 12.1759    & -15 38 41.041     & 0.6628  &1.0 & $-$    & $-$ & $-$ & $-$    & $-$ & $-$ & $-$  &990    & 0.89  &$<$140& -1.04\\      

\end{longtable}
\tablefoot{ 
Columns 1 and 2: right ascension and declination of the 
`component-1' (see column 5) fitted to the quasar, respectively; 
Column 3: absorption redshift; 
Column 4: rms in the map in mJy\,beam$^{-1}$; 
Column 5: component id; 
Column 6: flux of the component in mJy;
Columns 7 and 8: radius and position angle of the component with respect to 
`component-1', respectively;
Column 9: major axis in mas of the Gaussian component fitted to characterise the milliarcsecond scale structure;
Columns 10 and 11: axial ratio and position angle of the Gaussian component, respectively; 
Column 12: flux density in mJy from FIRST/NVSS (S$_T$); 
Column 13: ratio of total flux density detected in VLBA map and S$_T$ (called $f_{\rm VLBA}$).  
In case of quasars with the multiple Gaussian components the $c_f$ i.e. the ratio of the flux density of the strongest 
component detected in the VLBA map and the S$_T$ is also given.;  
Column 14: largest linear size (LS) i.e. separation between the farthest components of the quasar in pc at absorber redshift.  
For the quasars represented by a single Gaussian component we take major axis of the deconvolved component as the upper limit on 
quasar size; 
Column 15: spectral index, S$\propto\nu^{-\alpha}$, between the 20-cm and 50-cm/90-cm using flux densities from the arcsecond scale maps.  
}
\end{landscape}
}

\Online

\begin{appendix}

\begin{twocolumn}
\section{Log of the 21-cm absorption observations.}
\label{sec:21cmlog}
\begin{table}[h]
\caption{Log for the GBT, GMRT and WSRT observations. }
\begin{center}
\begin{tabular}{cccc}
\hline
\hline
{\large \strut} Quasar      & Telescope & Date       & Time      \\
            &           &            & (hrs)     \\
\hline
{\large \strut} J0457$-$2324   &  GBT  &  2010\,Mar\,12  &  0.8        \\ 
~J0800$+$5010   & GMRT  &  2009\,Jun\,05  &  4.4        \\ 
~J0817$+$2352   & GMRT  &  2009\,Jun\,23  &  2.4        \\ 
~J0930$+$4644   & WSRT  &  2009\,Aug\,22  &  22         \\
                &       &  2009\,Aug\,23  &             \\
                &       &  2009\,Sep\,06  &             \\
                &       &  2009\,Sep\,08  &             \\
~J0956$+$4046   & GMRT  &  2009\,Aug\,09  &  12         \\ 
                &       &  2009\,Aug\,24  &             \\ 
~J1007$+$2251   &  GBT  &  2009\,Oct\,15  &  4.4        \\ 
                & WSRT  &  2009\,Oct\,24  &  12         \\ 
~J1148$+$5254   &  GBT  &  2009\,Oct\,16  &  4.3        \\ 
                &       &  2009\,Nov\,19  &             \\ 
                &       &  2010\,Jan\,08  &             \\ 
~J1216$+$5843   &  GBT  &  2009\,Nov\,19  &  1.9        \\ 
                &       &  2010\,Feb\,18  &             \\ 
                & WSRT  &  2009\,Aug\,20  &  8.8        \\ 
~J1252$+$4427   &  GBT  &  2010\,Jan\,19  &  2.3        \\ 
~J1329$+$1053   &  GBT  &  2010\,Mar\,12  &   4.1       \\ 
                &       &  2010\,Mar\,13  &             \\ 
                &       &  2010\,Mar\,15  &             \\ 
                &       &  2010\,Mar\,22  &             \\ 
                & WSRT  &  2009\,Sep\,16  &   9.8       \\ 
~J1333$+$1649   &  GBT  &  2010\,Jan\,09  &   2.4       \\ 
                &       &  2010\,Mar\,12  &             \\ 
~J1408$-$0755   & GMRT  &  2009\,Jun\,21  &   2.5       \\
                &       &  2010\,Jan\,22\tablefootmark{\dag} &  4.5       \\ 
~J1410$+$6141   &  GBT  &  2009\,Sep\,16  &   5.0       \\ 
                &       &  2009\,Sep\,17  &             \\ 
                &       &  2009\,Oct\,16  &             \\ 
                &       &  2009\,Oct\,19  &             \\ 
~J1501$+$5619   & GMRT  &  2009\,Aug\,24  &   3.1       \\ 
~J1636$+$2112   &  GBT  &  2009\,Sep\,25  &   2.4       \\ 
~J2031$+$1219   & GMRT  &  2009\,Jun\,05  &   2.8       \\ 
~J2129$-$1538   & WSRT  &  2006\,Jul\,12  &   27        \\  
                &       &  2006\,Jul\,13  &             \\  
                &       &  2006\,Jul\,17  &             \\  
\hline
\end{tabular}
\end{center}
\tablefoot{ 
The last column gives the total on-source time.  \tablefoottext{\dag} Higher resolution spectrum of 
the absorber towards J1408$-$0755 to study the variability.  
}
\label{obslog}
\end{table}

\section{Determining flux densities and 21-cm optical depths}
\label{sec:conf}
\begin{table}
\caption{Summary of our 21-cm absorption measurements. 
Listed from left to right are the quasar name, flux density in mJy at the redshifted 
21-cm frequency, spectral resolution in \kms, spectral rms in mJy\,beam$^{-1}$\,channel$^{-1}$ 
at the resolution given in the previous column, 
3$\sigma$ optical depth limit for the spectra smoothed to 10\,\kms, and the integrated 21-cm 
optical depth or in case of non-detections the 3$\sigma$ upper limit for a spectral resolution 10\,\kms.
}
\begin{center}
\begin{tabular}{ccccccc}
\hline
\hline
{\Large \strut} Quasar             &      Flux  & $\delta v$     &  rms        & $\tau_{3\sigma,10}$         & $\int\tau$dv    \\
\hline
{\large \strut} J0457$-$2324       &  2400     & 3.9   &     3.2    & 0.002    & 0.20$\pm$0.02 \\
~J0800$+$5010       &  126      & 4.0   &     2.1    & 0.033    & $<$0.33 \\
~J0817$+$2352       &  244      & 3.8   &     2.4    & 0.019    & $<$0.19 \\
~J0930$+$4644       &  415      & 3.4   &     5.3    & 0.024    & $<$0.24 \\
~J0956$+$4046(P1)   &  53       & 3.8   &     0.9    & 0.035    & $<$0.35 \\
~J0956$+$4046(P2)   &  23       & 3.8   &     0.9    & 0.080    & $<$0.80 \\
~J1007$+$2251       &  397      & 3.2   &     1.7    & 0.008    & $<$0.08 \\
~J1148$+$5254       &  122      & 3.8   &     1.3    & 0.021    & $<$0.21 \\
~J1216$+$5843       &  537      & 3.6   &     2.7    & 0.010    & $<$0.10 \\ 
~J1252$+$4427       &  497      & 3.9   &     2.0    & 0.008    & 2.06$\pm$0.13 \\ 
~J1329$+$1053       &  138      & 3.5   &     1.2    & 0.017    & $<$0.17 \\
~J1333$+$1649       &  486      & 3.6   &     2.1    & 0.008    & $<$0.08 \\
~J1408$-$0755       &  744      & 3.8   &     1.9    & 0.005    & 0.51$\pm$0.05 \\ 
~                   &  685      & 1.9   &     1.9    & 0.004    & 0.48$\pm$0.04 \\ 
~J1410$+$6141       &  93       & 3.6   &     1.5    & 0.031    & $<$0.31 \\
~J1501$+$5619       &  223      & 3.8   &     2.1    & 0.018    & $<$0.18 \\
~J1636$+$2112       &  456      & 3.7   &     2.1    & 0.009    & $<$0.09 \\
~J2031$+$1219       &  595      & 3.5   &     2.7    & 0.009    & 2.11$\pm$0.15 \\
~J2129$-$1538       &  341      & 3.4   &     3.2    & 0.017    & $<$0.17 \\
\hline
\end{tabular}
\end{center}
\label{mg2obsres}
\end{table}

For the quasars observed with the GBT\footnote{GBT beam FWHM at 800\,MHz, 
corresponding to the central frequency of PF1-800 receiver, 
is 15$^\prime$. The rms confusion corresponding to this is 90\,mJy.} 
the rms confusion due to other radio sources in the beam can be the dominant effect that limits 
the accuracy with which flux density of the background quasar 
and therefore the 21-cm optical depth can be determined.  
We use interferometric images from our WSRT observations and 
literature to address this issue. 
First consider the two quasars with 21-cm absorption detections from the GBT data.  
For J1252+4427, in the FIRST catalogue, there are no strong (\gapp10\,mJy) sources within 15$^{\prime}$ and the 
total flux density due to {\it other} sources within the beam FWHM is only $\sim$20\,mJy.  
This suggests that the confusion is not a serious issue for this quasar, and indeed the flux density  
estimated from our GBT observations matches within $\sim$5\% with the estimate from the interpolation 
of flux densities from the FIRST (at 20-cm) and WENSS (at 92-cm) surveys.   
The other quasar with 21-cm detection i.e. J0457$-$2324 is a blazar and known to exhibit variability 
(see Section~\ref{sec:new21}). We measure the flux density of 2.72\,Jy from our GBT observations. The contribution 
due to other sources within the beam FWHM at 20-cm based on the NVSS is 384\,mJy (no beam-correction). 
The 90\% of contribution comes from a source that has a steep spectral index of 0.9.  
Assuming that this other source is not variable would imply that the flux density of J0457$-$2324 is 2.4\,Jy 
(beam correction applied) at the redshifted 21-cm absorption frequency.        

Now consider systems towards the quasars with no 21-cm detections from the GBT.
The flux density estimates for J1007+2251, J1216+5843 and J1329+1053 at the redshifted 21-cm frequency 
are estimated from our WSRT observations which took place within 6 months of the GBT observations (Table~\ref{obslog}).  
For J1148+5254 and J1410+6141, the flux densities at the redshifted 
21-cm frequency have been estimated by interpolating between the flux densities from the FIRST and 
WENSS surveys at 20-cm and 92-cm respectively.  
The flux densities for J1333+1649 and J1636+2112 have been determined using the FIRST and TXS surveys. 
Table~\ref{mg2obsres} lists the flux densities and optical depth values for all the \mgii\ systems observed.

\end{twocolumn}

\newpage
\begin{onecolumn}

\begin{landscape}
\section{Strong \mgii\ systems selected from \citet{Kanekar09mg2}}
\begin{table}[h]
\caption{Details of the strong \mgii\ systems ($W_{\rm r}\ge$1\AA) and 21-cm absorption measurements taken from \citet{Kanekar09mg2}. 
Listed from left to right are the quasar names, their emission redshifts, redshifts of the intervening \mgii\ systems, 
rest frame equivalent widths of the \mgiia, \mgiib, \mgia\ and \feiia\ absorption lines, integrated 21-cm optical 
depth or in case of non-detection the 3$\sigma$ upper limit to it for a spectral resolution of 10\,\kms, and quasar morphology 
at 1.4\,GHz.  The quasars are classified as compact (C) when deconvolved size in the FIRST is less than 2$^{\prime\prime}$ and 
resolved (R) when the deconvolved size $>$ 2$^{\prime\prime}$.  When the FIRST is not available, the classification is either 
based on the radio morphology from the references provided below or on the NVSS/SUMSS when nothing else is available. 
The last three columns give $f_{VLBA}$, $c_f$ and the largest linear size of the quasars estimated from the milliarcsecond scale resolution maps.
}
\begin{center}
\begin{tabular}{ccccccccccccc}
\hline
\hline
{\Large \strut} Quasar      &   \zem   & \zabs  &  $W_{\rm r}$(\mgiia)  & $W_{\rm r}$(\mgiib)  & $W_{\rm r}$(\mgia) & $W_{\rm r}$(\feiia) & $\int\tau$dv & Mor. & $f_{VLBA}$ & $c_f$ & LS \\
            &          &        &        (\AA)       &       (\AA)      &     (\AA)      &      (\AA)      &   (\kms)         &   & &  & (pc)  \\
\hline
\multicolumn{9}{c}{\bf {\large \strut} Systems at $0.5<z<1$}\\
~B0039$-$407 & 2.478  &  0.8485  &  2.47$\pm$0.01\tablefootmark{V}      & 2.34$\pm$0.01 &    0.56$\pm$0.01 & 1.90$\pm$0.01         &$<$0.49  & C\tablefootmark{1}   & $-$     &  $-$     &  $-$       \\   
~B0109+176   & 2.157  &  0.8392  &  1.75\tablefootmark{B}               & 1.20          & $<$0.20          & 1.09                  &$<$0.039 & R\tablefootmark{2}   & $-$     &  $-$     &  $-$       \\               
~B0240$-$060 & 1.805  &  0.5810  &  1.90$\pm$0.02\tablefootmark{V}      & 1.38$\pm$0.02 &    0.39$\pm$0.03 & 1.70$\pm$0.02         &$<$0.075 & C\tablefootmark{1}   & 1.00    &  1.00    &  $<$79     \\   
             & 1.805  &  0.7550  &  1.67$\pm$0.01\tablefootmark{V}      & 1.48$\pm$0.01 &    0.78$\pm$0.01 & 1.25$\pm$0.04\tablefootmark{E1}  &$<$0.061 & C\tablefootmark{1}   & 1.00    &  1.00    &  $<$88     \\     
~B0244$-$128 & 2.201  &  0.8282  &  1.66$\pm$0.01\tablefootmark{V}      & 1.56$\pm$0.01 &    0.49$\pm$0.01 & 1.33$\pm$0.01         &$<$0.13  & R\tablefootmark{3}   & 1.00    &  0.92    &     162    \\     
~B0409$-$045 & 1.684  &  0.8797  &  1.58$\pm$0.22          & 1.16$\pm$0.21 & $<$0.23          & 1.21$\pm$0.20         &$<$0.14  & C\tablefootmark{1}   & $-$     &  $-$     &  $-$       \\              
~B0445+097   & 2.108  &  0.8392  &  3.17\tablefootmark{B}               & 2.19          &    0.91          & 1.97                  &$<$0.054 & R\tablefootmark{2}   & 0.22    &  0.17    &     80     \\             
~B0812+332   & 2.426  &  0.8518  &  2.67$\pm$0.28          & 2.19$\pm$0.31 &    0.55$\pm$0.15 & 1.12$\pm$0.27         &$<$0.11  & C       & 0.07    &  0.07*   &  $<$30     \\        
~B0957+003   & 0.905  &  0.6722  &  1.96$\pm$0.16          & 1.57$\pm$0.17 &    0.32$\pm$0.16 & 1.15$\pm$0.24         &$<$0.06  & R       & $-$     &  $-$     &  $-$       \\        
~B1012+022   & 1.375  &  0.7632  &  1.53$\pm$0.05          & 1.28$\pm$0.05 &    0.33$\pm$0.05 & 0.71$\pm$0.05         &$<$0.06  & R       & $-$     &  $-$     &  $-$       \\                 
~B1200+068   & 2.182  &  0.862   &  5.29$\pm$0.27          & 4.88$\pm$0.26 &    3.04$\pm$0.27 & 4.08$\pm$0.31         &$<$0.08  & C       & 1.00    &  0.66*   &     170    \\          
~B1210+134   & 1.139  &  0.7717  &  1.22$\pm$0.09          & 1.14$\pm$0.09 &    0.38$\pm$0.10 & 0.87$\pm$0.10         &$<$0.052 & C?      & 0.20    &  0.20    &  $<$74     \\   
~B1222+438   & 1.075  &  0.7033  &  1.01$\pm$0.25          & 0.70$\pm$0.25 & $<$0.16          & 1.22$\pm$0.26         &$<$0.42  & R       & 0.70    &  0.70    &  $<$72     \\  
~B1324$-$047 & 1.882  &  0.7850  &  2.62$\pm$0.02\tablefootmark{V}      & 2.36$\pm$0.01 &    0.78$\pm$0.02 & 1.87$\pm$0.02         &$<$0.48  & C?      & 0.38    &  0.38    &  $<$75     \\  
~B1343+386   & 1.852  &  0.8076  &  1.61$\pm$0.12          & 1.45$\pm$0.12 & $<$0.14          & 1.20$\pm$0.12         &$<$0.06  & R       & 0.70    &  0.30*   &     81     \\          
~B1402$-$012 & 2.518  &  0.8901  &  1.14$\pm$0.09          & 1.03$\pm$0.09 &    0.30$\pm$0.10 & 1.08$\pm$0.09         &$<$0.09  & C       & 0.65    &  0.65    &  $<$78     \\  
~B1611+343   & 1.397  &  0.6672  &  1.21$\pm$0.06          & 1.09$\pm$0.06 & $<$0.07          & 0.58$\pm$0.07         &$<$0.09  & C       & 1.00    &  1.00    &  $<$70     \\   
~B1701+593   & 1.798  &  0.7238  &  1.67$\pm$0.22          & 1.17$\pm$0.22 &    0.80$\pm$0.20 & 1.24$\pm$0.24         &$<$0.10  & R       &  $-$    &  $-$     &  $-$       \\         
\multicolumn{9}{c}{\bf {\large \strut} Systems at $1.1<z<1.5$}\\                                                                                                                                
~B1005$-$333 & 1.837  & 1.3734   &  0.93\tablefootmark{E2}            & ...           &...               & 0.84                  &$<$0.11  & C$^1$   & 1.00    &  0.78    &     280    \\ 
~B1136+408   & 2.366  & 1.3702   &  1.37$\pm$0.21          & 0.69$\pm$0.23 & $<$0.23          & 0.60$\pm$0.23         &$<$0.13  & C       & 1.00    &  1.00?   &     102    \\ 
~~~B1142+052\tablefootmark{\dag} & 1.345  & 1.3431   &  2.15$\pm$0.11       & 1.61$\pm$0.12 &    1.29$\pm$0.14 & 1.40$\pm$0.17       &   0.557 & R       & 0.60    &  0.32    &     254    \\ 
~B1204+399   & 1.518  & 1.3254   &  1.41$\pm$0.11          & 1.10$\pm$0.11 & $<$0.11          & 0.76$\pm$0.14         &$<$0.23  & C       & 1.00    &  0.72*   &     564    \\ 
~B2003$-$025 & 1.457  & 1.2116   &  2.65$\pm$0.14\tablefootmark{A}      & 2.17$\pm$0.15 & $<$0.31          & 1.27$\pm$0.14         &$<$0.022 & R       & $-$     &  $-$     &     $-$    \\ 
\hline
\end{tabular}
\end{center}
\tablefoot{
\tablefoottext{\dag} This system is common with G09.  
\tablefoottext{V} Equivalent widths measured using VLT archival spectra. 
\tablefoottext{B} Equivalent widths from \citet{Barthel90}. 
\tablefoottext{Y} Equivalent widths from \citet{York07}. 
\tablefoottext{A} Equivalent widths from \citet{Aldcroft94}. 
\tablefoottext{E1} \feiia~ is not covered in the VLT UVES spectrum, the $W_{\rm r}$(\feiia) is taken from \citet{Ellison04}.
\tablefoottext{E2} Equivalent widths taken from \citet{Ellison04}.
\tablefoottext{1} Compact on the basis of NVSS/SUMSS. 
\tablefoottext{2} Projected linear size $>$2$^{\prime\prime}$ at 5\,GHz \citep{Barthel88}. 
\tablefoottext{3} Projected linear size $>$2$^{\prime\prime}$ at 5\,GHz \citep{Reid99}. 
}
\label{kansamp}
\end{table}
\end{landscape}
\end{onecolumn}

\newpage
\begin{onecolumn}

\begin{landscape}
\section{Strong \mgii\ systems selected from \citet{Lane_phd}}

The Table~6 of G09 lists strong \mgii\ systems with 21-cm measurements from \citet{Lane_phd}. 
Only systems at \zabs$>$0.5 are considered here (see Section~\ref{sec:unbiased}).
The systems towards B0109+176 and B0957+003 are common with the sample of \citet{Kanekar09mg2} and 
are listed in the Table~\ref{kansamp} with the more sensitive optical depth limits.
In addition to the absorber towards B1622+238, we have excluded the system towards the quasar 
B0827+243 as it is non-\mgii\ selected \citep[see][for the details]{Lane_phd}. 

\begin{table}[h]
\caption{Sample of \mgii\ systems with $W_{\rm r}\ge$1\AA~ from \citet{Lane_phd}. The column listings are same as in Table~\ref{kansamp}.}  
\begin{center}
\begin{tabular}{cccccccccccc}
\hline
\hline
{\Large \strut} Source      & \zem    & \zabs   &  $W_{\rm r}$(\mgiia)  & $W_{\rm r}$(\mgiib)  & $W_{\rm r}$(\mgia) & $W_{\rm r}$(\feiia) & $\int\tau$dv & Mor. & $f_{VLBA}$ & $c_f$ &  LS \\
            &          &        &  (\AA) & (\AA) & (\AA) & (\AA) & (km\,s$^{-1}$) &  &   &   & (pc) \\
\hline
{\large \strut} B0109$+$200 &  0.746     & 0.5346  & 2.26 &   1.71 &    $-$   &   $-$    &  $<$0.27        &   C\tablefootmark{1} &    $-$       &     $-$      &    $-$      \\ 
~B0229$+$341 &  1.240     & 0.7754  & 1.92 &   2.02 & $<$1.13  &   $-$    &  $<$0.43        &   R\tablefootmark{3} &    $-$       &     $-$      &    $-$      \\ 
~B0235$+$164 &  0.940     & 0.5238  & 2.42 &   2.34 &    0.91  &   1.79   &     13.0\tablefootmark{\dag}  &   R\tablefootmark{4} &    0.3-1     &     0.3-1    &  $<$77      \\ 
~B0420$-$014 &  0.915     & 0.6330  & 1.02 &   0.86 & $<$0.36  &   $-$    &  $<$1.59        &   C\tablefootmark{5} &    $-$       &     $-$      &    $-$      \\ 
~B0454$+$039 &  1.343     & 0.8597  & 1.53 &   1.40 &    0.37  &   1.11   &  $<$0.11        &   C     &    0.88      &     0.55     &     122     \\ 
~B0805$+$046 &  2.876     & 0.9598  & 1.01 &   0.83 & $<$0.60  &   0.24   &  $<$0.71        &   R\tablefootmark{2} &    $-$       &     $-$      &    $-$      \\ 
~B1218$+$339 &  1.519     & 0.7423  & 1.34 &   1.08 & $<$0.70  &   1.00   &  $<$0.20        &   R     &    0.07      &     0.07     &  $<$73      \\ 
~B1327$-$206 &  1.169     & 0.8530  & 2.11 &   1.48 & $<$0.40  &   0.76   &  $<$2.04        &   C\tablefootmark{1} &    $-$       &     $-$      &    $-$      \\ 
~B1354$+$258 &  2.004     & 0.8585  & 1.00 &   0.86 & $<$0.10  &$<$0.20   &  $<$1.05        &   R     &    $-$       &     $-$      &    $-$      \\ 
~B1556$-$245 &  2.815     & 0.7713  & 2.07 &   1.91 &    1.07  &$<$0.20   &  $<$0.68        &   C\tablefootmark{6} &    $-$       &     $-$      &    $-$      \\ 
~B1629$+$120 &  1.795     & 0.5313  & 1.40 &   1.35 &    0.31  &   0.70   &     0.49\tablefootmark{\ddag} &   C     &    0.30      &     0.10     &     7165\tablefootmark{7}\\ 
             &    ''      & 0.9004  & 1.06 &   0.67 &    0.44  &   0.63   &  $<$0.82        &   C     &    $-$       &     $-$      &    $-$      \\ 
~B2212$-$299 &  2.703     & 0.6329  & 1.26 &   1.00 &    0.36  &   $-$    &  $<$2.07        &   C$^1$ &    $-$       &     $-$      &    $-$      \\ 
\hline
\end{tabular}
\end{center}
\tablefoot{
\tablefoottext{\dag} 21-cm absorption data are from \citet{Wolfe78}.  
\tablefoottext{\ddag} 21-cm absorption data are from \citet{Kanekar03}.
\tablefoottext{1} Compact on the basis of NVSS. 
\tablefoottext{2} Projected linear size $>$2$^{\prime\prime}$ at 5\,GHz \citep{Barthel88}. 
\tablefoottext{3} Based on \citet{Swarup84}.
\tablefoottext{4} Based on the 1.64\,GHz image from \citet{Murphy93}.
\tablefoottext{5} Unresolved in the 3$^{\prime\prime}$ scale resolution VLA image at 1.4\,GHz \citep{Ulvestad81}.
\tablefoottext{6} Compact in the 5\,GHz images from \citet{Reid99} and \citet{Barthel00}.
\tablefoottext{7} Based on 2.3\,GHz image from \citet{Dallacasa98}; see \citet{Saikia90} for the subarcsecond scale image at 408\,MHz.
} 
\label{lanesamp}
\end{table}
\end{landscape}
\end{onecolumn}


\section{20\,cm VLBA maps of quasars from our sample}

\begin{figure*}[h]
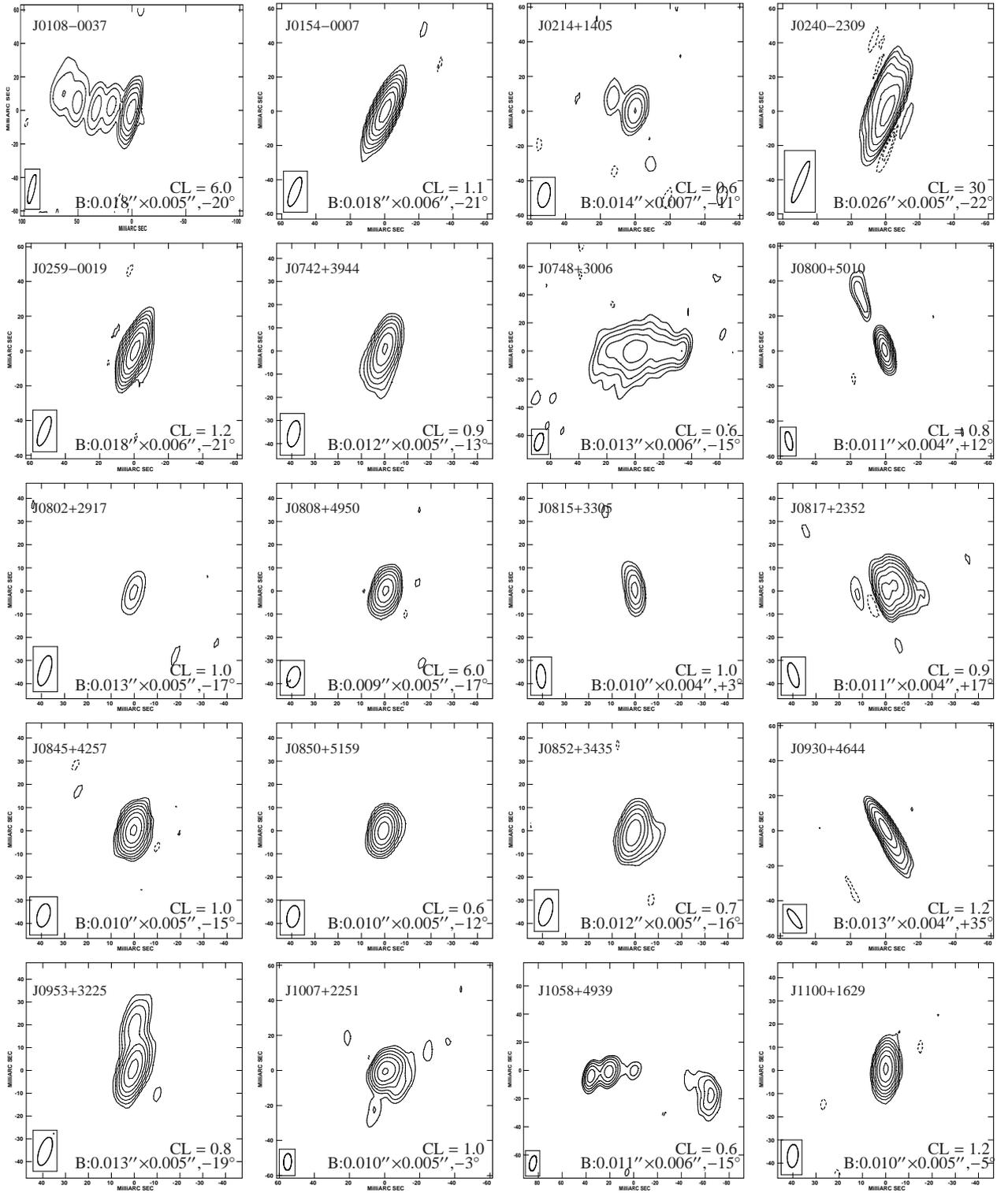

\centerline{
\vbox{
\hbox{
\psfig{figure=J0108MAP_NOLABELS.PS,height=4.0cm,width=4.1cm,angle=-90}
\psfig{figure=J0154MAP_NOLABELS.PS,height=4.0cm,width=4.1cm,angle=-90}
\psfig{figure=J0214MAP_NOLABELS.PS,height=4.0cm,width=4.1cm,angle=-90}
\psfig{figure=J0240MAP_NOLABELS.PS,height=4.0cm,width=4.1cm,angle=-90}
}
\hbox{
\psfig{figure=J0259MAP_NOLABELS.PS,height=4.0cm,width=4.1cm,angle=-90}
\psfig{figure=J0742MAP_NOLABELS.PS,height=4.0cm,width=4.1cm,angle=-90}
\psfig{figure=J0748MAP_NOLABELS.PS,height=4.0cm,width=4.1cm,angle=-90}
\psfig{figure=J0800MAP_NOLABELS.PS,height=4.0cm,width=4.1cm,angle=-90}
}
\hbox{
\psfig{figure=J0802MAP_NOLABELS.PS,height=4.0cm,width=4.1cm,angle=-90}
\psfig{figure=J0808MAP_NOLABELS.PS,height=4.0cm,width=4.1cm,angle=-90}
\psfig{figure=J0815MAP_NOLABELS.PS,height=4.0cm,width=4.1cm,angle=-90}
\psfig{figure=J0817MAP_NOLABELS.PS,height=4.0cm,width=4.1cm,angle=-90}
}
\hbox{
\psfig{figure=J0845MAP_NOLABELS.PS,height=4.0cm,width=4.1cm,angle=-90}
\psfig{figure=J0850MAP_NOLABELS.PS,height=4.0cm,width=4.1cm,angle=-90}
\psfig{figure=J0852MAP_NOLABELS.PS,height=4.0cm,width=4.1cm,angle=-90}
\psfig{figure=J0930MAP_NOLABELS.PS,height=4.0cm,width=4.1cm,angle=-90}
}
\hbox{
\psfig{figure=J0953MAP_NOLABELS.PS,height=4.0cm,width=4.1cm,angle=-90}
\psfig{figure=J1007MAP_NOLABELS.PS,height=4.0cm,width=4.1cm,angle=-90}
\psfig{figure=J1058MAP_NOLABELS.PS,height=4.0cm,width=4.1cm,angle=-90}
\psfig{figure=J1100MAP_NOLABELS.PS,height=4.0cm,width=4.1cm,angle=-90}
}
}
}
\caption[]{Contour plots of the VLBA images at 20\,cm. 
The rms in the images are listed in Table~\ref{vlbares} 
and the maps are centred at the Gaussian `component-1' fitted to characterise the 
milliarcsecond scale structure.  
The restoring beam, shown as the ellipse, and the first contour level (CL) in mJy\,beam$^{-1}$ are 
provided at the bottom of each image. The contour levels are plotted as 
CL$\times$($-$1, 1, 2, 4, 8,...)\,mJy\,beam$^{-1}$. 
}
\vskip -22.2cm
\begin{picture}(400,400)(0,0)
\put( 040,378){\scriptsize J0108$-$0037}
\put( 105,300){\tiny       CL = 6.0}
\put( 060,293){\tiny       B:0.018$^{\prime\prime}$$\times$0.005$^{\prime\prime}$,$-$20$^\circ$ }
\put( 160,378){\scriptsize J0154$-$0007}
\put( 225,300){\tiny       CL = 1.1}
\put( 180,293){\tiny       B:0.018$^{\prime\prime}$$\times$0.006$^{\prime\prime}$,$-$21$^\circ$ }
\put( 280,378){\scriptsize J0214$+$1405}
\put( 345,300){\tiny       CL = 0.6}
\put( 300,293){\tiny       B:0.014$^{\prime\prime}$$\times$0.007$^{\prime\prime}$,$-$11$^\circ$ }
\put( 400,378){\scriptsize J0240$-$2309}
\put( 465,300){\tiny       CL =  30}
\put( 420,293){\tiny       B:0.026$^{\prime\prime}$$\times$0.005$^{\prime\prime}$,$-$22$^\circ$ }
\put( 040,262){\scriptsize J0259$-$0019}
\put( 105,184){\tiny       CL = 1.2}
\put( 060,177){\tiny       B:0.018$^{\prime\prime}$$\times$0.006$^{\prime\prime}$,$-$21$^\circ$ }
\put( 160,262){\scriptsize J0742$+$3944}
\put( 225,184){\tiny       CL = 0.9}
\put( 180,177){\tiny       B:0.012$^{\prime\prime}$$\times$0.005$^{\prime\prime}$,$-$13$^\circ$ }
\put( 280,262){\scriptsize J0748$+$3006}
\put( 345,184){\tiny       CL = 0.6}
\put( 300,177){\tiny       B:0.013$^{\prime\prime}$$\times$0.006$^{\prime\prime}$,$-$15$^\circ$ }
\put( 400,262){\scriptsize J0800$+$5010}
\put( 465,184){\tiny       CL = 0.8}
\put( 420,177){\tiny       B:0.011$^{\prime\prime}$$\times$0.004$^{\prime\prime}$,$+$12$^\circ$ }
\put( 040,147){\scriptsize J0802$+$2917}
\put( 105,069){\tiny       CL = 1.0}
\put( 060,062){\tiny       B:0.013$^{\prime\prime}$$\times$0.005$^{\prime\prime}$,$-$17$^\circ$ }
\put( 160,147){\scriptsize J0808$+$4950}
\put( 225,069){\tiny       CL = 6.0}
\put( 180,062){\tiny       B:0.009$^{\prime\prime}$$\times$0.005$^{\prime\prime}$,$-$17$^\circ$ }
\put( 280,147){\scriptsize J0815$+$3305}
\put( 345,069){\tiny       CL = 1.0}
\put( 305,062){\tiny       B:0.010$^{\prime\prime}$$\times$0.004$^{\prime\prime}$,$+$3$^\circ$ }
\put( 400,147){\scriptsize J0817$+$2352}
\put( 465,069){\tiny       CL = 0.9}
\put( 420,062){\tiny       B:0.011$^{\prime\prime}$$\times$0.004$^{\prime\prime}$,$+$17$^\circ$ }
\put( 040,032){\scriptsize J0845$+$4257}
\put( 105,-45){\tiny       CL = 1.0}
\put( 060,-52){\tiny       B:0.010$^{\prime\prime}$$\times$0.005$^{\prime\prime}$,$-$15$^\circ$ }
\put( 160,032){\scriptsize J0850$+$5159}
\put( 225,-45){\tiny       CL = 0.6}
\put( 180,-52){\tiny       B:0.010$^{\prime\prime}$$\times$0.005$^{\prime\prime}$,$-$12$^\circ$ }
\put( 280,032){\scriptsize J0852$+$3435}
\put( 345,-45){\tiny       CL = 0.7}
\put( 300,-52){\tiny       B:0.012$^{\prime\prime}$$\times$0.005$^{\prime\prime}$,$-$16$^\circ$ }
\put( 400,032){\scriptsize J0930$+$4644}
\put( 465,-45){\tiny       CL = 1.2}
\put( 420,-52){\tiny       B:0.013$^{\prime\prime}$$\times$0.004$^{\prime\prime}$,$+$35$^\circ$ }
\put( 040,-84){\scriptsize J0953$+$3225}
\put( 105,-160){\tiny       CL = 0.8}
\put( 060,-167){\tiny       B:0.013$^{\prime\prime}$$\times$0.005$^{\prime\prime}$,$-$19$^\circ$ }
\put( 160,-84){\scriptsize J1007$+$2251}
\put( 225,-160){\tiny       CL = 1.0}
\put( 180,-167){\tiny       B:0.010$^{\prime\prime}$$\times$0.005$^{\prime\prime}$,$-$3$^\circ$ }
\put( 280,-84){\scriptsize J1058$+$4939}
\put( 345,-160){\tiny       CL = 0.6}
\put( 300,-167){\tiny       B:0.011$^{\prime\prime}$$\times$0.006$^{\prime\prime}$,$-$15$^\circ$ }
\put( 400,-84){\scriptsize J1100$+$1629}
\put( 465,-160){\tiny       CL = 1.2}
\put( 425,-167){\tiny       B:0.010$^{\prime\prime}$$\times$0.005$^{\prime\prime}$,$-$5$^\circ$ }
\end{picture}
\vskip +10.0cm
\label{vlbamap}
\end{figure*}

\begin{figure*}
\addtocounter{figure}{-1}
\centerline{
\vbox{
\hbox{
\psfig{figure=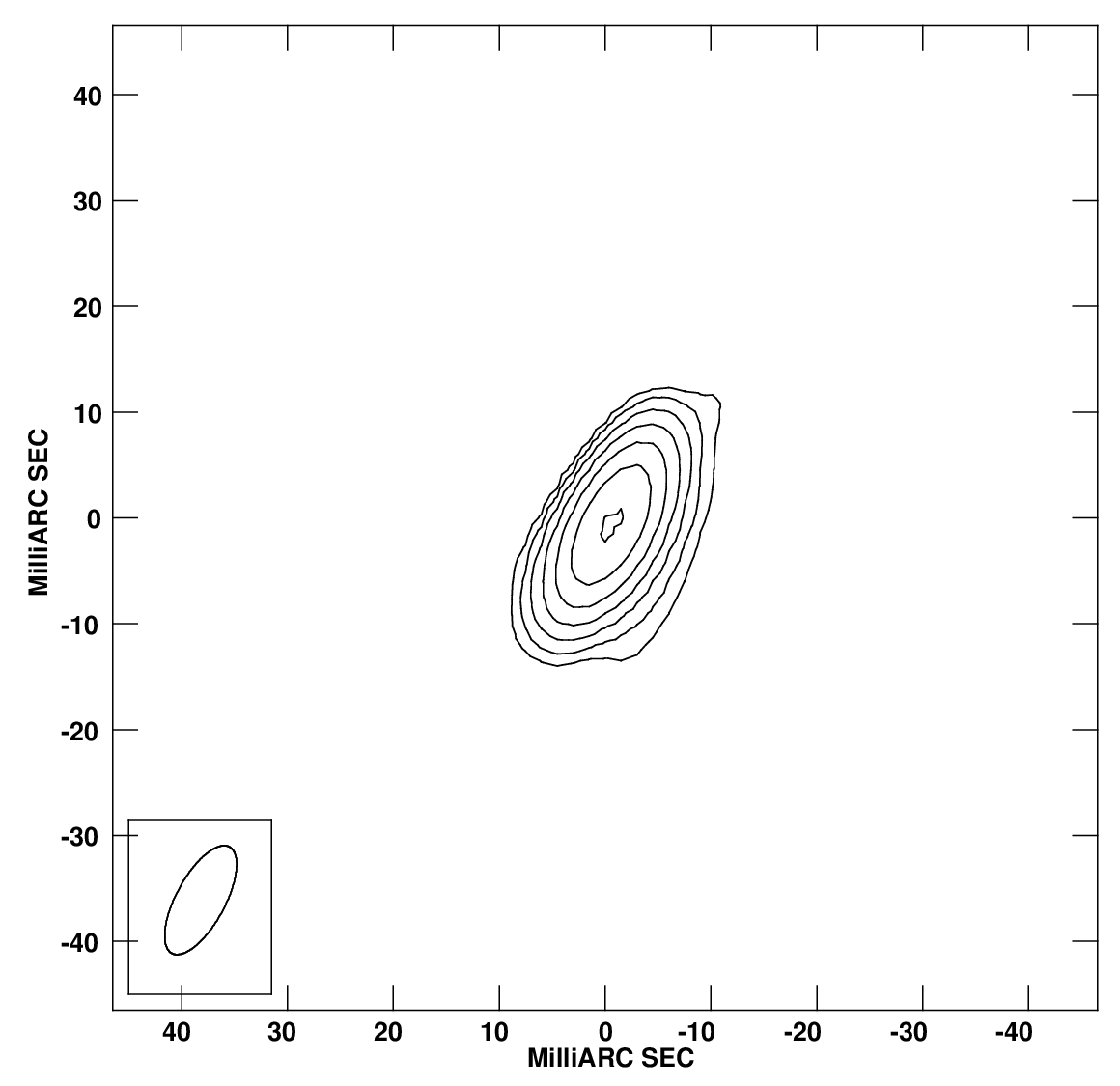,height=4.0cm,width=4.1cm,angle=-90}
\psfig{figure=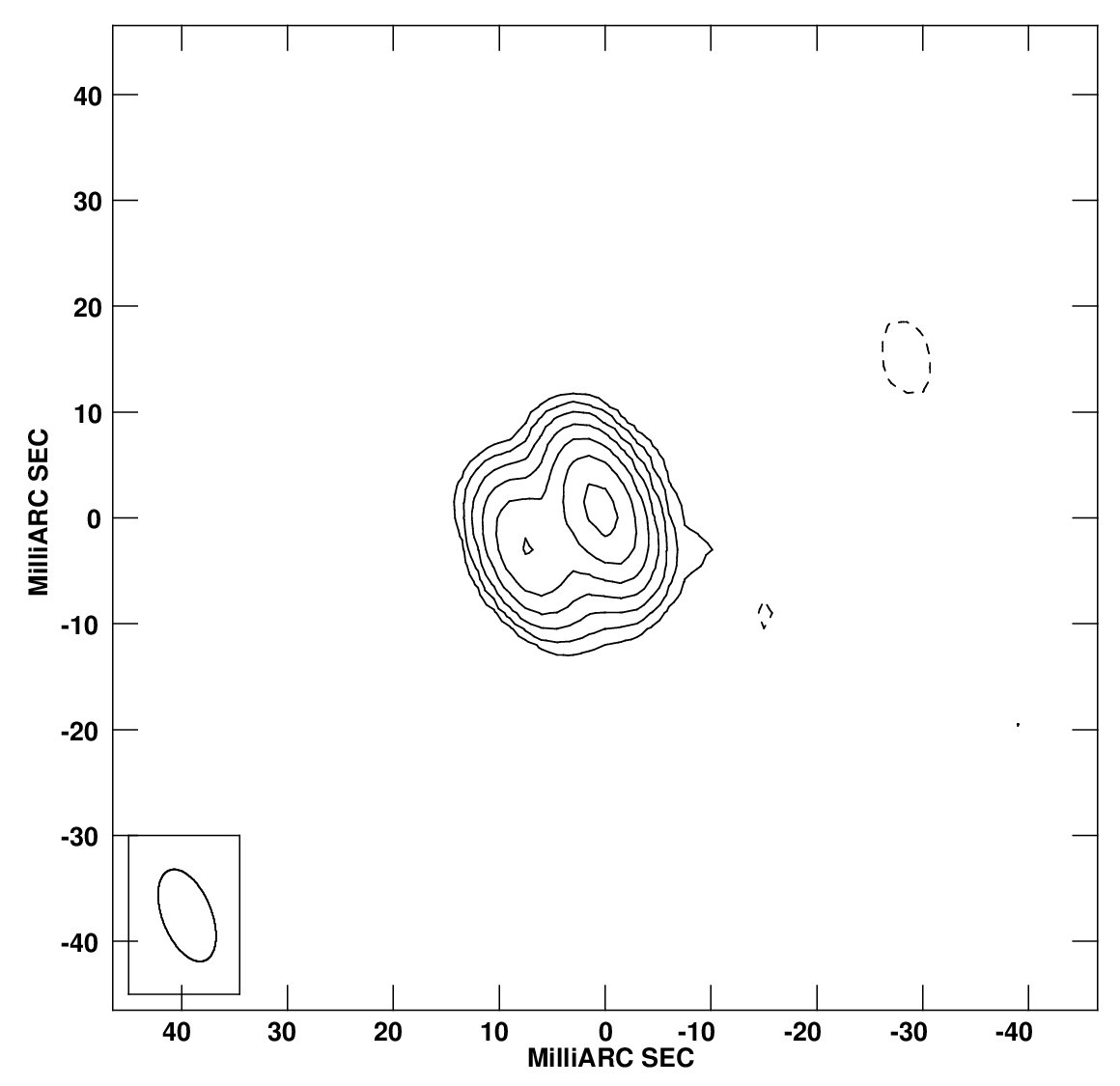,height=4.0cm,width=4.1cm,angle=-90}
\psfig{figure=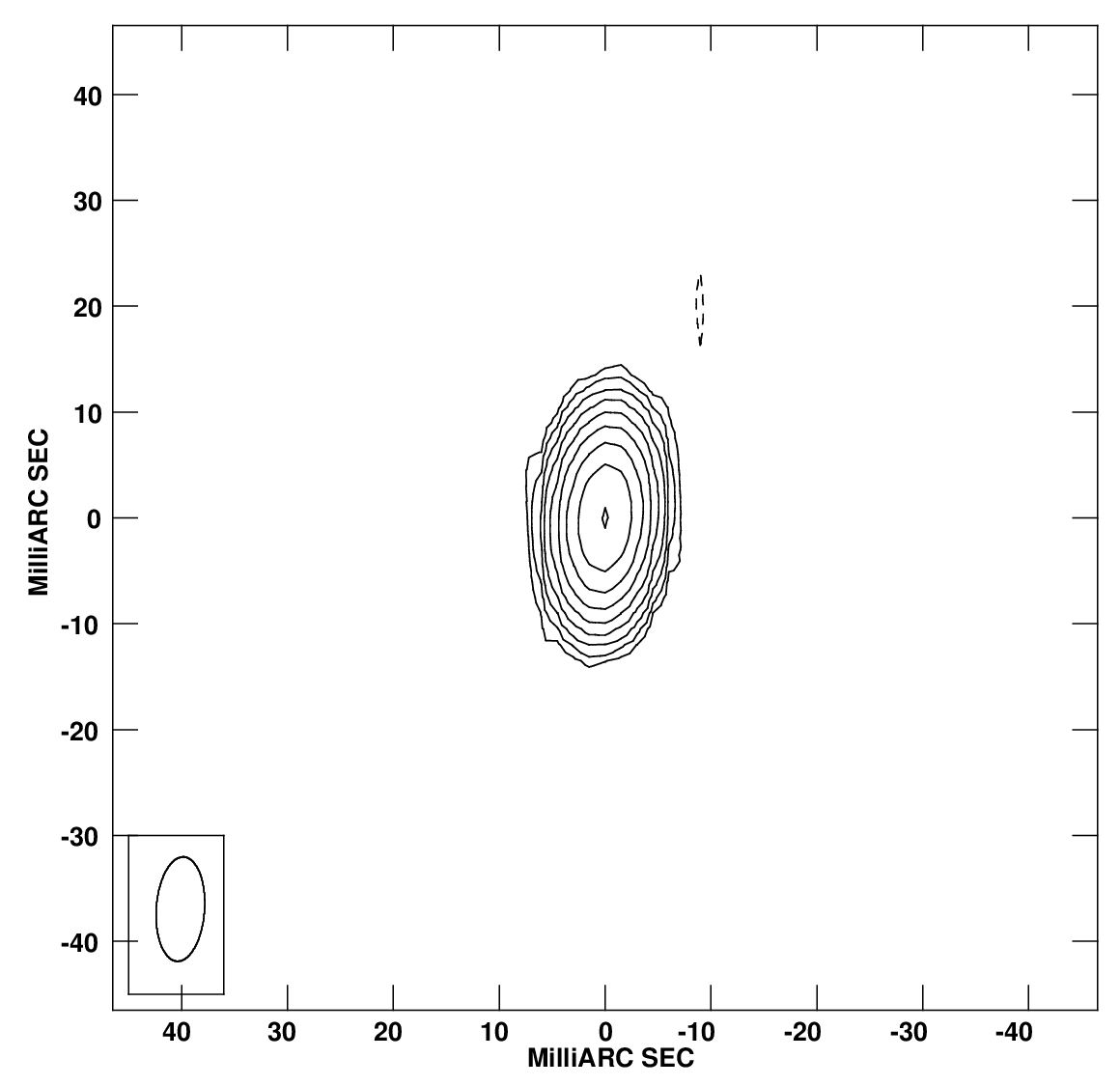,height=4.0cm,width=4.1cm,angle=-90}
\psfig{figure=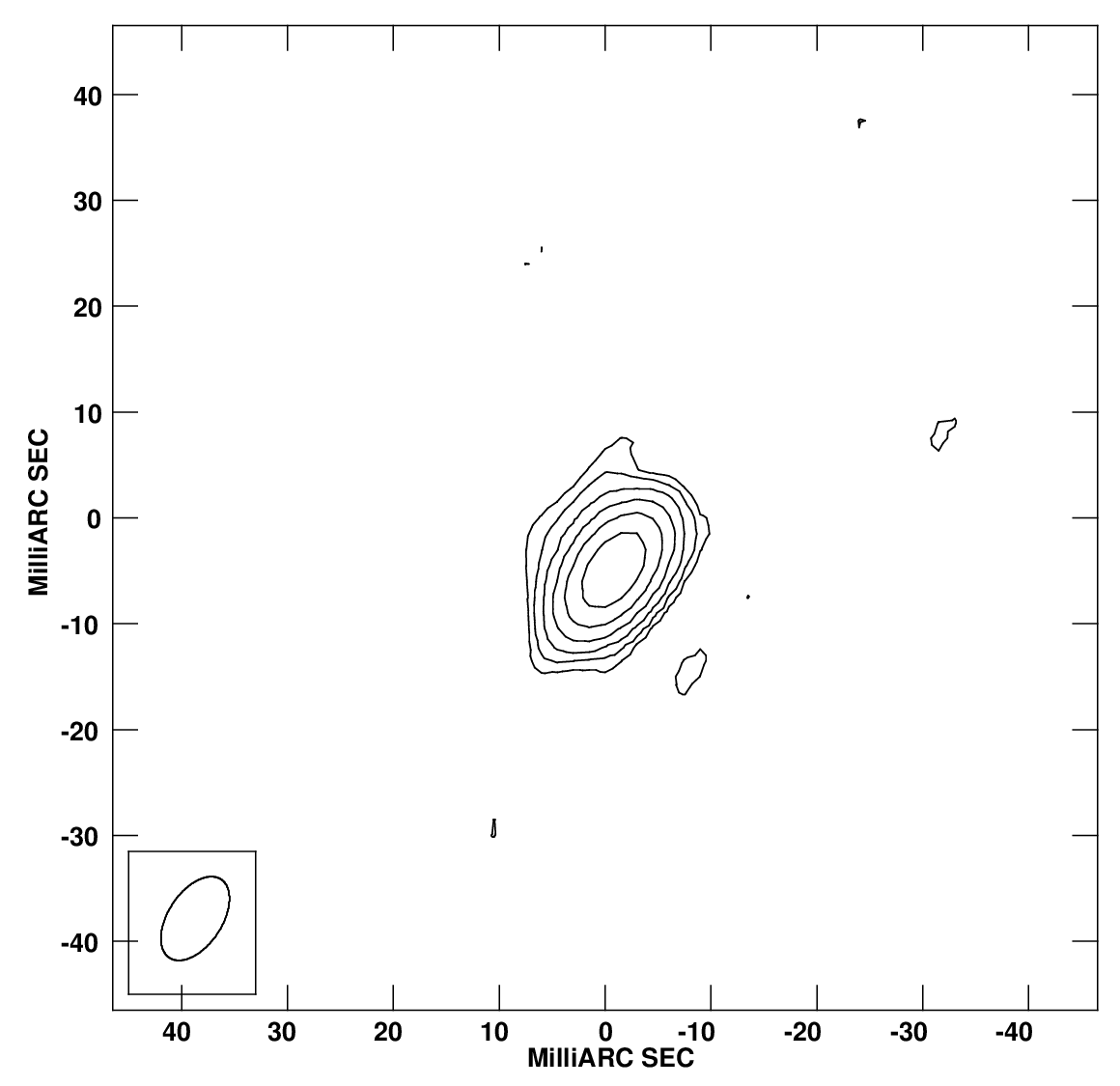,height=4.0cm,width=4.1cm,angle=-90}
}
\hbox{
\psfig{figure=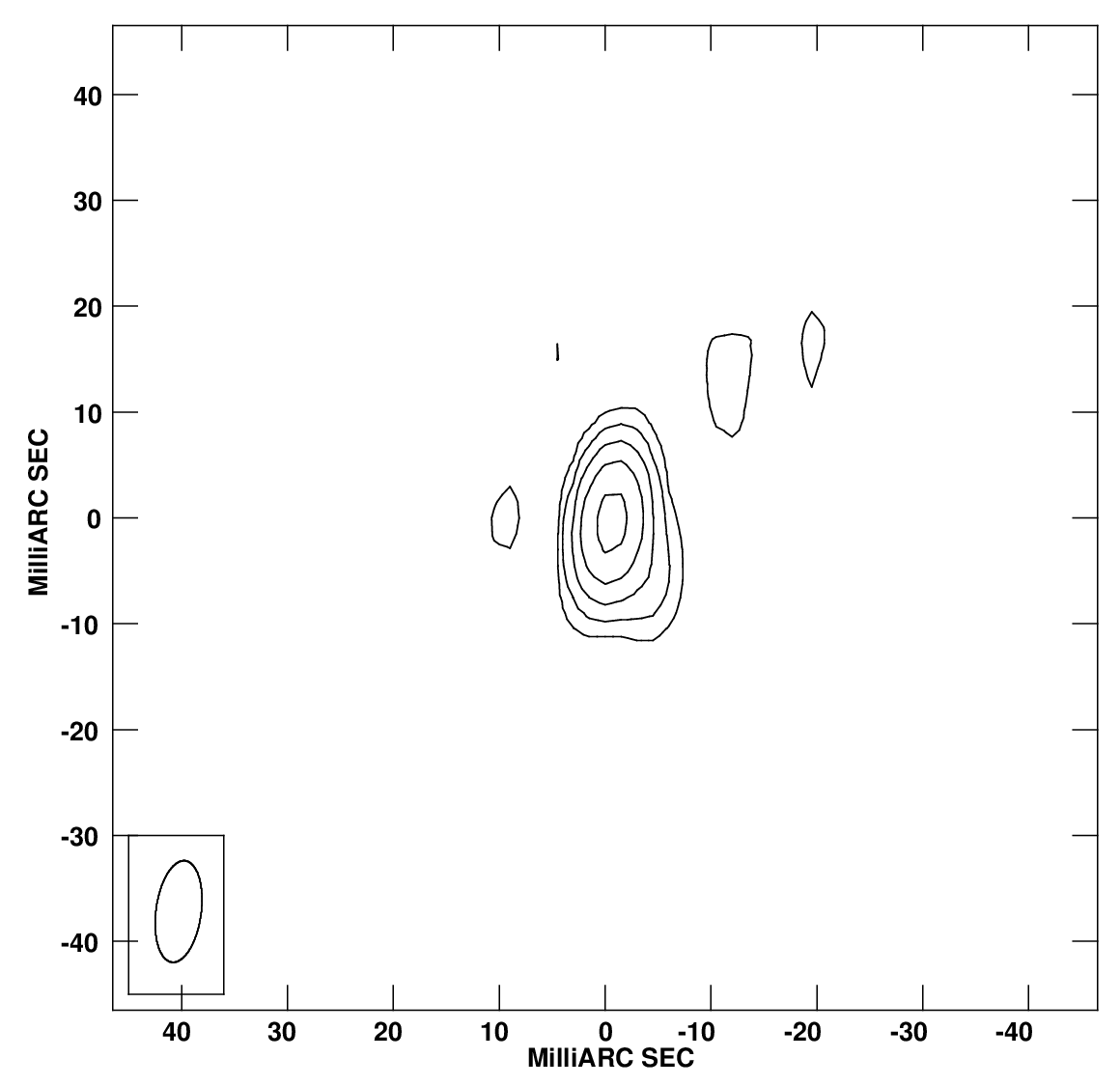,height=4.0cm,width=4.1cm,angle=-90}
\psfig{figure=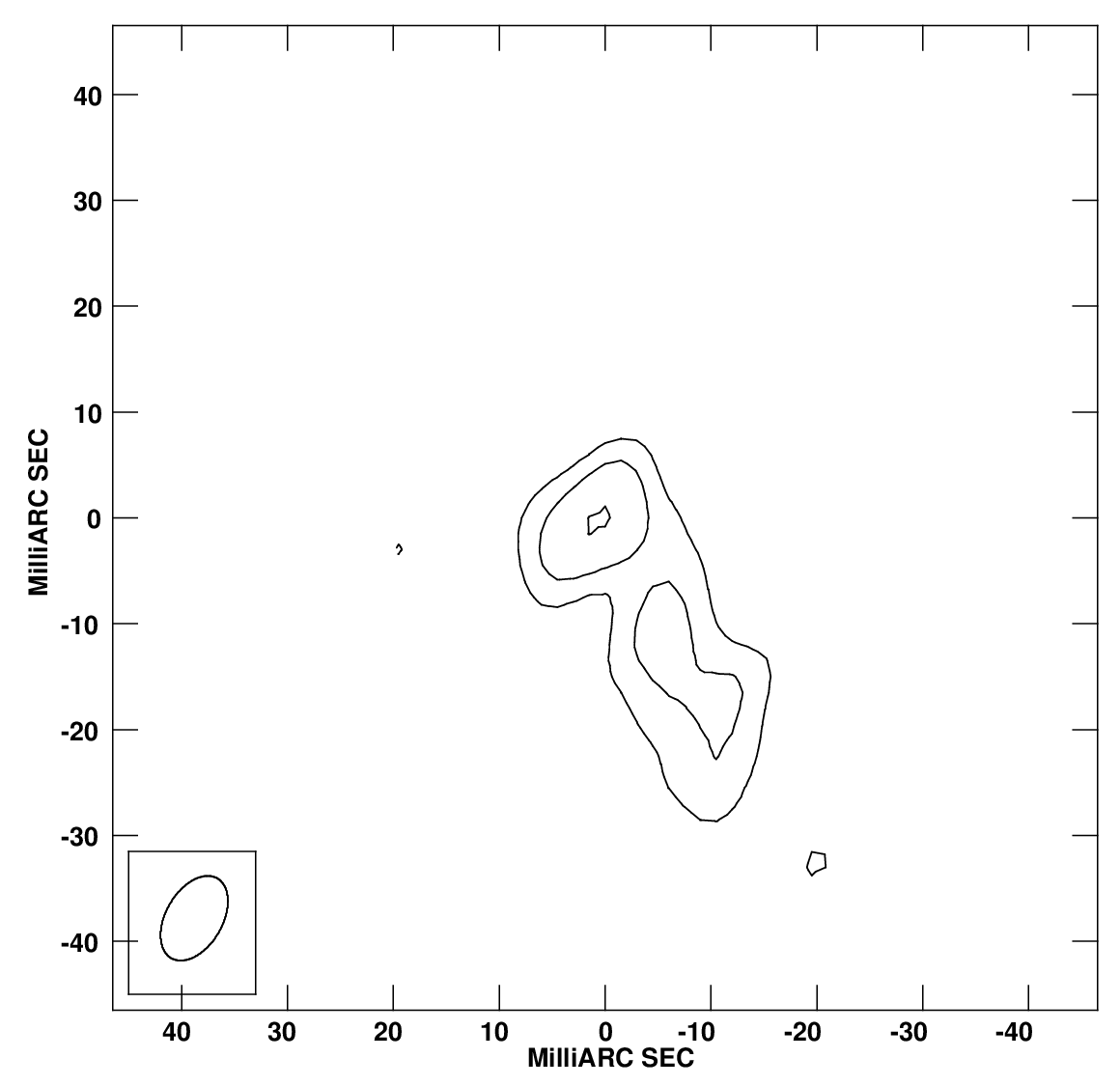,height=4.0cm,width=4.1cm,angle=-90}
\psfig{figure=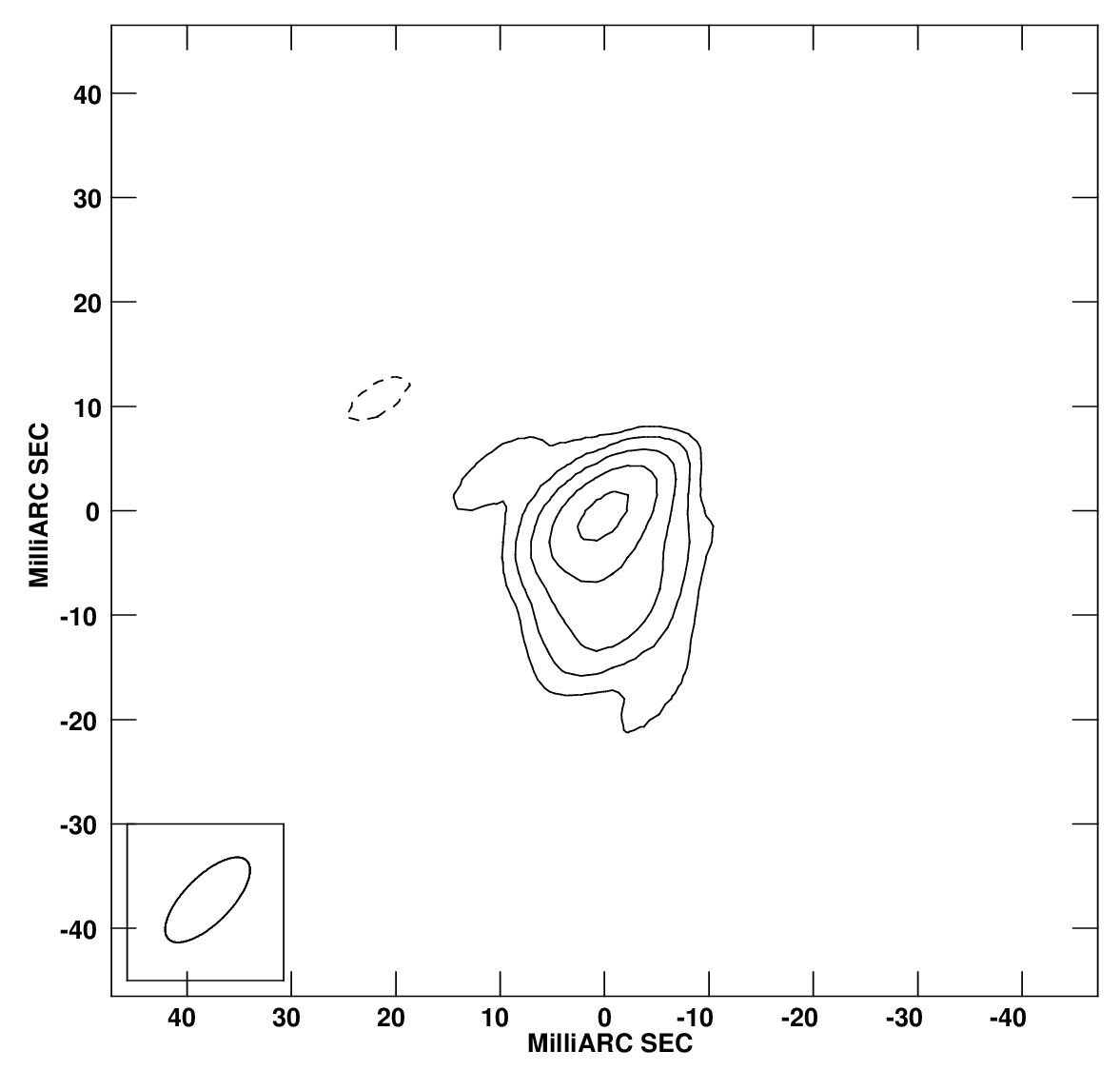,height=4.0cm,width=4.1cm,angle=-90}
\psfig{figure=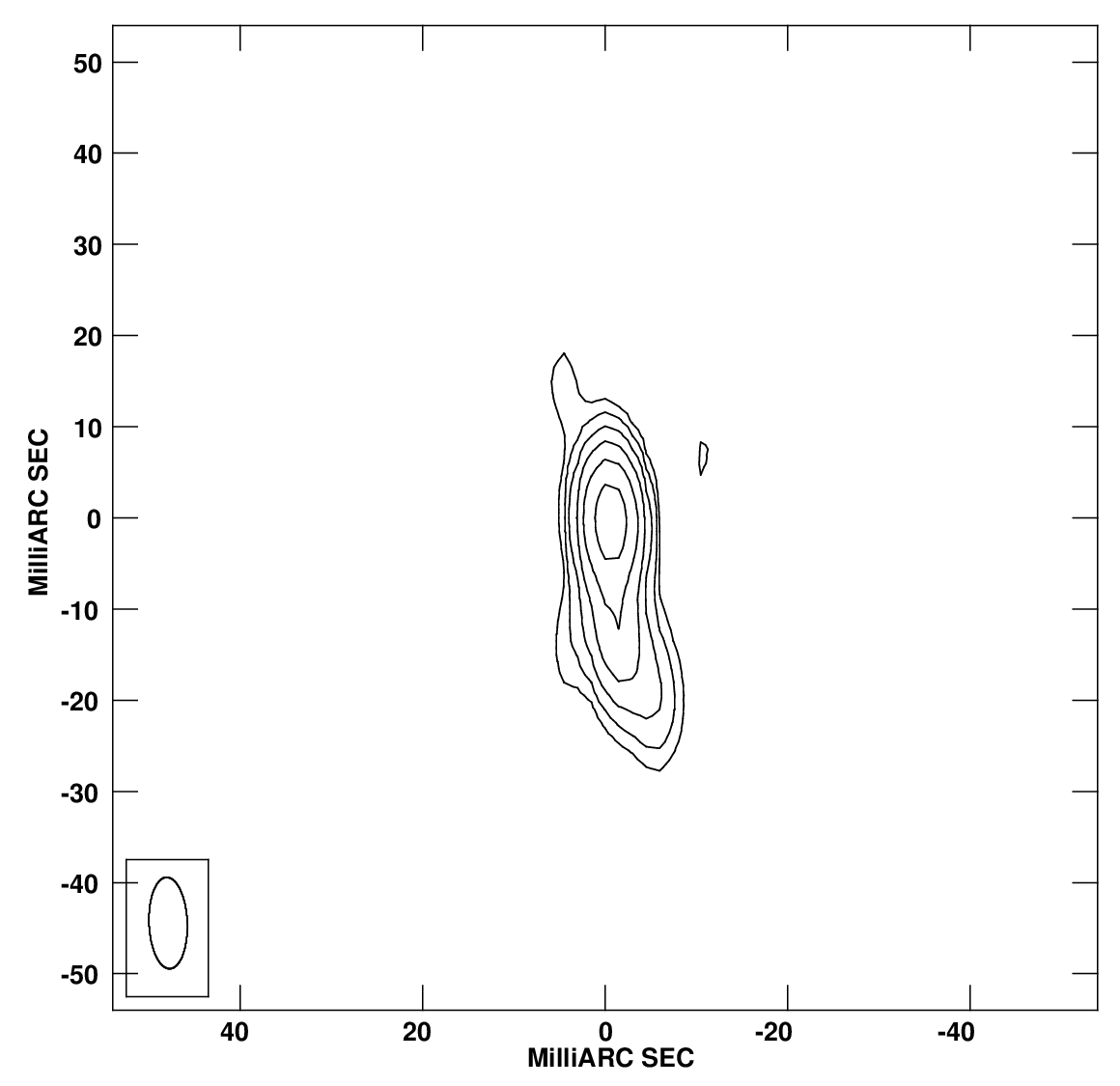,height=4.0cm,width=4.1cm,angle=-90}
}
\hbox{
\psfig{figure=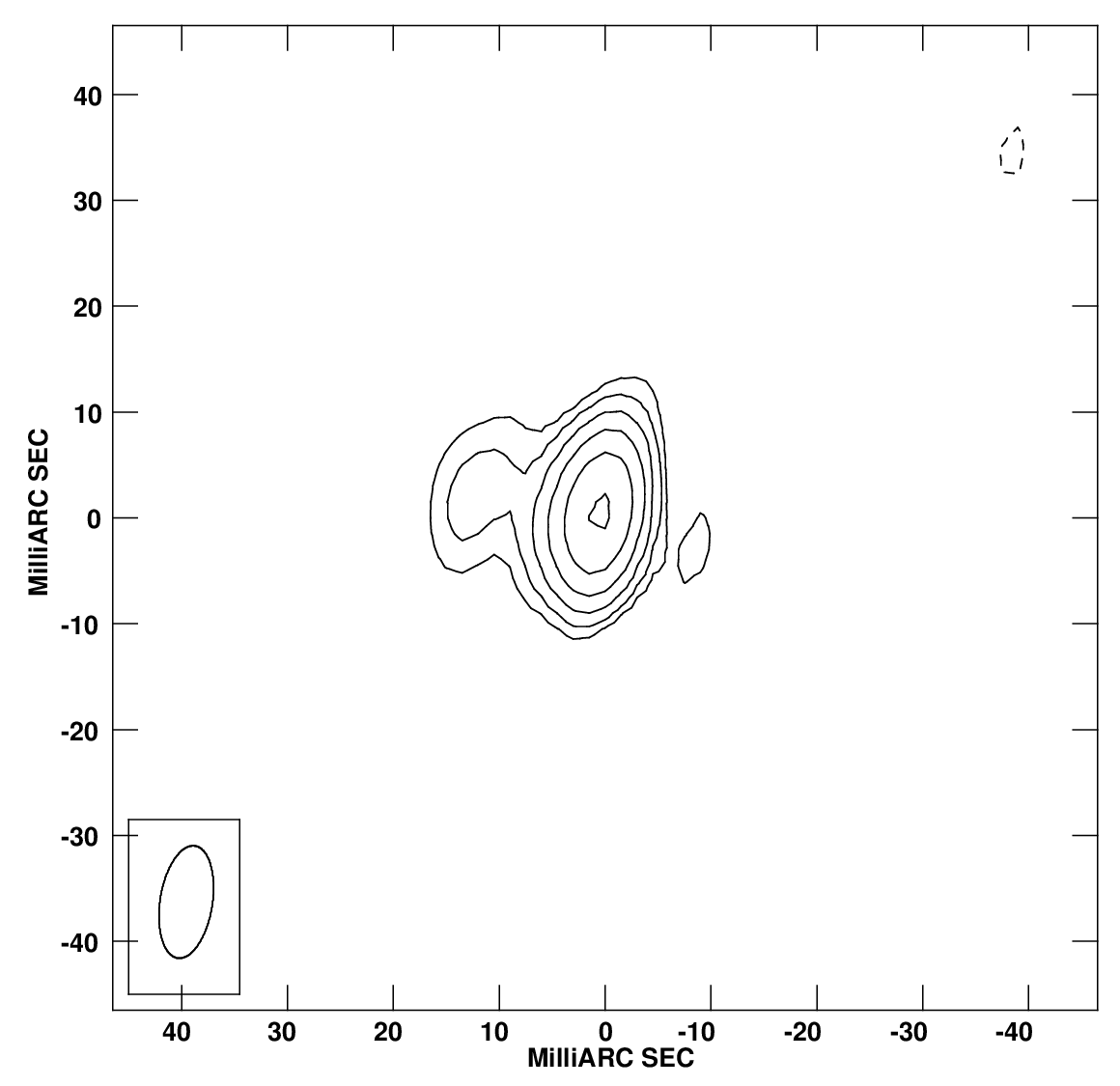,height=4.0cm,width=4.1cm,angle=-90}
\psfig{figure=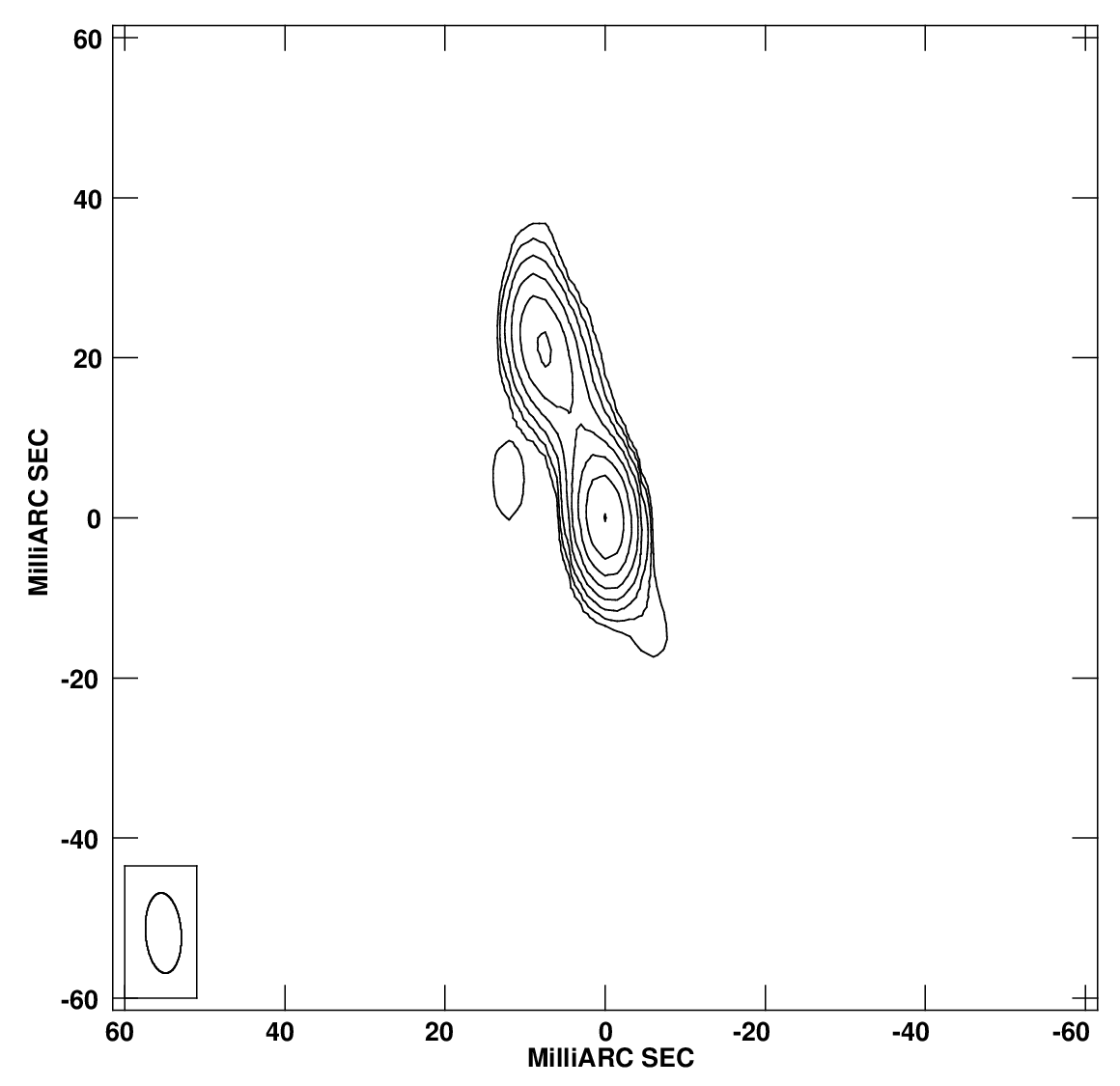,height=4.0cm,width=4.1cm,angle=-90}
\psfig{figure=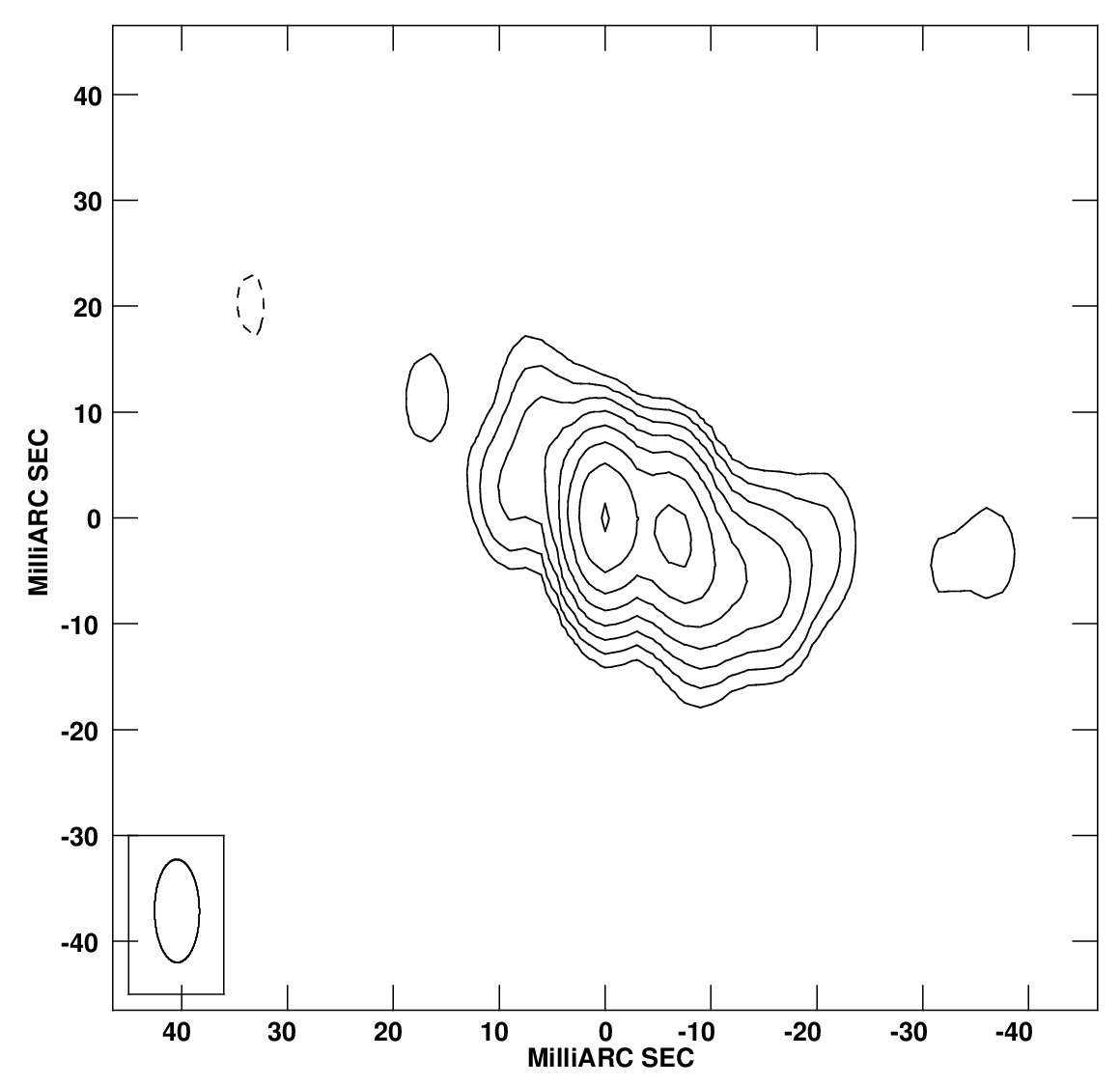,height=4.0cm,width=4.1cm,angle=-90}
\psfig{figure=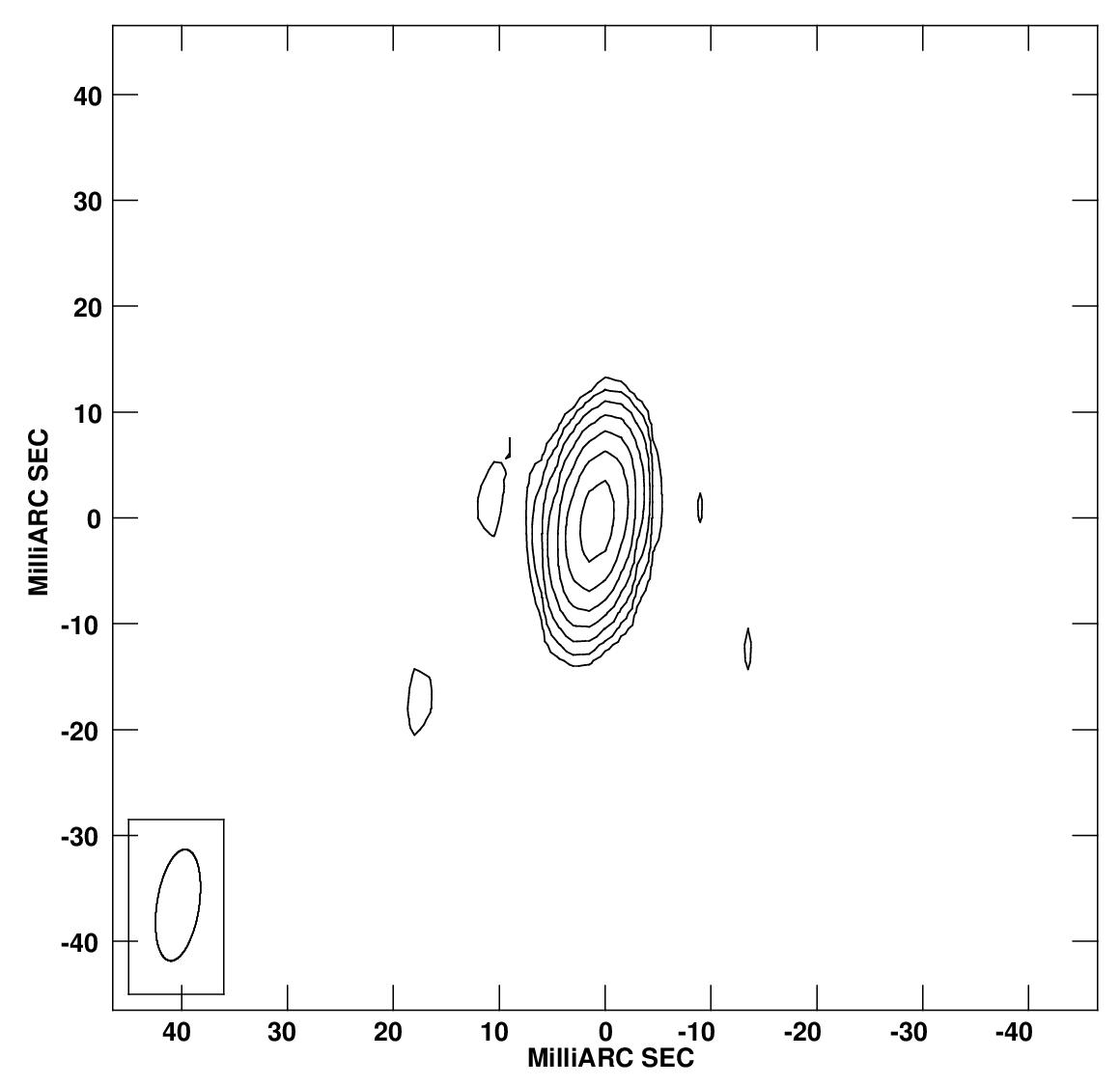,height=4.0cm,width=4.1cm,angle=-90}
}
\hbox{
\psfig{figure=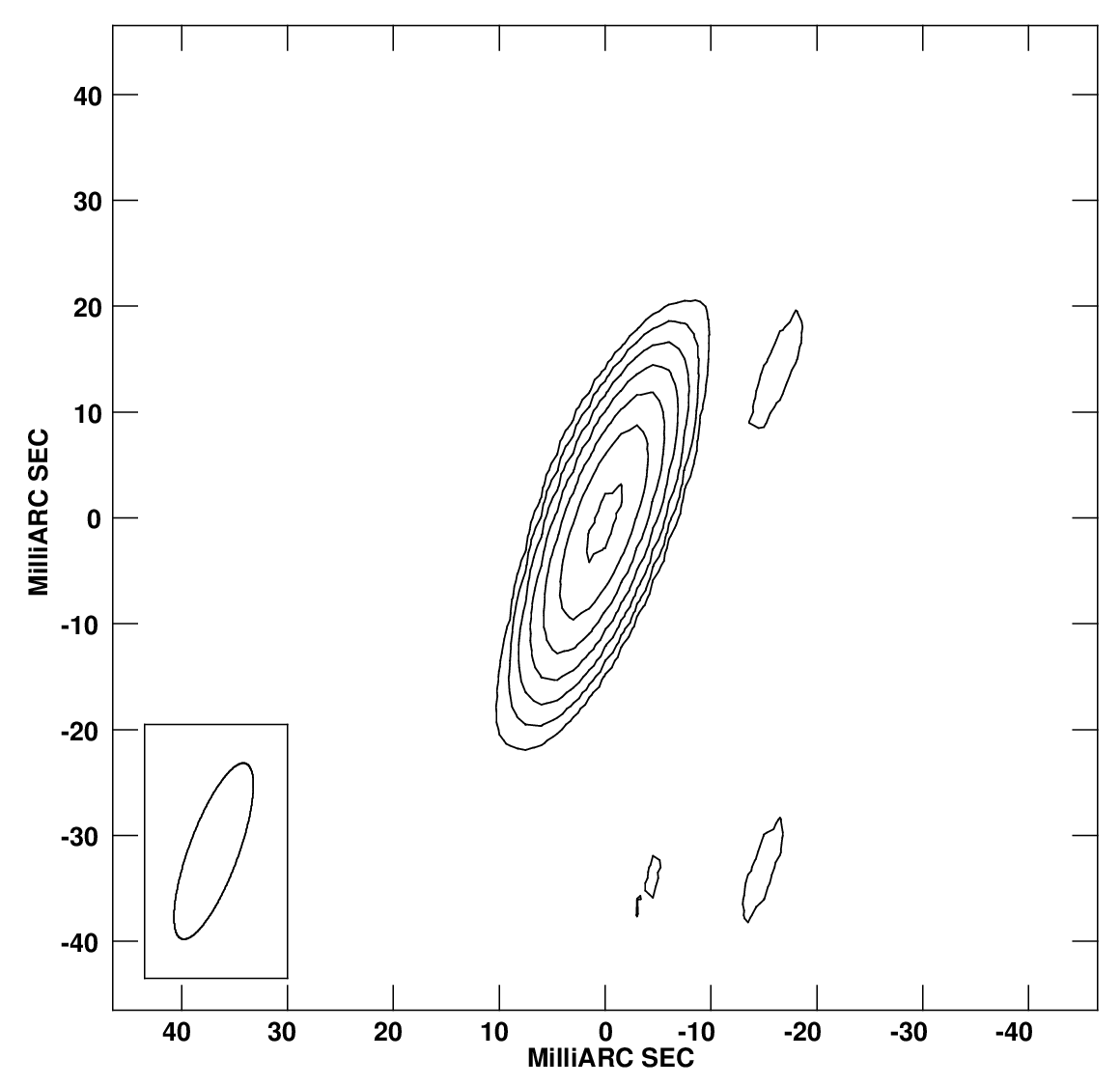,height=4.0cm,width=4.1cm,angle=-90}
\psfig{figure=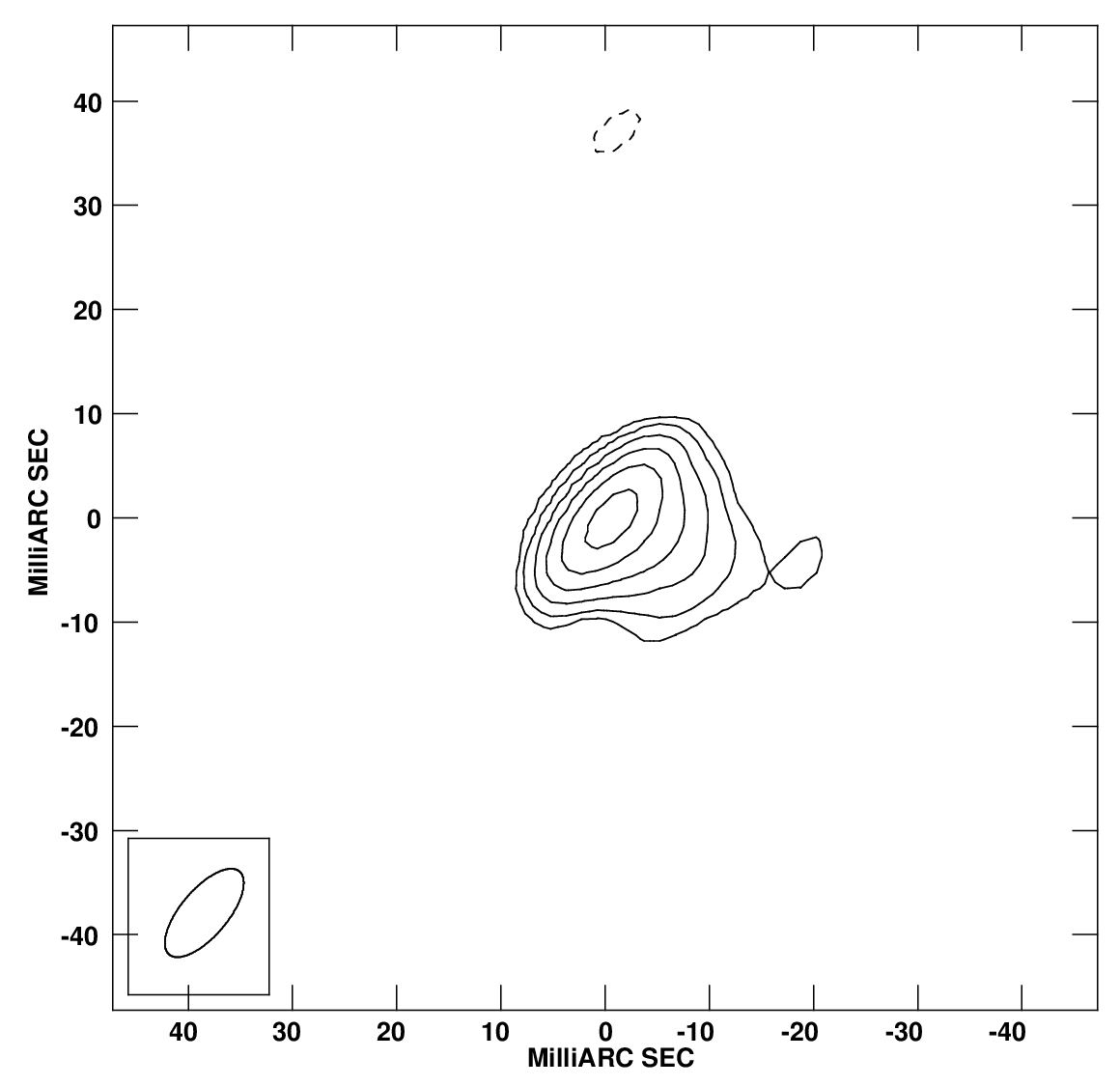,height=4.0cm,width=4.1cm,angle=-90}
\psfig{figure=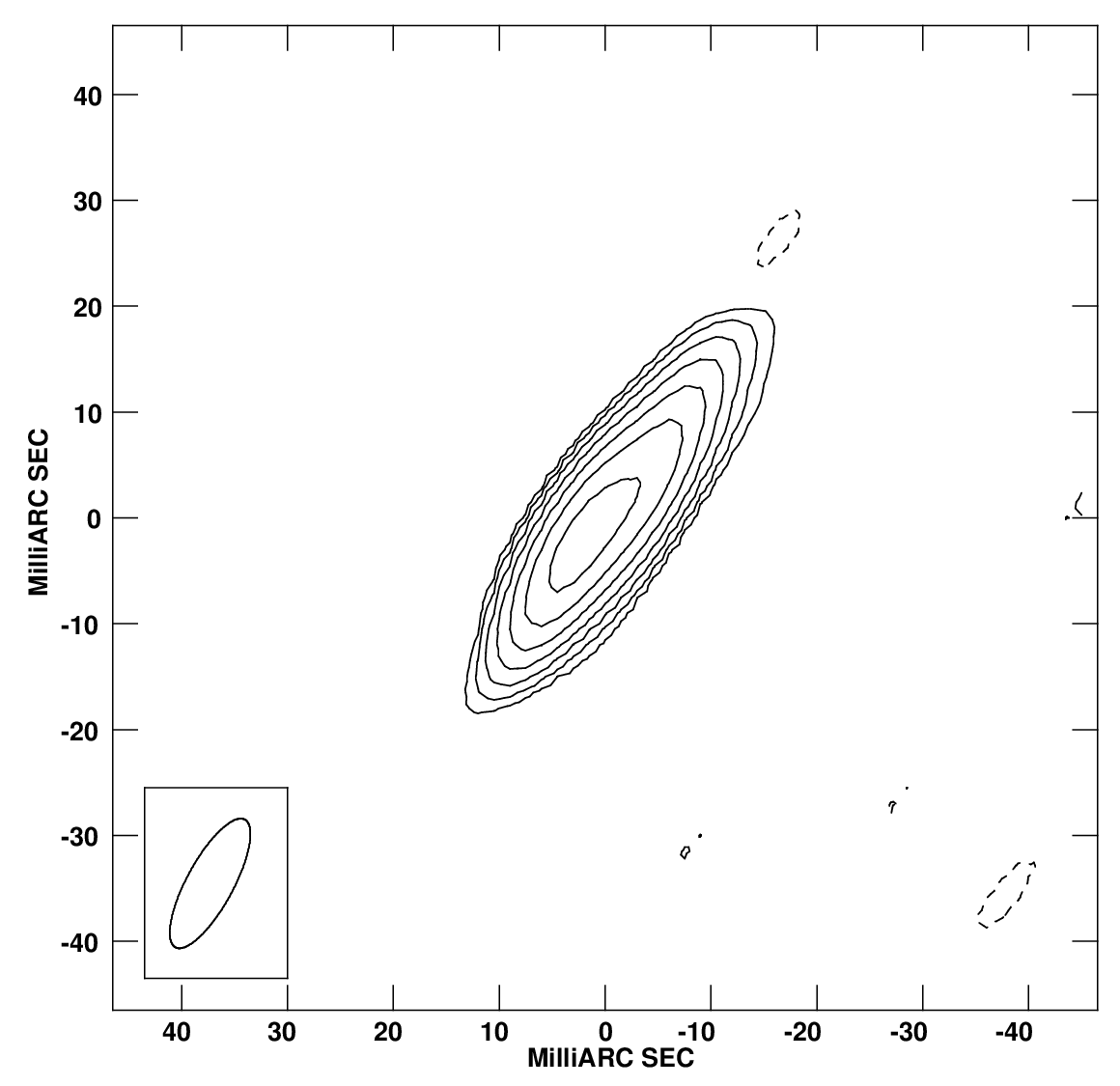,height=4.0cm,width=4.1cm,angle=-90}
\psfig{figure=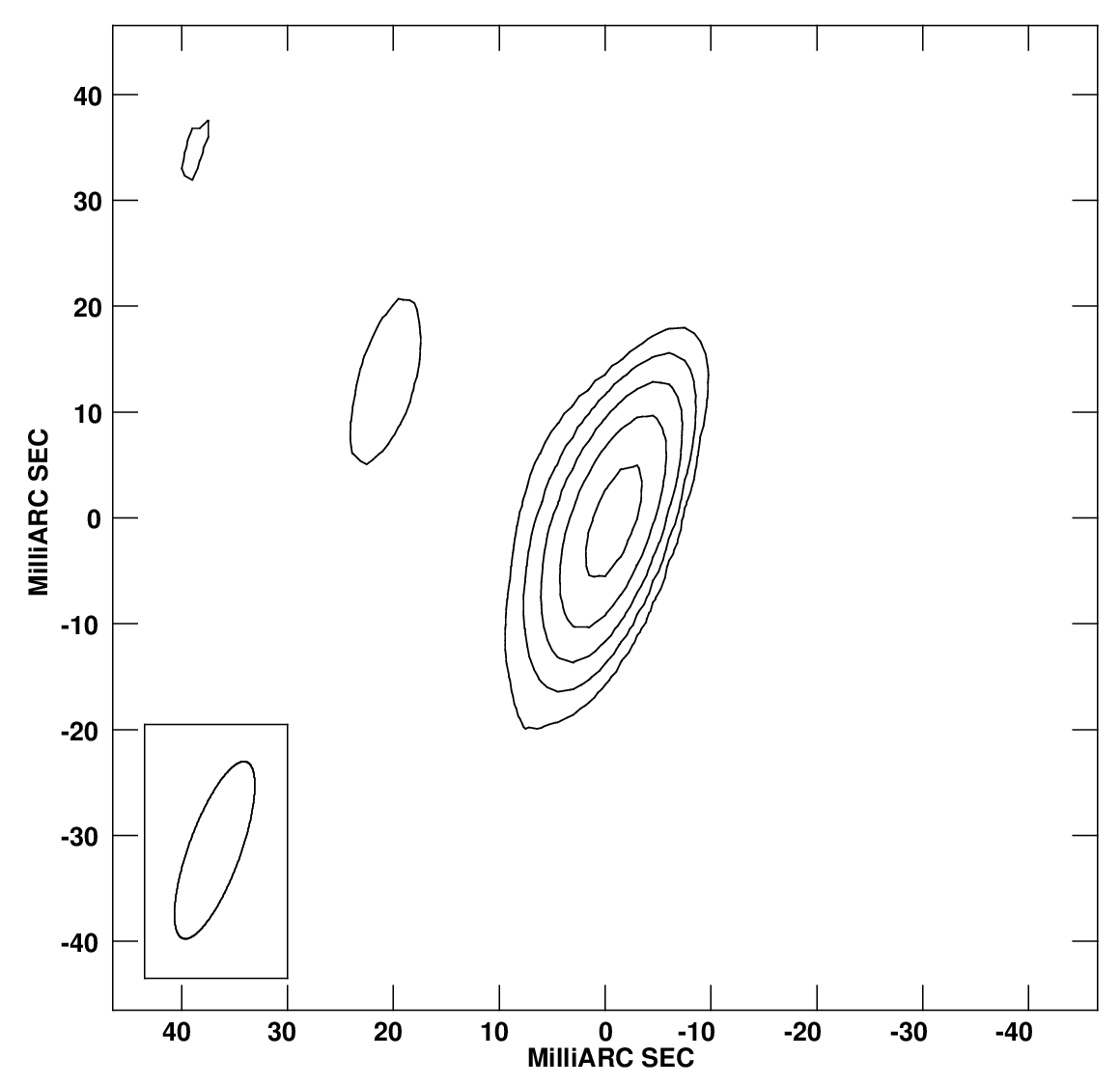,height=4.0cm,width=4.1cm,angle=-90}
}
\hbox{
\psfig{figure=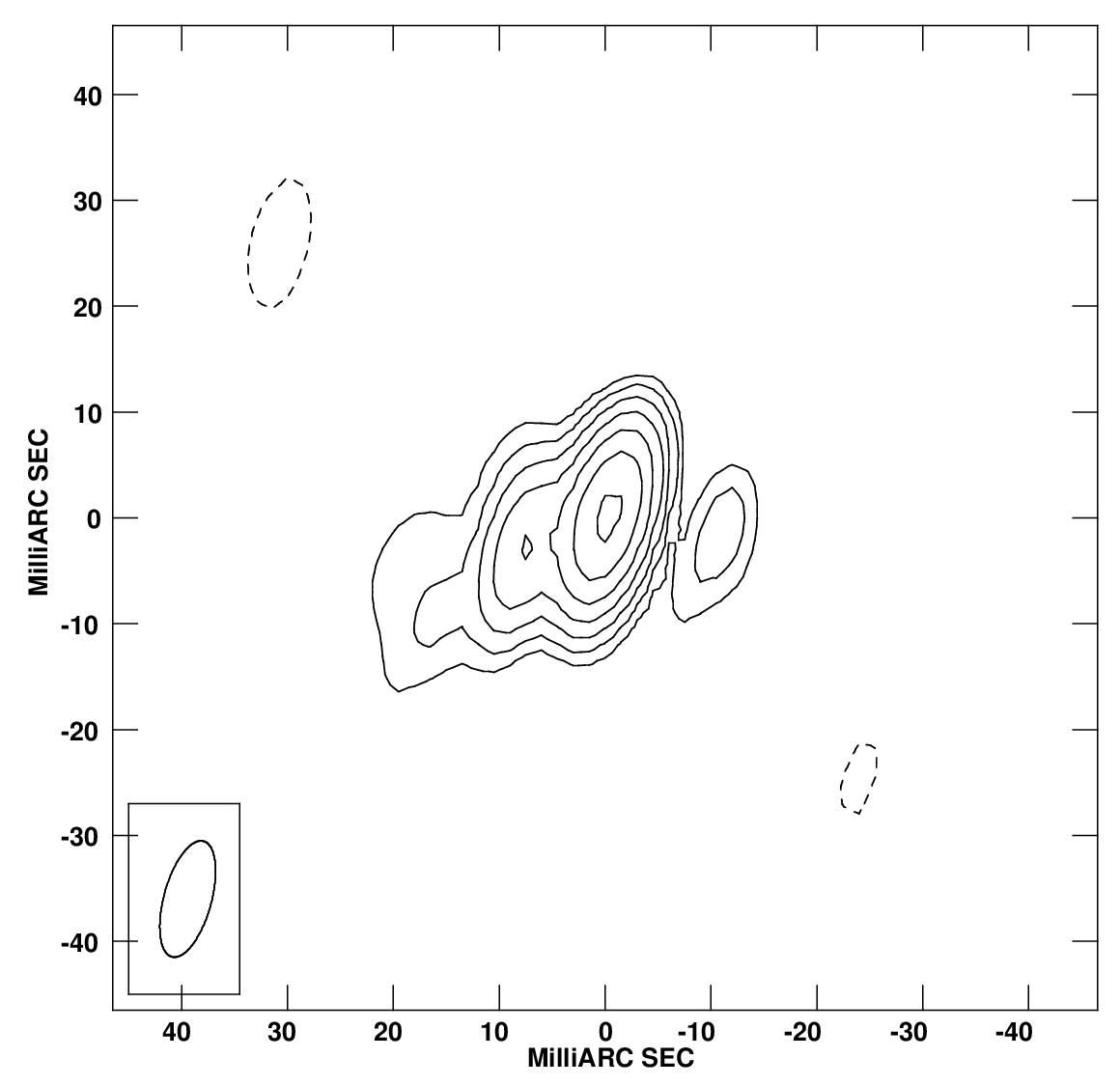,height=4.0cm,width=4.1cm,angle=-90}
\psfig{figure=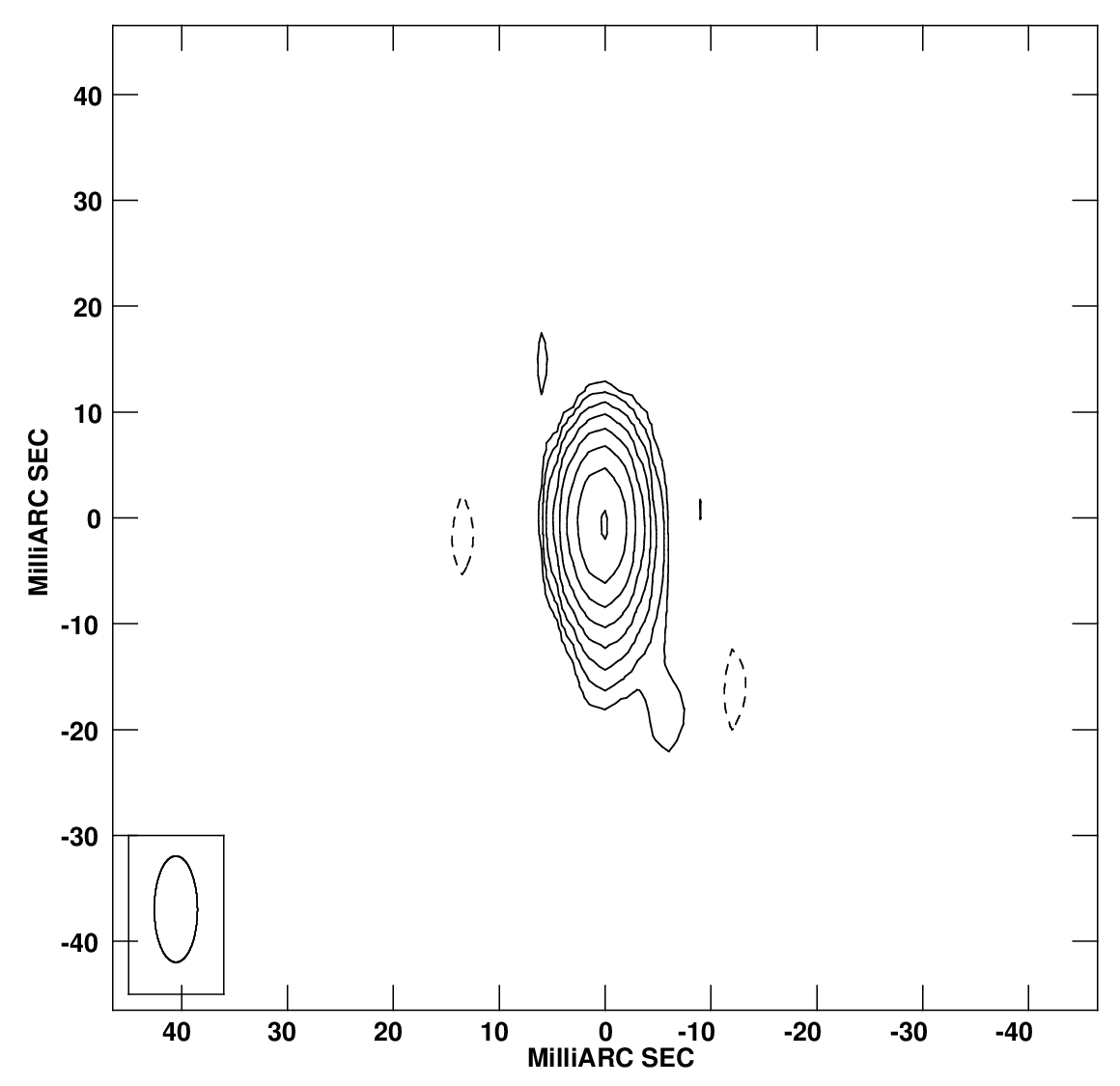,height=4.0cm,width=4.1cm,angle=-90}
\psfig{figure=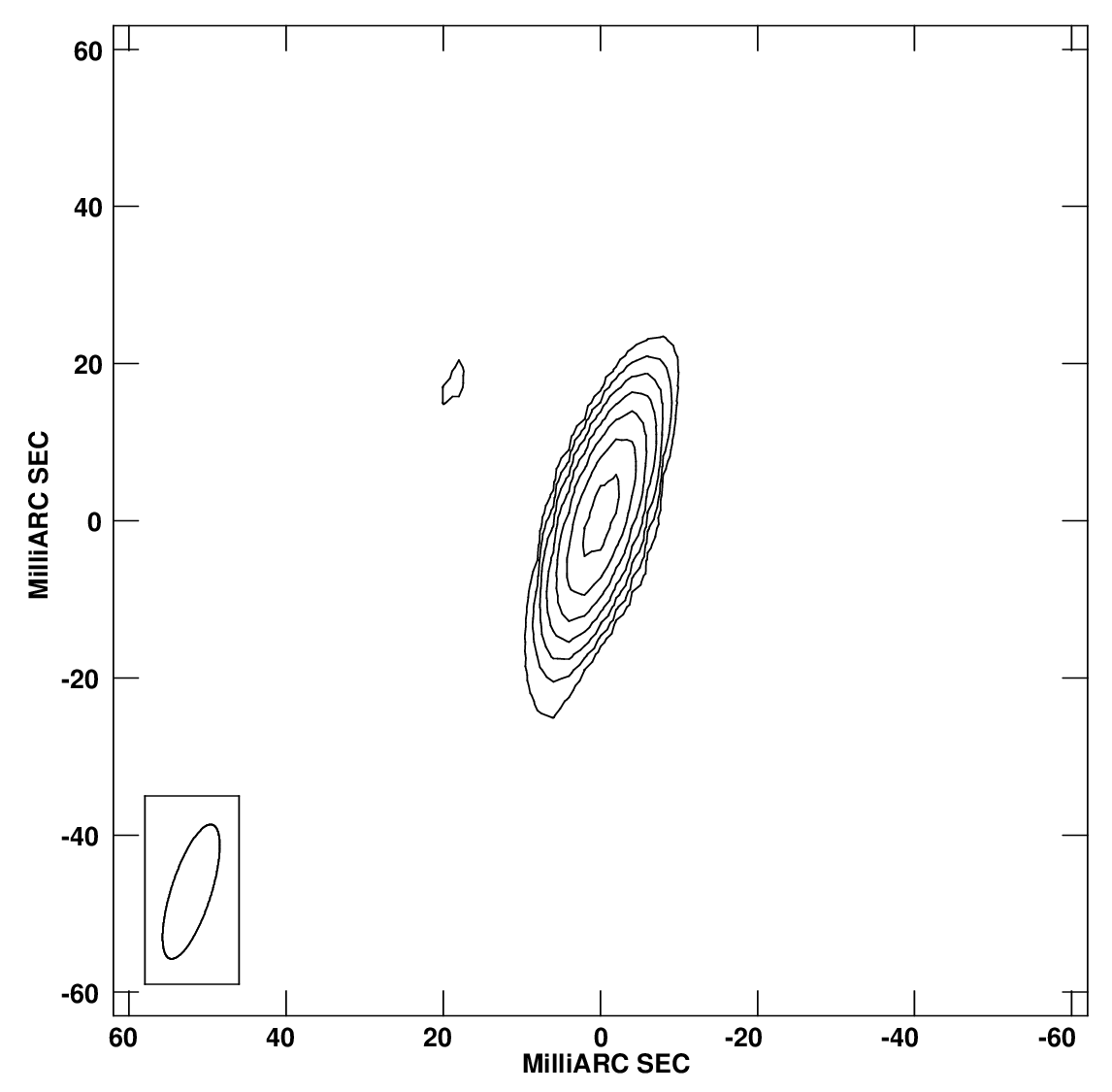,height=4.0cm,width=4.1cm,angle=-90}
\psfig{figure=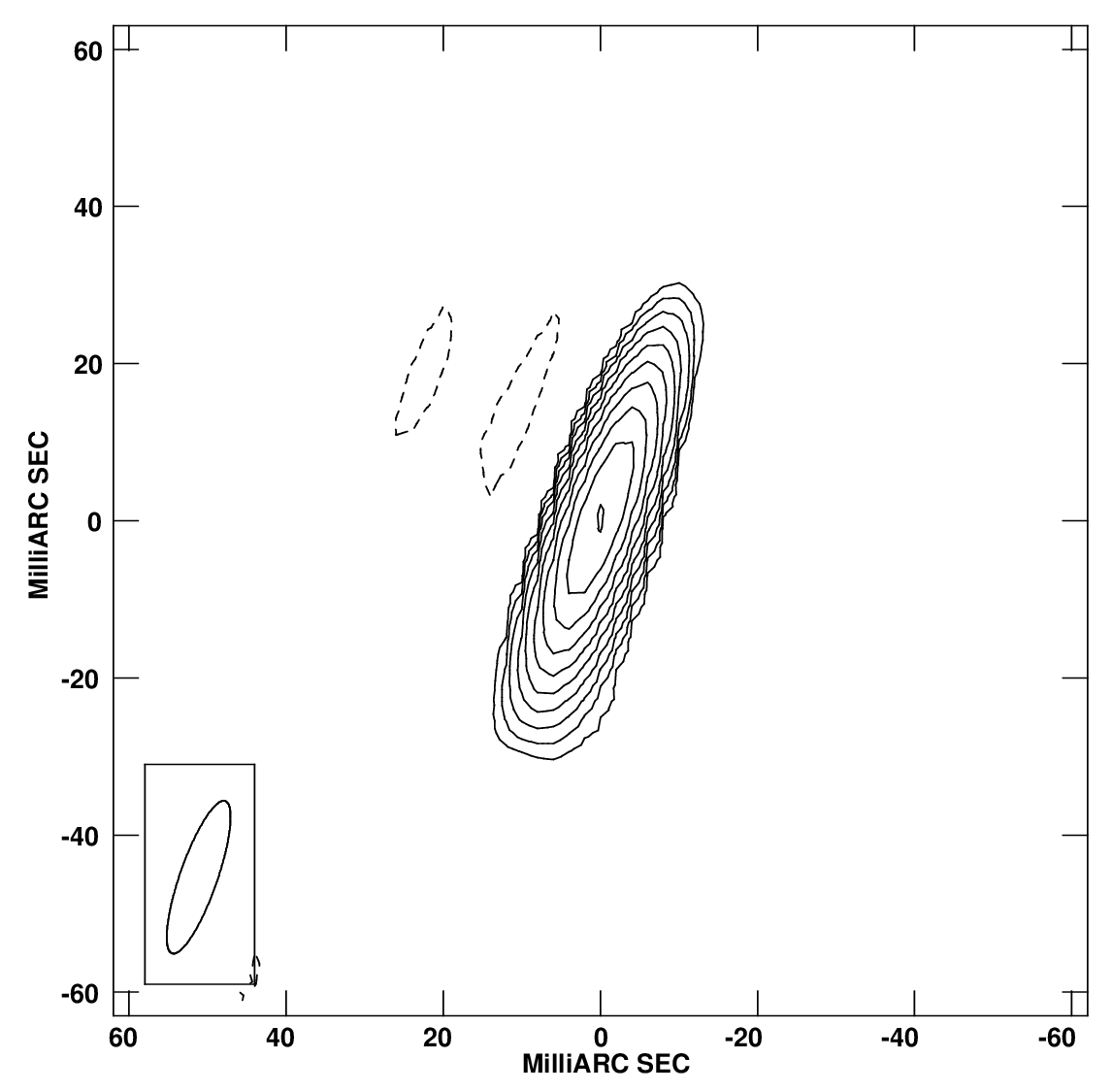,height=4.0cm,width=4.1cm,angle=-90}
}
}
}
\caption[]{ {\sl Continued}.
} 
\vskip -21.0cm
\begin{picture}(400,400)(0,0)
\put( 040,378){\scriptsize J1126$+$4516}
\put( 105,300){\tiny       CL = 4.0}
\put( 060,293){\tiny       B:0.011$^{\prime\prime}$$\times$0.005$^{\prime\prime}$,$-$28$^\circ$ }
\put( 160,378){\scriptsize J1148$+$5254}
\put( 225,300){\tiny       CL = 0.6}
\put( 180,293){\tiny       B:0.009$^{\prime\prime}$$\times$0.005$^{\prime\prime}$,$+$22$^\circ$ }
\put( 280,378){\scriptsize J1157$+$1638}
\put( 345,300){\tiny       CL = 3.0}
\put( 300,293){\tiny       B:0.010$^{\prime\prime}$$\times$0.005$^{\prime\prime}$,$-$4$^\circ$ }
\put( 400,378){\scriptsize J1208$+$5441}
\put( 465,300){\tiny       CL = 3.8}
\put( 420,293){\tiny       B:0.009$^{\prime\prime}$$\times$0.005$^{\prime\prime}$,$-$33$^\circ$ }
\put( 040,262){\scriptsize J1213$+$1307}
\put( 105,184){\tiny       CL = 5.0}
\put( 065,177){\tiny       B:0.010$^{\prime\prime}$$\times$0.004$^{\prime\prime}$,$-$8$^\circ$ }
\put( 160,262){\scriptsize J1232$+$5722}
\put( 225,184){\tiny       CL = 0.8}
\put( 180,177){\tiny       B:0.009$^{\prime\prime}$$\times$0.005$^{\prime\prime}$,$-$30$^\circ$ }
\put( 280,262){\scriptsize J1234$+$6455}
\put( 345,184){\tiny       CL = 0.9}
\put( 300,177){\tiny       B:0.011$^{\prime\prime}$$\times$0.005$^{\prime\prime}$,$-$45$^\circ$ }
\put( 400,262){\scriptsize J1300$+$0828}
\put( 465,184){\tiny       CL = 0.9}
\put( 425,177){\tiny       B:0.010$^{\prime\prime}$$\times$0.004$^{\prime\prime}$,$+$2$^\circ$ }
\put( 040,147){\scriptsize J1329$+$1053}
\put( 105,069){\tiny       CL = 0.8}
\put( 065,062){\tiny       B:0.011$^{\prime\prime}$$\times$0.005$^{\prime\prime}$,$-$9$^\circ$ }
\put( 160,147){\scriptsize J1333$+$1649}
\put( 225,069){\tiny       CL = 1.6}
\put( 185,062){\tiny       B:0.010$^{\prime\prime}$$\times$0.005$^{\prime\prime}$,$+$4$^\circ$ }
\put( 280,147){\scriptsize J1408$-$0752}
\put( 345,069){\tiny       CL = 1.6}
\put( 300,062){\tiny       B:0.010$^{\prime\prime}$$\times$0.004$^{\prime\prime}$,$-$0$^\circ$ }
\put( 400,147){\scriptsize J1410$+$6141}
\put( 465,069){\tiny       CL = 1.0}
\put( 425,062){\tiny       B:0.011$^{\prime\prime}$$\times$0.004$^{\prime\prime}$,$-$9$^\circ$ }
\put( 040,032){\scriptsize J1430$+$1043}
\put( 105,-45){\tiny       CL = 3.8}
\put( 060,-52){\tiny       B:0.018$^{\prime\prime}$$\times$0.005$^{\prime\prime}$,$-$20$^\circ$ }
\put( 160,032){\scriptsize J1501$+$5619}
\put( 225,-45){\tiny       CL = 0.9}
\put( 180,-52){\tiny       B:0.010$^{\prime\prime}$$\times$0.005$^{\prime\prime}$,$-$40$^\circ$ }
\put( 280,032){\scriptsize J1508$+$3347}
\put( 345,-45){\tiny       CL = 0.8}
\put( 300,-52){\tiny       B:0.014$^{\prime\prime}$$\times$0.004$^{\prime\prime}$,$-$29$^\circ$ }
\put( 400,032){\scriptsize J1623$+$0718}
\put( 465,-45){\tiny       CL = 1.0}
\put( 420,-52){\tiny       B:0.018$^{\prime\prime}$$\times$0.005$^{\prime\prime}$,$-$20$^\circ$ }
\put( 040,-84){\scriptsize J1636$+$2112}
\put( 105,-160){\tiny       CL = 1.8}
\put( 060,-167){\tiny       B:0.011$^{\prime\prime}$$\times$0.005$^{\prime\prime}$,$-$15$^\circ$ }
\put( 160,-84){\scriptsize J2031$+$1219}
\put( 225,-160){\tiny       CL = 3.0}
\put( 185,-167){\tiny       B:0.010$^{\prime\prime}$$\times$0.004$^{\prime\prime}$,$-$0$^\circ$ }
\put( 280,-84){\scriptsize J2340$-$0053}
\put( 345,-160){\tiny       CL = 1.2}
\put( 300,-167){\tiny       B:0.018$^{\prime\prime}$$\times$0.005$^{\prime\prime}$,$-$18$^\circ$ }
\put( 400,-84){\scriptsize J2358$-$1020}
\put( 465,-160){\tiny       CL = 1.0}
\put( 420,-167){\tiny       B:0.021$^{\prime\prime}$$\times$0.005$^{\prime\prime}$,$-$19$^\circ$ }
\end{picture}
\vskip +10.0cm
\label{vlbamap}
\end{figure*}

\end{appendix}

\end{document}